\documentclass[prc,aps,eqsecnum,amssymb,floatfix]{revtex4}
\usepackage{bm}
\usepackage{graphicx}
\tighten
\newcommand{\bna}{{\bm\nabla}}
\newcommand{\bfr}{{\bm r}}
\newcommand{\bfR}{{\bm R}}
\newcommand{\bfk}{{\bm k}}
\newcommand{\bfq}{{\bm q}}
\newcommand{\bfp}{{\bm p}}
\newcommand{\bfA}{{\bm A}}
\newcommand{\bfj}{{\bm j}}
\newcommand{\bfL}{{\bm L}}
\newcommand{\bfs}{{\bm s}}
\newcommand{\bfS}{{\bm S}}
\newcommand{\bfss}{{\bf S}}
\newcommand{\bfx}{{\bm x}}
\newcommand{\bfy}{{\bm y}}
\def\a{\alpha}
\begin{document}
\draft
\title{Electromagnetic structure of $A$=2 and 3 nuclei and
the nuclear current operator}
\author{
L.E. Marcucci$^{(a,b)}$, M. Viviani$^{(b,a)}$, R. Schiavilla$^{(c,d)}$, 
A. Kievsky$^{(b,a)}$, and S. Rosati$^{(a,b)}$}
\affiliation{
(a) Department of Physics ``E. Fermi'', University of Pisa, 
I-56127 Pisa, Italy\\
(b) INFN, Sezione di Pisa, I-56100 Pisa, Italy\\
(c) Department of Physics, Old Dominion University, Norfolk, Virginia 23529, 
USA \\
(d) Jefferson Lab, Newport News, Virginia 23606, USA 
}
\date{\today}

\begin{abstract}

Different models for conserved two- and three-body
electromagnetic currents are constructed from two- and
three-nucleon interactions, using either meson-exchange 
mechanisms or minimal substitution in the 
momentum dependence of these interactions. 
The connection between these two different schemes is 
elucidated. A number of low-energy electronuclear 
observables, including (i) $np$ radiative capture at 
thermal neutron energies and deuteron photodisintegration 
at low energies, (ii) $nd$ and $pd$  radiative capture 
reactions, and (iii) isoscalar and isovector magnetic form 
factors of $^3$H and $^3$He, are calculated in order to 
make a comparative study of these models for the current 
operator. The realistic Argonne $v_{18}$ two-nucleon and
Urbana IX or Tucson-Melbourne three-nucleon 
interactions are taken as a case study.
For $A$=3 processes, the bound and continuum wave functions, 
both below and above deuteron breakup threshold, are obtained 
with the correlated hyperspherical-harmonics method. 
Three-body currents give small but significant 
contributions to some of the polarization observables
in the $^2$H($p,\gamma$)$^3$He process and the  
$^2$H($n,\gamma$)$^3$H cross section at thermal 
neutron energies.
It is shown that the use of a current which did not exactly 
satisfy current conservation with the two- and three-nucleon 
interactions in the Hamiltonian was responsible 
for some of the discrepancies reported in previous studies 
between the experimental and theoretical polarization 
observables in $p$$d$ radiative capture.
\end{abstract}

\pacs{25.10.+s,25.40.Lw,24.70.+s,25.30.Bf}

\maketitle

\section{Introduction}
\label{sec:intro}

The present study investigates a number of different models
for the nuclear electromagnetic current derived from realistic
interactions.  The emphasis is on constructing two- and 
three-body currents which satisfy the current conservation relation 
(CCR) with the corresponding two- and three-nucleon interactions.  
Two different methods are 
adopted to achieve this goal: one is based on meson-exchange mechanisms,
the other uses minimal substitution in the explicit and 
implicit--through
the isospin exchange operator--momentum dependence of the interactions.
A by-product of this analysis is, in particular, the elucidation of
the sense in which these two different methods are related to each other.

A variety of electromagnetic observables involving the $A$=2 and 3 nuclei
are taken as case study for these current operator models, including
the $n$$p$ radiative capture at thermal neutron energy, the deuteron
photodisintegration at low energy, 
the magnetic form factors of $^3$He and $^3$H, and the
$n$$d$ and $p$$d$ radiative captures.  These processes have been
extensively studied in the past by several research groups (for
a review, see Ref.~\cite{Car98}).  Most recently, the authors of
the present paper (and collaborators) have investigated the $A$=3
radiative capture reactions below deuteron breakup threshold in
Refs.~\cite{Viv96,Viv00}, and the trinucleon form factors in
Ref.~\cite{Mar98}.  Below, we briefly review those aspects of
these earlier works that are more pertinent to the present study.

The $A$=3 bound- and scattering-state wave functions were
obtained using the pair-correlated hyperspherical harmonics
(PHH) method~\cite{Kie93,Kie94,Kie01} from a realistic Hamiltonian
model consisting of the Argonne $v_{18}$ two-nucleon~\cite{Wir95}
and Urbana IX three-nucleon interaction~\cite{Pud95} (AV18/UIX).
This technique allows for the inclusion of the Coulomb
interaction in both the bound and scattering states.
The nuclear electromagnetic current operator included, in addition
to one-body convection and spin-magnetization terms, two- and
three-body terms.  The dominant two-body terms were constructed using
meson-exchange mechanisms~\cite{Ris85a} from the momentum-independent
part of the AV18, including the long-range pion-exchange component,
and coincide with those derived here within the same approach.
They satisfy the CCR with this part of the interaction.

The two-body currents originating from the spin-orbit
components of the AV18 were constructed using again meson-exchange
mechanisms~\cite{Car90} ($\sigma$ and $\omega$ exchanges for the
isospin-independent, and $\rho$ exchange for the isospin dependent
terms), while those from the quadratic momentum-dependent components
were obtained by gauging only the momentum operators~\cite{Sch89},
but ignoring the implicit momentum dependence which comes through
the isospin exchange operator.  The resulting currents are not
strictly conserved.  This lack of current conservation was pointed
out in Ref.~\cite{Sch89}, but has not been sufficiently emphasized
in subsequent papers, mostly because of the short-range character
of these currents and their generally small associated contributions
to photo- and electronuclear processes, for example see
Refs.~\cite{Viv96,Car90,Sch89,Sch91}.  The overcome of this limitation
is one of the main aims of this work.

Earlier studies, as well as the present one, also take into
account the two-body currents, associated with the $\rho\pi\gamma$
and $\omega\pi\gamma$ transition mechanisms and with the excitation of
intermediate $\Delta$ resonances (for a review, see again
Ref.~\cite{Car98}).  However, these currents are purely transverse, and
therefore unconstrained by the CCR.  They are not the focus of the
present work.  

The effects of $\Delta$-isobar degrees of freedom in nuclear electroweak
processes were studied more thoroughly in Refs.~\cite{Mar98,Sch92},
using two different approximations.  One was based on first-order
perturbation theory (referred to above), while the other retained 
explicit one- and two-$\Delta$ admixtures in the nuclear wave functions
via the transition-correlation-operator (TCO) method~\cite{Sch92}. 
This latter approach is inherently non-perturbative.  In particular,
it generates three-body currents~\cite{Mar98}, which are strictly not
consistent with the three-nucleon interaction in the Hamiltonian.
In the present work these currents will be derived directly from
the three-nucleon interaction, and will satisfy by construction
the CCR with it.

The derived new models for the electromagnetic current
are tested in this paper on a variety of $A$=2 and 3 processes. 
The predictions for the $np$ radiative capture and deuteron 
electrodisintegration
cross sections at low energies remain practically unchanged and
are in agreement with the experimental data. For $A$=3
the situation is more interesting since the two-body currents
play a very important role. For example,
in Refs.~\cite{Viv96,Viv00} it was found that two-body currents
play a crucial role
in reproducing the cross section and polarization observables
measured in $p$$d$ and $n$$d$ radiative captures.  However, some
significant discrepancies remained unresolved.  
In the $n$$d$ case, the theoretical prediction
for the total cross section at thermal energies exceeds the experimental
value by 14\%.  With the present model of the current, the overprediction
is reduced to 9\%.
The origin of this overestimate is still puzzling, particularly
in view of the fact that the astrophysical $S$-factor for the $p$$d$
radiative capture at zero energy is calculated to be
within 1\%~\cite{Mar04} of that extrapolated from cross section measurements
in the range $\simeq 2\div20$ keV~\cite{Lun02}.

In the $p$$d$ case, the calculated tensor observables $T_{20}$ and
$T_{21}$ at center-of-mass (c.m.) energy of 2 MeV~\cite{Viv00} were
found to be at variance with data.  In that same work, it was also
shown that  these observables are sensitive to the small (suppressed) 
contributions arising from electric dipole transitions between the
initial $p$$d$ $P$-wave scattering states with  spin channel
$S=3/2$ and the final $^3$He bound state.  When these contributions
were calculated in the long-wavelength-approximation (LWA) using
the Siegert form of the $E_1$ operator, the resulting tensor
observables were much closer to the experimental values.  
Since the year 2000, more accurate PHH wave functions have become available
for the $A$=3 nuclei, and the calculations for $pd$ radiative capture
could be extended at energies above deuteron breakup
threshold~\cite{Viv03,Mar03a,Mar03b}.  
In preliminary calculations~\cite{Viv03}, we found that also at 3.33
MeV the theory could not reproduce the precise data for 
the tensor polarization observables $T_{20}$ and $T_{21}$~\cite{Goe92}.
In the present work, it will be shown how the use of a conserved
current indeed removes the discrepancy between theory and experiment
for these observables. Furthermore, the calculation has been 
extended up to 20 MeV.
 
In Ref.~\cite{Mar98} it was shown that the theoretical predictions
for the magnetic form factors of $^3$He and $^3$H were in satisfactory
agreement with experimental data at low and moderate values of the
momentum transfer.  The first diffraction region, however, was poorly
reproduced by the theoretical calculation, especially in the $^3$He
case.  The three-body current operators, constructed within the TCO
approach, gave only very small contributions.  This discrepancy
is not resolved in the present study.

Alternative descriptions of the $A$=2 and 3 electromagnetic
processes have been also recently reported. A conserved current model
was developed by Arenoh\"ovel and collaborators~\cite{Buc85} and applied
to $A$=2 reactions~\cite{Aren00,Aren04}. Several groups are
studying electromagnetic processes in the three-nucleon
system. In Refs.~\cite{Golak00,Skib03}, the nucleons are taken
as interacting via two- and three-nucleon 
potentials. The electromagnetic currents are then
constructed using the meson exchange scheme for
satisfying the CCR, but only with a part of the
interaction. In Ref.~\cite{Schad01}, the meson
exchange currents are taken into account using the Siegert's
theorem~\cite{Sieg37}. No three-body currents are considered in these 
works. In Ref.~\cite{Delt04}, a nuclear model is employed which
allows for the excitation of a nucleon to a $\Delta$ isobar,
and the two-body forces and currents
are generated by the exchange of mesons.  The $\Delta$ excitation
yields also effective three-body forces and three-body currents.
However, this current model does not
satisfy exactly the CCR with the adopted Hamiltonian
as discussed in Ref.~\cite{Delt04}. 
Very recently, models of the currents derived from chiral
Lagrangians are starting to appear~\cite{BS04}. In all these
calculations, the Coulomb interaction between protons 
in the scattering state is disregarded. Note that, in spite
of the differences of the various approaches, the theoretical
predictions of Refs.~\cite{Skib03,Schad01,Delt04} and of this paper 
turn  out to be, for most of the observables, quantitatively quite 
similar. An example will be presented for $nd$ radiative
capture, where the theoretical results are free from
the uncertainty related to the omission of the
Coulomb interaction. Also other approaches, as the 
Lorentz integral transform technique~\cite{ELOT00}, have been 
applied to study the electromagnetic response of the
trinucleon systems.

This paper is organized into six sections and three appendices.
In Sec.~\ref{sec:j2b} and~\ref{sec:j3b} we discuss the model
for the two- and three-body current operators, respectively.
In Sec.~\ref{sec:wf}, we briefly review the PHH method for
the $pd$ and $nd$ scattering-state wave function, below and
above deuteron breakup threshold. In Sec.~\ref{sec:res} we
present results for the magnetic
structure of the $A$=3 nuclei, for the deuteron
photodisintegration cross section at low energy, for 
the $np$ radiative capture at thermal neutron energies, and
for $nd$ and $pd$ radiative captures at c.m. energies
up to 20 MeV.  Finally, in Sec.~\ref{sec:con}, we summarize
our conclusions. The connection between the meson-exchange and
minimal-substitution approaches is elaborated in 
Appendix~\ref{app:path}, while a collection of formulas for the two-body
current operators associated with the quadratic momentum-dependent
terms of the two-nucleon interaction, and for the three-body
current operators in configuration space, are given in the
Appendices~\ref{app:l2so2} and~\ref{app:j3br}.

\section{Two-body current}
\label{sec:j2b}

The nuclear electromagnetic charge, $\rho(\bfq)$, and current, 
$\bfj(\bfq)$, operators can be written as sums of one-, two-, and 
many-body terms that operate on the nucleon degrees of freedom:

\begin{eqnarray}
  \rho(\bfq)&=&\sum_i \rho_i(\bfq)+\sum_{i<j} \rho_{ij}(\bfq)+\ldots
   \ , \label{eq:charge} \\
  \bfj(\bfq)&=&\sum_i \bfj_i(\bfq)+\sum_{i<j} \bfj_{ij}(\bfq)+\ldots
   \ . \label{eq:current}
\end{eqnarray}
The one-body operators $\rho_i(\bfq)$ and 
$\bfj_i(\bfq)$ are derived from the non-relativistic reduction
of the covariant single-nucleon current, by expanding
in powers of $1/m$, $m$ being the nucleon mass.  In the
notation of Ref.~\cite{Car98}, the one-body charge operator in
configuration space is given by

\begin{equation}
  \rho_i(\bfq)=\rho_{i,{\rm NR}}(\bfq)+\rho_{i,{\rm RC}}(\bfq)\ ,
  \label{eq:cg_t}
\end{equation}
where the leading order term, labeled NR, is 

\begin{equation}
  \rho_{i,{\rm NR}}(\bfq)=\epsilon_i {\rm e}^{{\rm i} \bfq\cdot\bfr_i} \ ,
  \label{eq:chargenr}
\end{equation}
with

\begin{equation}
  \epsilon_i=\frac{1}{2}\left[G_E^S(q_\mu^2) 
 +G_E^V(q_\mu^2)\tau_{i,z}\right]\ ,\label{eq:ei}
\end{equation}
while the term labeled RC is proportional to $1/m^2$ and
is explicitly listed in Ref.~\cite{Car98}.  In Eq.~(\ref{eq:ei})
$G_E^S(q_\mu^2)$ and $G_E^V(q_\mu^2)$ are the 
isoscalar and isovector combinations of the nucleon
electric Sachs form factors, respectively, evaluated at the
four-momentum transfer $q_\mu^2$.

The electromagnetic current operator must satisfy the
current conservation relation (CCR)

\begin{equation}
  \bfq\cdot\bfj(\bfq)= [H,\rho(\bfq)]\ ,\label{eq:ccr}
\end{equation}
where the nuclear Hamiltonian $H$ is taken to consist, quite
generally, of two- and three-body interactions, denoted as
$v_{ij}$ and $V_{ijk}$ respectively,

\begin{equation}
  H=\sum_i {\bfp_i^2\over 2m}+\sum_{i<j} v_{ij} + \sum_{i<j<k}
     V_{ijk}\ .\label{eq:nucH}
\end{equation}
Realistic models for these interactions contain isospin- and 
momentum-dependent terms which do not commute with the charge
operators.  To lowest order in $1/m$, Eq.~(\ref{eq:ccr})
separates into

\begin{eqnarray}
  \bfq\cdot\bfj_i(\bfq)&=& \biggl[{\bfp_i^2\over 2m},
   \rho_{i,{\rm NR}}(\bfq)\biggr]
    \ ,\label{eq:ccr1}\\
  \bfq\cdot\bfj_{ij}(\bfq)&=& [v_{ij},\rho_{i,{\rm NR}}(\bfq)
   +\rho_{j,{\rm NR}}(\bfq)]
 \ , \label{eq:ccr2}
\end{eqnarray}
and similarly for the three-body current $\bfj_{ijk}(\bfq)$.
It has been tacitly assumed that two-body terms in $\rho(\bfq)$
are of order $1/m^2$.  The one-body current
is easily shown to satisfy Eq.~(\ref{eq:ccr1}).
However, it is rather difficult to construct conserved two- and
three-body currents.

It is useful to adopt the classification scheme of Ref.~\cite{Ris89},
and separate the current $\bfj_{ij}(\bfq)$ into model-independent (MI)
and model-dependent (MD) parts,

\begin{equation}
   \bfj_{ij}(\bfq)=\bfj_{ij}^{MI}(\bfq)+\bfj_{ij}^{MD}(\bfq)\ .
   \label{eq:jfull}
\end{equation}
The MI two-body current $\bfj_{ij}^{MI}(\bfq)$ has a longitudinal
component, constructed so as to satisfy the CCR of Eq.~(\ref{eq:ccr2}) 
(see following subsections), while the MD two-body current
$\bfj_{ij}^{MD}(\bfq)$ is purely transverse and therefore is
un-constrained by the CCR.  The latter will not be discussed any
further in the present section; it suffices to say that it is taken
to consist of the isoscalar $\rho\pi\gamma$
and isovector $\omega\pi\gamma$ transition currents, as well as of the
isovector current associated with excitation of intermediate
$\Delta$ resonances~\cite{Viv96,Viv00,Mar98}.

A method to derive $\bfj_{ij}^{MI}(\bfq)$ was developed by Riska
and collaborators~\cite{Ris85a,Sch89,Ris85b,Ris85c} and
Arenh\"ovel and collaborators~\cite{Buc85} 
(see Ref.~\cite{Car98} for a review). An alternative approach, which
we will revisit and generalize in the present work, is
based on ideas first proposed by Sachs in Ref.~\cite{Sac48}, 
and later applied by Nyman in Ref.~\cite{Nym67} to derive
the magnetic-dipole transition operator due to the
one-pion-exchange potential.  We will refer to these
two different approaches as the ``meson-exchange'' (ME) 
and ``minimal-substitution'' (MS) scheme, respectively.
To appreciate the differences and similarities between them, 
they are discussed in the two following subsections. 

In the rest of the paper, we will use the following notation:
a generic nucleon-nucleon interaction will be written as

 \begin{equation}
  v_{ij}=v^{IC}_{ij}+v^{IB}_{ij}\ , \qquad
  v^{IC}_{ij}=v^{0}_{ij}+v^{p}_{ij}\ ,
  \label{eq:vijtot}
\end{equation}
where $v^{IC}_{ij}$ and $v^{IB}_{ij}$ are the isospin-symmetry
conserving ($I$$C$) and breaking ($I$$B$) parts of the potential, respectively.
In turn, $v^{0}_{ij}$ and $v^{p}_{ij}$ are the momentum-independent
and momentum-dependent components of $v^{IC}_{ij}$, respectively.
The next two subsections deal with the $v^{IC}_{ij}$ part of the potential.
The two-body current associated with its $v^{IB}_{ij}$ part 
will be considered in Sec.~\ref{subsec:ist}.

For later reference in Sec.~\ref{sec:res}, a summary of the 
models for two-body current operators used in the present work is 
given in Sec.~\ref{subsec:sumj2}.

\subsection{The two-body current operator in the meson-exchange scheme }
\label{subsec:j2me}

First consider the isospin-conserving momentum-independent 
part of the potential $v^0_{ij}$, which can be written as

\begin{equation}
  v^0_{ij}=v_{1,ij}+v_{2,ij}\,{\bm \tau}_i\cdot{\bm \tau}_j \ ,
  \label{eq:vij}
\end{equation}
where ${\bm \tau}_i$ and ${\bm \tau}_j$ are the isospin Pauli matrices, 
and $v_1$ and $v_2$ are in general functions of the positions
and spin operators of the two nucleons; $v_2$ includes the long-range
one-pion-exchange component.  In particular, the isospin-dependent terms
are given by

\begin{equation}
  v_{2,ij}\,{\bm\tau}_i\cdot{\bm\tau}_j=
  [ v_\tau(r_{ij})+
  v_{\sigma\tau}(r_{ij})\,{\bm\sigma}_i\cdot{\bm\sigma}_j
  +v_{t\tau}(r_{ij})\,S_{ij}] \,{\bm\tau}_i\cdot{\bm\tau}_j\ ,
\label{eq:v2ij}
\end{equation}
where $S_{ij}$ is the standard tensor operator,
${\bm\sigma}_i$ and ${\bm\sigma}_j$ are the spin Pauli matrices, 
and the notation of Ref.~\cite{Wir84} is used.
In momentum-space, $v_2$ reads

\begin{equation}
  v_2(\bfk)\,{\bm\tau}_i\cdot{\bm\tau}_j=
  [v_{\tau}(k)+v_{\sigma\tau}(k)\,k^2\,{\bm\sigma}_i\cdot{\bm\sigma}_j+
  v_{t\tau}(k)S_{ij}(\bfk)]{\bm\tau}_i\cdot{\bm\tau}_j \ ,
\label{eq:vnn}
\end{equation}
where $v_{\tau}(k)$, $v_{\sigma\tau}(k)$ and $v_{t\tau}(k)$ are 
related to their configuration-space counter-parts by the relations

\begin{eqnarray}
  v_\tau(k)&=&4\pi \int_0^\infty r^2 dr\, j_0(kr)v_\tau(r) 
  \label{eq:vt}\>\>, \\
  v_{\sigma \tau}(k)&=&\frac{4\pi}{k^2} \int_0^\infty r^2 dr \, 
  \left [ j_0(kr)-1 \right ] v_{\sigma \tau}(r) \label{eq:vst}\>\>, \\
  v_{t \tau}(k)&=&\frac{4\pi}{k^2} 
  \int_0^\infty r^2 dr\, j_2(kr)v_{t\tau}(r) \label{eq:vtt} \>\>.
\end{eqnarray}
The factor $[j_0(kr)-1]$ in the
expression for $v_{\sigma \tau}(k)$ ensures that the volume integral 
of $v_{\sigma \tau}(r)$
vanishes~\cite{Sch89}, and the tensor operator in momentum-space is 
defined as $S_{ij}(\bfk)=k^{2}({\bm \sigma}_i\cdot{\bm\sigma}_j)-
3({\bm\sigma}_i\cdot\bfk)({\bm \sigma}_j\cdot\bfk)$.

If the isospin-dependent interaction $v_2(\bfk)$ is assumed to be
induced by $\pi$- and $\rho$-meson exchanges, as for example in the Bonn
model~\cite{Mac01}, then

\begin{equation}
v_2(\bfk)\,{\bm\tau}_i\cdot{\bm\tau}_j=
\Big[v_{\rho S}(k)+
   [2 v_\rho(k) + v_\pi(k)]\,k^2\,{\bm\sigma}_i\cdot{\bm\sigma}_j+
   [  v_\rho(k) - v_\pi(k)] S_{ij}(\bfk)\Big]\, {\bm\tau}_i\cdot{\bm\tau}_j \ ,
\label{eq:v6nn}
\end{equation}
with  

\begin{eqnarray}
  v_\pi(k) &=&
  -{\frac {f_{\pi NN}^2} {3m_{\pi}^2} }{\frac {f^2_\pi(k)} { k^2+
  m_{\pi}^2} } \ , \label{eq:vpi} \\
  v_\rho(k) &=& -{\frac {g_{\rho NN}^{2}
  (1+\kappa_{\rho NN})^2}{12 m^2} } {\frac {f^2_\rho(k)} { k^2+
  m_{\rho}^2} } \label{eq:vrho} \ , \\
  v_{\rho S} &=& g_{\rho NN}^{2}
  \frac{f^2_\rho(k)} { k^2+m_{\rho}^2} \ ,
\label{eq:vrhos} 
\end{eqnarray}
where $f_{\pi NN}$, and $g_{\rho NN}$ and $\kappa_{\rho NN}$
are the coupling constants of the $\pi$- and $\rho$-mesons,
$f_\pi(k)$ and $f_\rho(k)$ are the associated form factors
(usually, of monopole type), $m_{\pi}$ and $m_{\rho}$ are
their masses, and finally $m$ is the nucleon mass.

More generally, if one assumes that the interaction $v_2(\bfk)$ is
due to the exchange of a number of ``$\pi$-like'' pseudoscalar ($PS$)
and ``$\rho$-like'' vector ($V$) mesons, then one finds 

\begin{equation}
v_2(\bfk)\,{\bm\tau}_i\cdot{\bm\tau}_j=
\Big[v_{VS}(k)+
   [2 v_{V}(k) + v_{PS}(k)]\,k^2\,{\bm\sigma}_i\cdot{\bm\sigma}_j+
   [  v_{V}(k) - v_{PS}(k)] S_{ij}(\bfk)\Big]\, {\bm\tau}_i\cdot{\bm\tau}_j \ ,
\label{eq:v6vps}
\end{equation}
where the functions $v_{PS}(k)$, $v_{V}(k)$, and $v_{VS}(k)$ 
are given by

\begin{eqnarray}
  v_{PS}(k)&=& \sum_{a=1}^{N} f^2_{PS,a} 
                  \frac{1} { k^2+m_a^2} \>\>\>, 
\label{eq:vpsdef1} \\
  v_{V}(k) &=&  \sum_{a=1}^{N} f^2_{V,a} 
                \frac{1} { k^2+m_a^2}  \>\>\>,  
\label{eq:vvdef1} \\
  v_{VS}(k) &=& \sum_{a=1}^{N} f^2_{VS,a} 
                  \frac{1} { k^2+m_a^2}
              \>\>\> .
\label{eq:vvsdef1}
\end{eqnarray}
In the expressions above, $m_a$ is the mass
and $f^2_{PS,a}\equiv-f^2_{\pi NN,a}/3m^2_a$, 
$f^2_{V,a}\equiv -g_{\rho NN,a}^{2}(1+\kappa_{\rho NN,a})^2/12 m^2$
and $f^2_{VS,a}\equiv g_{\rho NN,a}^{2}$ are the coupling constants of
the exchanged $a$-meson. These parameters are fixed so that

\begin{eqnarray}
  v_{PS}(k)&=& [v_{\sigma \tau}(k)- 2 \> v_{t\tau}(k)]/3 \>\>\>, 
\label{eq:vpsdef} \\
  v_{V}(k) &=&  [v_{\sigma \tau}(k) + v_{t\tau}(k)]/3  \>\>\>,  
\label{eq:vvdef} \\
v_{VS}(k) &=&  v_{\tau}(k) \>\>\> ,
\label{eq:vvsdef}
\end{eqnarray}
where $v_{\sigma \tau}(k)$, $v_{t\tau}(k)$, and 
$v_{\tau}(k)$ are given in Eqs. (\ref{eq:vt})--(\ref{eq:vtt}).

The two-body currents due to these $PS$- and 
$V$-meson exchanges are then derived by minimal
substitution in the effective $PS$-$NN$ and $V$-$NN$
coupling Lagrangians.  The non-relativistic reduction of
the associated Feynman amplitudes in momentum space leads to:

\begin{eqnarray}
\noalign{\medskip}
\bfj_{ij}(\bfk_i,\bfk_j;PS) 
&=&3{\rm i}\,G_{E}^{V}(q_{\mu}^2)
   ({\bm\tau}_i \times {\bm\tau}_j)_z 
   \bigg[ v_{PS}(k_j) {\bm\sigma}_i ({\bm\sigma}_j \cdot \bfk_j) 
        - v_{PS}(k_i) {\bm\sigma}_j ({\bm\sigma}_i \cdot \bfk_i) 
    \nonumber \\
&&+ { \bfk_i - \bfk_j \over k_i^2 -k_j^2 } 
  \bigl [ v_{PS}(k_i)- v_{PS}(k_j) \bigr ] 
   ({\bm\sigma}_i \cdot \bfk_i)
   ({\bm\sigma}_j \cdot \bfk_j) \bigg] \ , \label{eq:jps} \\
\noalign{\medskip}
  \bfj_{ij}(\bfk_i,\bfk_j; V) 
   &=& -3{\rm i}\,G_{E}^{V}(q_{\mu}^2) 
   ({\bm\tau}_i \times {\bm\tau}_j)_z 
 \bigg[ 
   v_V(k_j) {\bm\sigma}_i \times ({\bm\sigma}_j \times \bfk_j) 
 - v_V(k_i) {\bm\sigma}_j \times ({\bm\sigma}_i \times \bfk_i) 
   \nonumber \\
&&- { v_V (k_i) - v_V (k_j) \over k_i^2 -k_j^2 } \bigl [ 
       (\bfk_i-\bfk_j)({\bm\sigma}_i \times \bfk_i)\cdot
                            ({\bm\sigma}_j \times \bfk_j) \nonumber \\
&&+ ({\bm\sigma}_i \times \bfk_i)\> 
  {\bm\sigma}_j \cdot (\bfk_i\times \bfk_j)
  +({\bm\sigma}_j \times \bfk_j)\> {\bm\sigma}_i \cdot 
  (\bfk_i\times \bfk_j) \bigr ] \bigg ] \ , \label{eq:jv} \\
\noalign{\medskip}
\bfj_{ij}(\bfk_i,\bfk_j; VS) &=& -{\rm i}\,G_{E}^{V}(q_{\mu}^2)
   ({\bm\tau}_i \times {\bm\tau}_j)_z\, 
{ \bfk_i - \bfk_j \over k_i^2 -k_j^2 }\,
 [ v_{VS}(k_i)- v_{VS}(k_j) ] \ , 
\end{eqnarray}
where $\bfk_i$ and $\bfk_j$ are the fractional momenta delivered
to nucleon $i$ and $j$, with $\bfk_i+\bfk_j=\bfq$, and
$G_E^V(q_\mu^2)$ is the isovector combination of the nucleon 
electric Sachs form factors~\cite{Car98}.  The current 

\begin{equation}
\bfj^{ME}_{ij}(\bfk_i,\bfk_j;v^0) = 
\bfj_{ij}(\bfk_i,\bfk_j;PS)+ 
\bfj_{ij}(\bfk_i,\bfk_j;V) +
\bfj_{ij}(\bfk_i,\bfk_j;VS)\ , \label{eq:jv6} 
\end{equation}
satisfies exactly the CCR 
with the potential given in Eq.~(\ref{eq:vij}). 

Configuration-space expressions are obtained from

\begin{equation}
  \bfj_{ij}(\bfq; B)=\int d{\bfx}\>
  {\rm e}^{ {\rm i} \bfq \cdot {\bfx} }
  \int \frac { d\bfk_i } {(2\pi)^3} \frac { d\bfk_j } {(2\pi)^3} 
  {\rm e}^{ {\rm i} \bfk_i \cdot (\bfr_i-{\bfx}) }
  {\rm e}^{ {\rm i} \bfk_j \cdot (\bfr_j-{\bfx}) } 
  \bfj_{ij}(\bfk_i,\bfk_j; B)\ , \label{eq:jrs}
\end{equation}
where $B$=$PS$, $V$ or $VS$, and
can be found in the Appendix of Ref.~\cite{Sch89}.
For later reference, we report below the configuration-space
expression for the current associated with the
isospin-dependent central potential $\bfj_{ij}(\bfq; VS)$:

\begin{eqnarray}
  \bfj_{ij}(\bfq; VS)&=&-G_E^V(q_\mu^2)\,
  ({\bm\tau}_i\times{\bm\tau}_j)_z
  \,{\rm e}^{{\rm i}\bfq\cdot\bfR}\,
 \sum_{a=1}^{N} \frac{f^2_{VS,a}}{4\pi}
  \int_{-1/2}^{1/2}\,dx\,
  \biggl({ {\rm i} x \bfq \over L_a(x)} +\hat\bfr\biggr)
  {\rm e}^{-{\rm i}x\bfq\cdot\bfr}\, {\rm e}^{-r L_a(x)}\ , \nonumber \\
  L_a(x)&=&\sqrt{m_a^2+\frac{q^2}{4}\Big(1-4x^2\Big)} \ ,
\label{eq:jvs}
\end{eqnarray}
where $\bfr\equiv{\bfr}_{i}-{\bfr}_{j}$ and
$\bfR \equiv(\bfr_i+\bfr_j)/2$.

The construction of the two-body currents associated with the
isospin-conserving momentum-dependent part of the interaction
$v^p_{ij}$ is less straightforward.  A procedure similar to the
one just reviewed has been applied to the case of the currents
from the spin-orbit components of the interaction~\cite{Car90}.
It consists, in essence, of attributing these to exchanges
of ``$\sigma$-like'' and ``$\omega$-like'' mesons for the
isospin-independent terms, and to ``$\rho$-like'' mesons for
the isospin-dependent ones.  Explicit expressions for the
resulting currents can be found in Ref.~\cite{Car90}.

The two-body currents from the quadratic
momentum-dependent terms of the interaction are listed in
Ref.~\cite{Sch89} and were obtained by minimal
substitution, that is ${\bfp}_i\rightarrow{\bfp}_i
-\epsilon_i\,\bfA(\bfr_i)$.  While minimal substitution ensures
current conservation for the isospin-independent
(quadratic-momentum-dependent) components of the interaction,  
this prescription does not lead to a conserved current for
the isospin-dependent ones.  Indeed, the commutator in
Eq.~(\ref{eq:ccr2}) gives rise to an isovector term proportional to
$({\bm\tau}_i\times{\bm\tau}_j)_z$, which cannot
be generated by minimal substitution (for a discussion
of this point, see Ref.~\cite{Sch04}).  These isovector
currents were ignored in all previous work since, in
view of their short range, they
were expected to give negligible contributions.

The currents associated with $v^{IB}_{ij}$, which have
never been considered up until now, will be discussed 
in Sec.~\ref{subsec:ist}.

\subsection{The two-body current operator in the minimal-substitution scheme}
\label{subsec:j2ms}

Consider again the isospin-conserving 
momentum-independent part of the potential given 
in Eqs.~(\ref{eq:vij}) and~(\ref{eq:v2ij}).
The isospin operator ${\bm \tau}_i\cdot{\bm \tau}_j$ is formally 
equivalent to an implicit momentum dependence~\cite{Sac48}, since
it can be expressed in terms of the 
space-exchange operator, $P_{ij}$, using the formula

\begin{equation}
  {\bm \tau}_i\cdot{\bm \tau}_j = -1-
  (1+{\bm \sigma}_i\cdot{\bm \sigma}_j) P_{ij} \ ,
  \label{eq:tt}
\end{equation}
valid when operating on antisymmetric wave functions, as 
in the case of a fermionic system.  The space-exchange operator
is defined as 

\begin{equation}
  P_{ij} \, f(\bfr_i,\bfr_j)\equiv
   {\rm e}^{\bfr_{ji}\cdot{\bna}_i
  + \bfr_{ij}\cdot{\bna}_j} \, f(\bfr_i,\bfr_j)
  = f(\bfr_j,\bfr_i)\ ,
\label{eq:p}
\end{equation}
where the ${\bna}$-operators act only 
on the generic function $f(\bfr_i,\bfr_j)$ and not on 
the vectors $\bfr_{ij}=\bfr_i-\bfr_j=-\bfr_{ji}$
in the exponential.  In the presence of an electromagnetic
field, after performing minimal substitution, the operator
$P_{ij}$ becomes~\cite{Sac48}

\begin{eqnarray}
  P_{ij}\rightarrow{P}^\bfA_{ij}&=&
  {\rm e}^{\bfr_{ji}\cdot[{\bna}_i-{\rm i}\epsilon_i\bfA(\bfr_i)]
  +\bfr_{ij}\cdot[{\bna}_j-{\rm i}\epsilon_j\bfA(\bfr_j)]}
  \nonumber \\
  &\equiv&
  {\rm e}^{\bfr_{ji}\cdot{\bna}_i+g_i(\bfr_i)} \,
  {\rm e}^{\bfr_{ij}\cdot{\bna}_j+g_j(\bfr_j)} \ ,
  \label{eq:phat1} 
\end{eqnarray}
where $\bfA(\bfr)$ is the vector potential, and the functions
$g_i(\bfr_i)$ and $g_j(\bfr_j)$ have been defined as
$g_i(\bfr_i)\equiv -{\rm i}\,\epsilon_i\,\bfr_{ji}
\cdot\bfA(\bfr_i)$ and $g_j(\bfr_j)\equiv -{\rm i}\,\epsilon_j\,\bfr_{ij}
\cdot\bfA(\bfr_j)$.  The operator ${ P}^\bfA_{ij}$ is then
the product of two operators, each having the general form

\begin{equation}
P(\bfr)= {\rm e}^{{\bf a}\cdot{\bna}+g(\bfr)} \ ,
\label{eq:pgen} 
\end{equation}
where ${\bf a}$ is a vector independent of
$\bfr$.  It has been shown in Ref.~\cite{Sac48} 
that $P(\bfr)$ can expressed as 

\begin{equation}
  P(\bfr)=
  {\rm e}^{{1\over a} \int_\bfr^{\bfr+{\bf a}}ds\,g(\bfs)} \,
  {\rm e}^{{\bf a}\cdot{\bna}} \ ,
\label{eq:pgen1} 
\end{equation}
where $ds$ is an infinitesimal element of a straight line parallel to 
${\bf a}$, which goes from position $\bfr$ to position $\bfr+{\bf a}$.
Using this general result in Eq.~(\ref{eq:phat1}), we obtain

\begin{equation}
  {P}^\bfA_{ij} ={\rm e}^{
  -{\rm i}\epsilon_i\int_{\bfr_i}^{\bfr_j}d\bfs\cdot\bfA(\bfs)
  -{\rm i}\epsilon_j\int_{\bfr_j}^{\bfr_i}d\bfs'\cdot\bfA(\bfs')}
  P_{ij} \ ,
\label{eq:phat}
\end{equation}
with $d\bfs=ds\, {\hat\bfr}_{ji}$ and
$d\bfs'=ds\, {\hat\bfr}_{ij}$.  The line integrals are
performed on straight lines parallel to ${\hat\bfr_{ji}}$ 
and ${\hat\bfr_{ij}}$.

\begin{figure}[h]
\includegraphics[height=5cm]{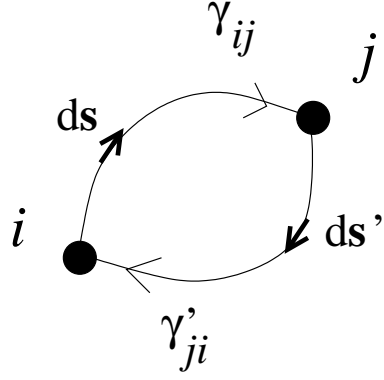}
\caption{Integration paths from position $i$ to position $j$ and vice versa
for the integral of Eq.~(\protect\ref{eq:phat}).}
\label{fig:path2b}
\end{figure}

The procedure of Ref.~\cite{Sac48} leading 
to Eq.~(\ref{eq:phat}) can be generalized and the two integrals 
can be performed on two generic paths $\gamma_{ij}$ and $\gamma'_{ji}$, that 
go from position $i$ to position $j$ and {\it vice versa}, as shown 
in Fig.~\ref{fig:path2b}.  Indeed, for a gauge transformation

\begin{eqnarray}
  \bfA(\bfr) &\rightarrow& \bfA(\bfr) -\bna G(\bfr)\ , \label{eq:gt1}\\
  \Psi &\rightarrow& {\rm e}^{{\rm i} \sum_i \epsilon_i G(\bfr_i)}\; \Psi\ ,
  \label{eq:gt2}
\end{eqnarray}
where $G(\bfr)$ is a generic function, it can be shown that the state

\begin{equation}
  {\bm \tau}_i\cdot{\bm \tau}_j \Psi = \bigg[-1-
  (1+{\bm \sigma}_i\cdot{\bm \sigma}_j)P^\bfA_{ij} \bigg]
  \; \Psi   \ ,
  \label{eq:tt2}   
\end{equation}
where $P^\bfA_{ij}$ is given in Eq.~(\ref{eq:phat}) with the generic 
integration paths of Fig.~\ref{fig:path2b}, transforms as

\begin{eqnarray}
   \bigg[ -1-
  (1+{\bm \sigma}_i\cdot{\bm \sigma}_j) P^\bfA_{ij} \bigg]
  \; \Psi &\rightarrow&
  \bigg[-1- (1+{\bm \sigma}_i\cdot{\bm \sigma}_j)
   P^{\bfA-\bna G}_{ij} \bigg]
  \; {\rm e}^{ {\rm i} \sum_i \epsilon_i G(\bfr_i)}\;\Psi  \nonumber \\
  &=& {\rm e}^{ {\rm i} \sum_i \epsilon_i G(\bfr_i)}\;
    \bigg[-1-
   ( 1+{\bm \sigma}_i\cdot{\bm \sigma}_j) P^\bfA_{ij} \bigg]
   \; \Psi\ , \label{eq:gt3}
\end{eqnarray}
in conformity with the requirements of gauge invariance of the theory.
A conserved current can then be derived by considering an infinitesimal
gauge transformation, as discussed in Ref.~\cite{Sac48}.

Rather than following the general procedure of Ref.~\cite{Sac48}, we
obtain the current in the limit of weak electromagnetic fields, since
calculations of photo- and electronuclear observables are typically
carried out in first-order perturbation theory in these fields.
Then, by retaining only linear terms in the vector potential, we find

\begin{eqnarray}
  v_{ij}&\rightarrow& v_{1,ij} + v_{2,ij}
  \bigg[-1-(1+{\bm \sigma}_i\cdot{\bm \sigma}_j) {P}^\bfA_{ij}\bigg]  
\nonumber \\
  &\simeq& v^0_{ij}+v_{2,ij}\, 
  \bigg[ -{\rm i}\epsilon_i\int_{\gamma_{ij}}d\bfs\cdot\bfA(\bfs)
         -{\rm i}\epsilon_j\int_{\gamma'_{ji}}d\bfs'\cdot\bfA(\bfs')\bigg]
  (1+{\bm \tau}_i\cdot{\bm \tau}_j) \nonumber \\
  &\equiv& v^0_{ij} -\int \bfj_{ij}({\bfx})\cdot\bfA({\bfx}) d{\bfx} \ ,
  \label{eq:vms}
\end{eqnarray}
where the paths $\gamma_{ij}$ and $\gamma'_{ji}$ are those illustrated 
in Fig.~\ref{fig:path2b}, and in the second and third lines of the
equation above $v^0_{ij}$ is as defined in Eq.~(\ref{eq:vij}).  The
current density operator is then given by

\begin{equation}
  \bfj_{ij}({\bfx})={\rm i}\,v_{2,ij}\, 
  \bigg[ \epsilon_i\int_{\gamma_{ij}}d\bfs\,\delta({\bfx}-\bfs)
  +\epsilon_j\int_{\gamma'_{ji}}d\bfs'\,\delta({\bfx}-\bfs')\bigg]
  \,(1+{\bm\tau}_i\cdot{\bm\tau}_j) \ ,
\label{eq:jx}
\end{equation}
and its Fourier transform reads

\begin{equation}
  \bfj_{ij}(\bfq) ={\rm i}\,v_{2,ij}\bigg(
  \epsilon_i\int_{\gamma_{ij}}d\bfs\,{\rm e}^{{\rm i}\bfq\cdot\bfs}
 +\epsilon_j\int_{\gamma'_{ji}}d\bfs'\,{\rm e}^{{\rm i}\bfq\cdot\bfs'}\bigg)
  \,(1+{\bm\tau}_i\cdot{\bm\tau}_j) \ .
\label{eq:jq}
\end{equation}

A number of comments are now in order.  Firstly,
the current in Eq.~(\ref{eq:jq}) by construction satisfies the CCR

\begin{equation}
  \bfq\cdot \bfj_{ij}(\bfq)=
  \biggl [ v^0_{ij},\rho_{i,{\rm NR}}(\bfq)
   +\rho_{j,{\rm NR}}(\bfq) \biggr ]\ .
\label{eq:ccrv0}
\end{equation}
This is easily verified by observing that
along any path from $\bfr_i$ to $\bfr_j$

\begin{equation} 
{\rm i} \bfq \cdot \int_{\bfr_i}^{\bfr_j}
\! d\bfs\; {\rm e}^{{\rm i}\bfq\cdot\bfs}=
{\rm e}^{{\rm i}\bfq\cdot\bfr_j}-{\rm e}^{{\rm i}\bfq\cdot\bfr_i} \ ,
\label{eq:intq}
\end{equation}
and that

\begin{equation}
{\rm i}\, (\epsilon_i-\epsilon_j)(1+{\bm \tau}_i\cdot{\bm \tau}_j)=
  G_E^V(q_\mu^2)\,({\bm \tau}_i\times{\bm \tau}_j)_z \ .
  \label{eq:tauz}
\end{equation}

Secondly, because of the arbitrariness of the two integration paths, 
the prescription just outlined does not lead to a unique two-body 
current.  Moreover, if $v_{2,ij}$ consists of a sum of different terms, then,
for each of these, different paths $\gamma_{ij}$ and $\gamma^\prime_{ji}$  
can be selected.  Hence, Eq.~(\ref{eq:jq}) can be interpreted as a 
{\it parameterization} of all possible two-body currents which satisfy
the CCR with the two-body potential given in Eq.~(\ref{eq:vij}).
In particular, it is interesting to note that the longitudinal 
part of the two-body currents of Eqs.~(\ref{eq:jps})--(\ref{eq:jvs}) 
obtained in the ME approach can also be derived in the MS scheme.  
This connection is shown in Appendix~\ref{app:path}.

Thirdly, the choice of a linear path ($LP$) for $\gamma_{ij}$
and $\gamma^\prime_{ji}$ in Eq.~(\ref{eq:jq})  (as in 
Ref.~\cite{Sac48}), namely

\begin{equation}
  \bfs=-\bfs^\prime=\bfr_i-x\bfr_{ij} \ , \hspace{1cm} 0\leq x\leq 1 \ ,
  \label{eq:slin} \\
\end{equation}
gives

\begin{equation}
  \bfj^{LP}_{ij}(\bfq)={\rm i}\,v_{2,ij}\, G_E^V(q_\mu^2)\, 
  ({\bm \tau}_i\times{\bm \tau}_j)_z\, \bfr_{ij}\,f_{ij}(\bfq) \ , 
  \label{eq:jql}
\end{equation}
with

\begin{equation}
  f_{ij}(\bfq)\equiv\frac{ {\rm e}^{ {\rm i}\bfq\cdot\bfr_i} 
        -{\rm e}^{ {\rm i}\bfq\cdot\bfr_j} }
  {\bfq\cdot\bfr_{ij} } \ .
\label{eq:fijq}
\end{equation}
Note that $f_{ij}(\bfq=0)={\rm i}$. 

Lastly, it is important to observe that, in the limit $q\rightarrow 0$,
the current operator $\bfj_{ij}(\bfq)$ becomes path-independent, i.e.
it is unique, and is given by

\begin{equation}
  \bfj_{ij}(\bfq=0)=-v_{2,ij}G_E^V(q_\mu^2)\, 
  ({\bm \tau}_i\times{\bm \tau}_j)_z\, \bfr_{ij} \ .
  \label{eq:jq0}
\end{equation}
This result can also be derived in a more direct way by considering
the following identities

\begin{eqnarray}
\bfj_{ij}(\bfq=0)=\int d\bfx\, \bfj_{ij}(\bfx)&=&-\int d\bfx \, \bfx\, 
 \bna \cdot \bfj_{ij}(\bfx) \nonumber \\
&=&{\rm i}\, \left[ v_{2,ij} \, , \, \int d\bfx
\, \bfx \left[ \rho_{i,{\rm NR}}(\bfx)+
               \rho_{j,{\rm NR}}(\bfx)\right] \right] \ ,
\label{eq:j_s}
\end{eqnarray}  
where in the first line the volume integral of $\bfj_{ij}(\bfx)$
has been re-expressed in terms of the divergence of the current,
ignoring vanishing surface contributions, and in the second line
use has been made of the CCR.  Evaluation of the commutator leads
to Eq.~(\ref{eq:jq0}) above.  Note that, {\it mutatis mutandis},
namely $\bfj_{ij}(\bfx) \rightarrow \bfj(\bfx)$
and $v_{2,ij} \rightarrow H$, etc., the second line of
Eq.~(\ref{eq:j_s}) is, in essence, the Siegert theorem
for the electric dipole operator, to which $\bfj(\bfq=0)$
is proportional.

We now derive the current operators 
associated with the momentum-dependent operators of the 
two-nucleon interaction, within the present scheme. 
In the case of the spin-orbit interactions, 
$v^p_1$ and $v^p_2$ of Eq.~(\ref{eq:vij}) are

\begin{eqnarray}
  v^p_{1,ij}&=&v_{b}(r)\, \bfL\cdot\bfS \ , \nonumber \\
  v^p_{2,ij}&=&v_{b\tau}(r)\, \bfL\cdot\bfS\ ,
  \label{eq:so}
\end{eqnarray}
where the notation of Refs.~\cite{Wir95,Wir84} is used, and 
$\bfL=\bfr_{ij}\times({\bfp}_i - {\bfp}_j )/2$, 
${\bfp}_i$ and ${\bfp}_j$ being the particles' momentum 
operators.  Performing minimal substitution in $v_1$, we obtain 

\begin{equation}
  \bfj_{ij}(\bfq;b)=
  \frac{1}{2} v_b(r) \biggl( \epsilon_i\,{\rm e}^{{\rm i}\bfq\cdot\bfr_i}
  -\epsilon_j\, {\rm e}^{{\rm i}\bfq\cdot\bfr_j}\biggr)\,
  \bfS\times \bfr_{ij} \ .
  \label{eq:jsoq}
\end{equation}

For the isospin-dependent term $v^p_2$, we first symmetrize as

\begin{equation}
v^p_{2,ij}\,{\bm\tau}_i\cdot{\bm\tau}_j=
\frac{1}{2}v_{b\tau}(r) \left( \bfL\cdot\bfS
\, {\bm \tau}_i\cdot{\bm\tau}_j + {\bm \tau}_i\cdot{\bm\tau}_j\,
\bfL\cdot\bfS \right) \ ,
\label{eq:v2t}
\end{equation}
and then perform minimal substitution in both $\bfL\cdot\bfS$ and 
${\bm\tau}_i\cdot{\bm\tau}_j$.  
When only terms linear in the vector potential  
$\bfA$ are kept and the path $\gamma_{ji}^\prime=-\gamma_{ij}$ is taken,
the associated current is found to be

\begin{eqnarray}
  \bfj_{ij}(\bfq;b\tau)&=&\frac{1}{4} v_{b\tau}(r)
  \bfS\times \bfr_{ij}\,
  \biggl( \eta_j {\rm e}^{{\rm i}\bfq\cdot\bfr_i} -
   \eta_i {\rm e}^{{\rm i}\bfq\cdot\bfr_j}\biggr) \nonumber \\
  &+&\frac{1}{2} v_{b\tau}(r) G_E^V(q_\mu^2) 
  ({\bm\tau}_i\times{\bm \tau}_j)_z
  \biggl(\,\bfL\cdot\bfS
  \,\int_{\gamma_{ij}}d\bfs\,{\rm e}^{{\rm i}\bfq\cdot\bfs} 
  +\int_{\gamma_{ij}}d\bfs\,{\rm e}^{{\rm i}\bfq\cdot\bfs} \,\,
   \bfL\cdot\bfS\,\biggr) \ ,
  \label{eq:jsotq}
\end{eqnarray}
with $\eta_i=G_E^S(q_\mu^2)
\,{\bm\tau}_i\cdot{\bm\tau}_j+G_E^V(q_\mu^2)\tau_{i,z}$.
In particular, the linear path of Eq.~(\ref{eq:slin}) leads to

\begin{eqnarray}
  \bfj_{ij}^{LP}(\bfq;b\tau)&=&\frac{1}{4} v_{b\tau}(r)
  \bfS\times \bfr_{ij}\,
  \biggl( \eta_j {\rm e}^{{\rm i}\bfq\cdot\bfr_i} -
   \eta_i {\rm e}^{{\rm i}\bfq\cdot\bfr_j}\biggr) \nonumber \\
  &+&\frac{\rm i}{2} v_{b\tau}(r) G_E^V(q_\mu^2) 
  ({\bm\tau}_i\times{\bm \tau}_j)_z
  \biggl[\,\bfL\cdot\bfS\,\, \bfr_{ij}\,f_{ij}(\bfq)
   +\bfr_{ij}\,f_{ij}(\bfq)\,
   \bfL\cdot\bfS\,\biggr] \ ,
\label{eq:jsotl}
\end{eqnarray}
with $f_{ij}(\bfq)$ defined in Eq.~(\ref{eq:fijq}).

The current operators associated with the quadratic 
momentum-dependent terms of the interaction can be derived
in a similar fashion.  Their explicit expressions are listed
in Appendix~\ref{app:l2so2}.

\subsection{Two-body current associated with the isospin-symmetry
breaking interactions}
\label{subsec:ist}

The current operators constructed so far in the ME and MS
schemes satisfy the CCR with the isospin symmetric component
of the two-nucleon potential.  However, the latest realistic 
models of the nucleon-nucleon interaction contain isospin-symmetry
breaking terms.  In the notation of Ref.~\cite{Wir95},
this part is written as 

\begin{equation}
  v^{IB}_{ij}=\sum_{p=15}^{18} v_p(r_{ij}) O^p_{ij} \ ,
  \label{eq:vcsb}
\end{equation}
and the four isospin-symmetry breaking operators have the form:

\begin{equation}
  O^{p=15,\ldots,18}_{ij}=T_{ij}\>\>\> ,\,
  {\bm\sigma}_i\cdot{\bm\sigma}_j T_{ij}\>\>\>,\,
  S_{ij}\, T_{ij}\>\>\>, \, (\tau_{i,z}+\tau_{j,z}) \ ,
\label{eq:ocsb}
\end{equation}
where $S_{ij}$ is the standard tensor operator and the isotensor operator 
$T_{ij}$ is defined as $T_{ij}=3\tau_{i,z}\,\tau_{j,z}-{\bm\tau}_i\cdot
{\bm\tau}_j$. The dependence on
${\bm\tau}_i\cdot{\bm\tau}_j$ generates two-body currents which can
be taken into account by modifying the isospin-dependent 
central, spin-spin and tensor terms of the potential as

\begin{eqnarray}
  {\hat v_{\tau}}(r)&=& [\, v_{\tau}(r)-v_{15}(r) \,] \ , \nonumber \\
  {\hat v_{\sigma\tau}}(r)&=& [\, v_{\sigma\tau}(r)-v_{16}(r) \,] \ ,
  \nonumber \\
  {\hat v_{t\tau}}(r)&=& [\, v_{t\tau}(r)-v_{17}(r) \,] \ , 
  \label{eq:vcsb2}
\end{eqnarray}
and by using ${\hat v_{\tau}}(r)$, ${\hat v_{\sigma\tau}}(r)$, and 
${\hat v_{t\tau}}(r)$, instead of $v_{\tau}(r)$, $v_{\sigma\tau}(r)$, and 
$v_{t\tau}(r)$ in $v_{2,ij}$ of Eq.~(\ref{eq:v2ij}).
However, the contributions 
associated with the currents from these isospin-symmetry 
breaking terms have been found to be negligibly small 
for all the observables of interest here.

\subsection{Summary of two-body current models}
\label{subsec:sumj2}

For the sake of clarity and for later reference, we summarize
here the salient features of the three different models
for the model-independent current
$\bfj_{ij}^{MI}(\bfq)$ corresponding to the AV18
interaction~\cite{Wir95}, considered in the present paper.

\begin{itemize}

\item Old-ME model. This model is that introduced in
Refs.~\cite{Viv96,Viv00,Mar98} and discussed in
Sec.~\ref{subsec:j2me}.  It is given by

\begin{equation}
   \bfj^{MI,{\rm old}}_{ij}(\bfq)=
   \bfj_{ij}^{ME}(\bfq;v^0)+\bfj_{ij}^{ME}(\bfq;v^p)\ .
\label{eq:jmiold}
\end{equation}
We re-emphasize that, while $\bfj_{ij}^{ME}(\bfq;v^0)$ satisfies
the CCR with $v^0_{ij}$ (which includes the long-range one-pion-exchange
term), $\bfj_{ij}^{ME}(\bfq;v^p)$ is not strictly conserved,
as discussed in Sec.~\ref{subsec:j2me}.

\item New-ME model.  This model retains $\bfj_{ij}^{ME}(\bfq;v^0)$
for the momentum-independent interaction, as in the ``old-ME'' model.
For the momentum-dependent interaction, it uses instead 
the two-body current obtained in the MS scheme with a linear
path, explicitly

\begin{equation}
   \bfj^{MI,{\rm new}}_{ij}(\bfq)=
   \bfj_{ij}^{ME}(\bfq;v^0)+\bfj_{ij}^{LP}(\bfq;v^p)\ ,
\label{eq:jminew}
\end{equation}
where

\begin{eqnarray}
  \bfj_{ij}^{LP}(\bfq;v^p)=
  \bfj_{ij}(\bfq;b)+
  \bfj_{ij}^{LP}(\bfq;b\tau)&+&
  \bfj_{ij}(\bfq;LL)+
  \bfj_{ij}^{LP}(\bfq;LL\tau)\nonumber \\
  &+&
  \bfj_{ij}(\bfq;bb)+
  \bfj_{ij}^{LP}(\bfq;bb\tau) \ , \label{eq:jslvp}
\end{eqnarray}
and $\bfj_{ij}(\bfq;b)$, $\bfj_{ij}^{LP}(\bfq;b\tau)$, 
$\bfj_{ij}(\bfq;LL)$, $\bfj_{ij}^{LP}(\bfq;LL\tau)$, $\bfj_{ij}(\bfq;bb)$
and $\bfj_{ij}^{LP}(\bfq;bb\tau)$ are listed respectively in 
Eqs.~(\ref{eq:jsoq}),~(\ref{eq:jsotl}),~(\ref{eq:jllq}),~(\ref{eq:jllqt}),
(\ref{eq:jbbq}),~(\ref{eq:jbbqt}).  In addition, the
isospin-symmetry-breaking contributions are included via
Eq.~(\ref{eq:vcsb2}).  The two-body current operator
given above satisfies {\it exactly} the CCR with the AV18
potential. 

It is important to stress that the longitudinal component
of $\bfj_{ij}^{ME}(\bfq;v^0)$ 
can also be obtained in the MS scheme, as discussed in
the previous section and in Appendix~\ref{app:path}. 

\item Linear path MS ($LP$-MS) model.  This model uses
a two-body current obtained in the MS scheme using a linear path,
explicitly

\begin{equation}
   \bfj^{MI,LP}_{ij}(\bfq)=
   \bfj_{ij}^{LP}(\bfq;v^0)+\bfj_{ij}^{LP}(\bfq;v^p)\ ,
\label{eq:jmilp}
\end{equation}
where $\bfj_{ij}^{LP}(\bfq;v^0)$ is the current given in 
Eq.~(\ref{eq:jql}) and $\bfj_{ij}^{LP}(\bfq;v^p)$ is the same as
in Eq.~(\ref{eq:jslvp}).

\end{itemize}

\section{Three-body Current}
\label{sec:j3b}

Three-body currents involving the excitation
of an intermediate $\Delta$ resonance were
derived recently  in the context of a study of explicit
$\Delta$ components in the trinucleon wave functions~\cite{Mar98,Delt04}.
In addition, the three-body current associated
with $S$-wave $\pi$$N$ scattering on an intermediate
nucleon was also included in Ref.~\cite{Mar98}.  The conclusion of that
work was that these three-body mechanisms 
give a small contribution to the magnetic form factors
of $^3$H and $^3$He over a wide range of momentum
transfer.  However, the three-body currents considered
in Ref.~\cite{Mar98} were not strictly consistent with
the three-nucleon interaction (TNI) included in the
Hamiltonian.

In this section we generalize the meson-exchange (ME)
and minimal substitution (MS) approaches to the
case of the three-body current induced by a
TNI $V_{ijk}$.  The resulting current satisfies, by
construction, the CCR with $V_{ijk}$.  To be specific,
we consider below the Urbana-IX model~\cite{Pud95}, but the
methods that are developed are applicable to other
phenomenological models of TNI's, such as the Tucson-Melbourne~\cite{Coo79}
and Brazil~\cite{Rob} ones.

\subsection{The three-body current in the meson-exchange scheme}
\label{subsec:j3me}

The Urbana-type TNI is written as sum of
a short-range spin- and isospin-independent
term and a term involving the excitation of an
intermediate $\Delta$ resonance.  The central term 
is irrelevant to the following discussion and is therefore
ignored, while the $\Delta$-excitation term is
given by~\cite{Pud95}

\begin{eqnarray}
V_{ijk} &=& \sum_{{\rm cyclic}\, ijk} V_{j;ki} \label{eq:tnif} \ , \\
V_{j;ki}&=&A_{2\pi}\bigg( \{X_{ij}\,,\,X_{jk}\}\,
  \{{\bm\tau}_i\cdot{\bm\tau}_j\,,\,{\bm\tau}_j\cdot{\bm\tau}_k\}\,
  +\frac{1}{4}\,[X_{ij}\,,\,X_{jk}]\,
  [{\bm\tau}_i\cdot{\bm\tau}_j\,,\,{\bm\tau}_j\cdot{\bm\tau}_k]\
  \bigg) \ ,\label{eq:vijkr} 
\end{eqnarray}
where $\{\ldots\}$ ($[\ldots]$) denotes the anticommutator (commutator),

\begin{equation}
  X_{ij}=v_{\sigma\tau}^{II}(r){\bm\sigma}_i\cdot{\bm\sigma}_j
  +v_{t\tau}^{II}(r)S_{ij} \ , \label{eq:xij}\\
\end{equation}
and $v_{\sigma\tau}^{II}(r)$ and $v_{t\tau}^{II}(r)$
are the standard spin-isospin and tensor-isospin
functions occurring in the one-pion-exchange interaction,
modified by a short-range gaussian cutoff.
The parameter $A_{2\pi}$ as well as the strength of the
central term are determined by reproducing the triton
binding energy in a Green's function
Monte Carlo calculation, and the nuclear matter equilibrium
density in an approximate hypernetted-chain variational
calculation~\cite{Pud95}.

In momentum space, $V_{j;ki}$ can be conveniently expressed as

\begin{equation}
V_{j;ki}(\bfk_j,\bfk_i,\bfk_k) =
   \frac{9}{2}\, A_{2\pi}\,
  \biggl[v_{jk}^\dagger(\bfk_k;\Delta N\rightarrow NN)\,
           v_{ij}(\bfk_i;NN\rightarrow N\Delta)\,+ {\rm h.\,c.}\biggr]
        \delta(\bfk_i+\bfk_j+\bfk_k) \ ,
  \label{eq:vijk}
\end{equation}
where the $N\Delta$-transition interaction is defined as

\begin{equation}
  v_{ij}(\bfk;NN\rightarrow N\Delta)=\biggl[
  v_{\sigma\tau}^{II}(k)\, k^2\,{\bm\sigma}_i\cdot\bfss_j
  +v_{t\tau}^{II}(k)S_{ij}^{II}(\bfk)
   \biggr]\,{\bm\tau}_i\cdot{\bf T}_j \ .
\label{eq:vNNDN} 
\end{equation}
Here $\bfss_j$ and ${\bf T}_j$ are the spin- and isospin-transition 
operators that convert nucleon $j$ into a $\Delta$-isobar, and 
$S_{ij}^{II}(\bfk)$ is the momentum-space tensor operator 
in which the Pauli spin operator of particle $j$ is replaced 
by $\bfss_j$.  The functions $v_{\sigma \tau}^{II}(k)$ and 
$v_{t\tau}^{II}(k)$ are related to their configuration-space 
counter-parts by similar relations to those in Eqs.~(\ref{eq:vst})
and~(\ref{eq:vtt}).  The momentum transfers to nucleons $i$, $j$,
$k$, respectively $\bfk_i$, $\bfk_j$, and $\bfk_k$, sum up to zero.
Manipulation of products of transition spin and/or isospin operators
is facilitated by making use of the following identity:

\begin{equation}
  \bfss_\alpha^\dagger\,\,\bfss_\beta = 
  \frac{2}{3}\delta_{\alpha\,\beta}\,-\,
  \frac{\rm i}{3}\epsilon_{\alpha\,\beta\,\gamma}{\bm\sigma}_\gamma \ .
\end{equation}

\begin{figure}[h]
\includegraphics[height=4cm]{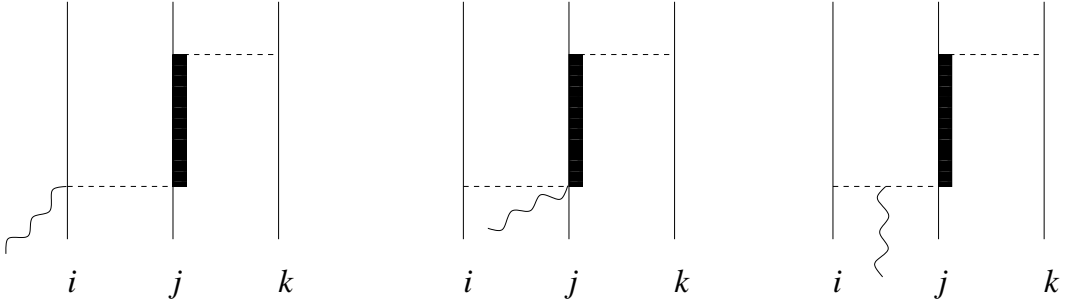}
\caption{Feynman diagram representation of the three-body currents associated 
with $PS$- and $V$-meson exchanges. Thin, thick, dashed and 
wavy lines denote nucleons, $\Delta$-isobars, 
$a$-mesons, and photons, respectively.}
\label{fig:j3fig}
\end{figure}

The $N\Delta$-transition interaction is assumed to originate from
exchanges of ``$\pi$-like'' ($PS$) and ``$\rho$-like'' ($V$) mesons, with
the associated components $v_{PS}^{II}(k)$ and $v_V^{II}(k)$
related to $v_{\sigma\tau}^{II}(k)$ and $v_{t\tau}^{II}(k)$
by relations identical to those in Eqs.~(\ref{eq:vpsdef})
and~(\ref{eq:vvdef}).
Thus, the $PS$- and $V$-exchange three-body currents, illustrated
in Fig.~\ref{fig:j3fig}, in momentum space read 

\begin{eqnarray}
\bfj^{ME}_{j;ki}(\bfq)&=&
 \frac{9}{2}\,A_{2\pi}\,
  \biggl[v_{jk}^{\dagger}(\bfk_k;\Delta N\rightarrow NN)\,
  \bigl[ \bfj_{ij}^{II}(\bfk_i, \bfk_j; PS)\,+
         \bfj_{ij}^{II}(\bfk_i, \bfk_j; V)\bigr]\nonumber \\
&+& \bigl[ \bfj_{jk}^{II}(\bfk_j, \bfk_k; PS)\,+
           \bfj_{jk}^{II}(\bfk_j, \bfk_k; V)\bigr]^\dagger
  v_{ij} (\bfk_i;NN\rightarrow N \Delta) +{\rm h.\,c.} \biggr]
 \ ,\label{eq:jijk}
\end{eqnarray}
where the $PS$ and $V$ currents
$\bfj_{ij}^{II}(\bfk_i,\bfk_j; PS)$ 
and $\bfj_{ij}^{II}(\bfk_i,\bfk_j;V)$ involving
$\Delta$ excitation are obtained from 
Eqs.~(\ref{eq:jps}) and~(\ref{eq:jv}) with
the replacements ${\bm\sigma}_j\,({\bm\tau}_j)\rightarrow
  \bfss_j\,({\bf T}_j)$.
Configuration-space expressions are listed in Appendix~\ref{app:j3br}.
Finally, the current above
satisfies the CCR with the TNI of Eq.~(\ref{eq:vijk}).

\subsection{The three-body current in the minimal-substitution scheme}
\label{subsec:j3ms}

\begin{figure}[h]
\includegraphics[height=5cm]{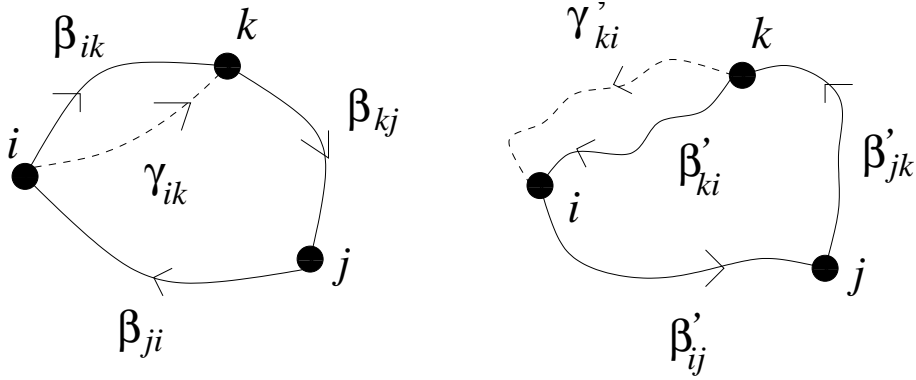}
\caption{Integration paths chosen to derive the three-body current 
operator of Eqs.~(\protect\ref{eq:j3ba}) and~(\protect\ref{eq:j3ms}).}
\label{fig:path3b}
\end{figure}

Consider the isospin dependence of the TNI.
The anticommutator term is first expressed as 

\begin{equation}
 \{ {\bm\tau}_i\cdot{\bm\tau}_j\, , \,{\bm\tau}_j\cdot{\bm\tau}_k\}
  = 2\, {\bm\tau}_i\cdot{\bm\tau}_k \ ,
  \label{eq:ttanti}
\end{equation}
and the associated current $\bfj^A_{j;ki}(\bfq)$ is then derived
with the same methods discussed in Sec.~\ref{subsec:j2ms} and is given by 
(see Eq.~(\ref{eq:jq}))

\begin{equation}
 \bfj^A_{j;ki}(\bfq) = 2 {\rm i}\, A_{2\pi}\,
  \{X_{ij}\, , \,X_{jk}\}
   \biggl( \epsilon_i\int_{\gamma_{ik}}d\bfs\,
    {\rm e}^{{\rm i}\bfq\cdot\bfs}+
    \epsilon_k\int_{\gamma_{ki}'}d\bfs'\,
    {\rm e}^{{\rm i}\bfq\cdot\bfs'} \biggr)
  \,(1+{\bm\tau}_i\cdot{\bm\tau}_k) \ ,
\label{eq:j3ba}
\end{equation}
where $\gamma_{ik}$ and $\gamma_{ki}'$ are generic paths
from $\bfr_i$ to $\bfr_k$ and $\bfr_k$ to $\bfr_i$ 
(see Fig.~\ref{fig:path3b}).

\begin{figure}[ht]
\includegraphics[height=3cm]{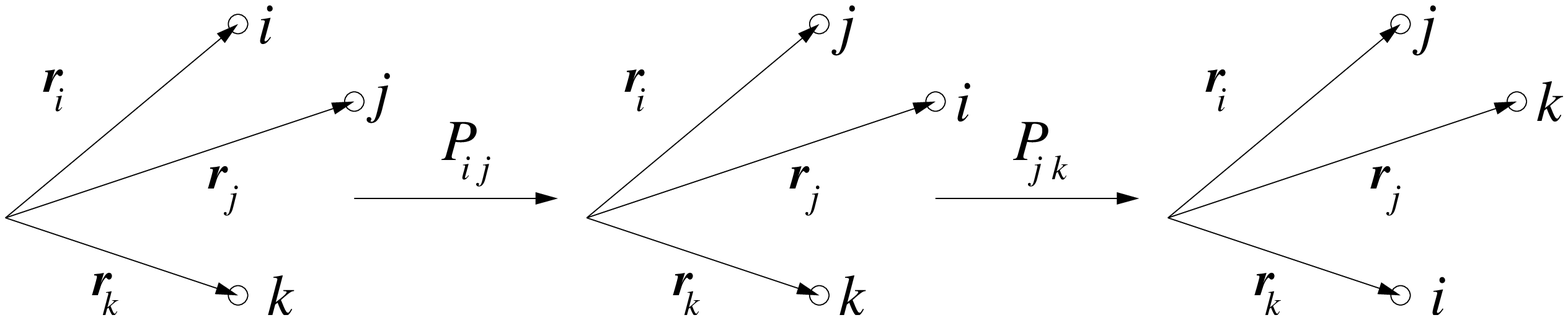}
\caption{Final position of particles $i$, $j$, and $k$ 
after that the product of the space exchange operators 
$P_{jk}P_{ij}$ is applied.}
\label{fig:pijk}
\end{figure}

In the case of the commutator term, we first note that

\begin{equation}
\frac{1}{4} [ {\bm\tau}_i\cdot{\bm\tau}_j\, ,\, {\bm\tau}_j\cdot{\bm\tau}_k]=
P_{ij}^\tau P_{jk}^\tau - P_{jk}^\tau  P_{ij}^\tau \ ,
  \label{eq:ttcommu}
\end{equation}
where $P_{ij}^\tau$=$(1+{\bm\tau}_i\cdot{\bm\tau}_j)/2$
is the isospin-exchange operator.  The product
$P_{ij}^\tau P_{jk}^\tau$, when acting on antisymmetric
wave functions, is equivalent to

\begin{equation}
P_{ij}^\tau P_{jk}^\tau=P_{jk}^\sigma  P_{ij}^\sigma P_{jk}  P_{ij} \ ,
\label{eq:rs1}
\end{equation}
where $P_{ij}^\sigma$ is the spin-exchange operator, defined similarly
as $P_{ij}^\tau$, and $P_{ij}$ is space-exchange operator introduced
in Eq.~(\ref{eq:p}).  Note the ordering of the operators on the
r.h.s. of the equation above.  Obviously, the product
$P_{jk}^\tau  P_{ij}^\tau$ is given by a relation similar
to Eq.~(\ref{eq:rs1}) in which the order of the $jk$ and $ij$
pairs is inverted.  The products of space exchange operators
$P_{jk}  P_{ij}$ and $P_{ij} P_{jk}$ are equivalent, respectively,
to the exchanges $(\bfr_i,\bfr_j,\bfr_k) \rightarrow (\bfr_k,\bfr_i,\bfr_j)$ 
and $(\bfr_i,\bfr_j,\bfr_k) \rightarrow (\bfr_j,\bfr_k,\bfr_i)$
(see Fig.~\ref{fig:pijk}), and
can formally be expressed by the operators

\begin{eqnarray}
P_{jk}  P_{ij}&=& {\rm e}^{\bfr_{ki}\cdot\bna_i+\bfr_{ij}\cdot\bna_j+
          \bfr_{jk}\cdot\bna_k}\ , \\
P_{ij}  P_{jk}&=& {\rm e}^{\bfr_{ji}\cdot\bna_i+\bfr_{kj}\cdot\bna_j+
          \bfr_{ik}\cdot\bna_k}\ ,
\label{eq:jtt6}
\end{eqnarray}
where, as before, the gradients do not act on the position coordinates
in the exponential.  The methods of Sec.~\ref{subsec:j2ms} can now
be applied to the present case.  Gauging the gradient operators and
retaining only linear terms in the vector potential (valid for
weak electromagnetic fields) lead to the following current
$\bfj^C_{j;ki}(\bfq)$ from the commutator term of the TNI:

\begin{eqnarray}
&&\bfj^C_{j;ki}(\bfq)=
   { {\rm i} \over 4} A_{2\pi} [X_{ij}\, ,\, X_{jk}] \nonumber \\
  && \biggl [ \Bigl (
  \epsilon_i\int_{\beta_{ik}}\! d\bfs \;{\rm e}^{{\rm i}\bfq\cdot\bfs}+
  \epsilon_j\int_{\beta_{ji}}\! d\bfs \;{\rm e}^{{\rm i}\bfq\cdot\bfs}+
  \epsilon_k\int_{\beta_{kj}}\! d\bfs \;{\rm e}^{{\rm i}\bfq\cdot\bfs}
 \Bigr ) (1+{\bm\tau}_i\cdot{\bm\tau}_j)
         (1+{\bm\tau}_j\cdot{\bm\tau}_k) \nonumber \\
  &-& \Bigl(
  \epsilon_i\int_{\beta_{ij}'}\! d\bfs \;{\rm e}^{{\rm i}\bfq\cdot\bfs}+
  \epsilon_j\int_{\beta_{jk}'}\! d\bfs \;{\rm e}^{{\rm i}\bfq\cdot\bfs}+
  \epsilon_k\int_{\beta_{ki}'}\! d\bfs \;{\rm e}^{{\rm i}\bfq\cdot\bfs}
 \Bigr ) (1+{\bm\tau}_j\cdot{\bm\tau}_k)
         (1+{\bm\tau}_i\cdot{\bm\tau}_j) \biggr ]\ , \label{eq:j3ms}
\end{eqnarray}
where $\beta_{ik}$ ($\beta'_{ki}$) is a generic path starting at 
$\bfr_i$ ($\bfr_k$) and ending at $\bfr_k$ ($\bfr_i$), and
so on (see Fig.~\ref{fig:path3b}).

The expressions for $\bfj^A_{j;ki}(\bfq)$ and
$\bfj^C_{j;ki}(\bfq)$ may be simplified by selecting
the following paths:

\begin{equation}
  \gamma_{ki}'=-\gamma_{ik}\ ,\quad
  \beta_{ij}'=-\beta_{ji}\ , \quad
  \beta_{jk}'=-\beta_{kj}\ , \quad
  \beta_{ki}'=-\beta_{ik}\ , \quad
  \beta_{ik}=-\beta_{ji}-\beta_{kj} \ ,
\label{eq:choice1}
\end{equation}
namely the path from $\bfr_k$ to $\bfr_i$ is 
taken to be the same as that from $\bfr_i$ to
$\bfr_k$ but in the opposite direction, and so on.
The last relation means that the path $\beta_{ik}$
from $\bfr_i$ to $\bfr_k$ is chosen to go through
the position $\bfr_j$ exactly along the 
paths $-\beta_{ji}$ and $-\beta_{kj}$ (the two latter
paths are still arbitrary).  We then obtain:

\begin{eqnarray}
\bfj_{j;ki}(\bfq)&=&\bfj^A_{j;ki}(\bfq)+\bfj^C_{j;ki}(\bfq) \nonumber \\
                &=&2\,A_{2\pi}\,G_E^V(q_\mu^2)\,
  \{X_{ij}\,,\,X_{jk}\}\,({\bm\tau}_i\times{\bm\tau}_k)_z
  \,\int_{\gamma_{ik}} d\bfs\,
  {\rm e}^{{\rm i}\bfq\cdot\bfs}
  \nonumber \\
  &+& \frac{\rm i}{4}\, A_{2\pi}\,G_E^V(q_\mu^2) 
     \,[X_{ij}\,,\,X_{jk}] \bigg [
  (\tau_{i,z}\, {\bm\tau}_j\cdot{\bm\tau}_k-
        \tau_{j,z}\, {\bm\tau}_i\cdot{\bm\tau}_k)
       \int_{\beta_{jk}} d\bfs\,
       {\rm e}^{{\rm i}\bfq\cdot\bfs}
   \nonumber \\
 &&\phantom{G_E^V(q_\mu^2)\,A_{2\pi}\, \frac{\rm i}{2}
     \,[X_{ij}\,,\,X_{jk}]}
   + (\tau_{k,z}\, {\bm\tau}_i\cdot{\bm\tau}_j-
           \tau_{j,z}\, {\bm\tau}_i\cdot{\bm\tau}_k)
          \int_{\beta_{ij}'} d\bfs\,
          {\rm e}^{{\rm i}\bfq\cdot\bfs} \bigg ]
      \ .\label{eq:j3ms2}
\end{eqnarray}

This current is easily shown to
satisfy the CCR with the TNI.
As in the case of two-body currents, the
limit $\bfq$=0 is path-independent,

\begin{equation}
\bfj_{j;ki}(\bfq=0)=
{\rm i}\, \left[ V_{j;ki} \, , \, \int d\bfx
\, \bfx \left[ \rho_{i,{\rm NR}}(\bfx)+
               \rho_{j,{\rm NR}}(\bfx)+
               \rho_{k,{\rm NR}}(\bfx)\right] \right] \ .
\end{equation}
Furthermore, when the paths $\gamma_{ik}$,
$\beta_{jk}$ and $\beta_{ij}'$ are taken as straight lines
as in Eq.(\ref{eq:slin}), then Eq.~(\ref{eq:j3ms2}) becomes

\begin{eqnarray}
  \bfj_{j;ki}^{LP}(\bfq)&=&2{\rm i}\,A_{2\pi}\,G_E^V(q_\mu^2)\,
  \{X_{ij}\,,\,X_{jk}\}\,({\bm\tau}_i\times{\bm\tau}_k)_z
  \,\bfr_{ik}\, f_{ik}(\bfq) 
  \nonumber \\
  &-& \frac{1}{2}\,A_{2\pi}\, G_E^V(q_\mu^2)
     \,[X_{ij}\,,\,X_{jk}] \bigg [
  (\tau_{i,z}\, {\bm\tau}_j\cdot{\bm\tau}_k-
        \tau_{j,z}\, {\bm\tau}_i\cdot{\bm\tau}_k )
        \bfr_{jk}\, f_{jk}(\bfq) 
   \nonumber \\
 &&\phantom{G_E^V(q_\mu^2)\,A_{2\pi}\, \frac{\rm i}{4}
     \,[X_{ij}\,,\,X_{jk}] } 
   + ( \tau_{k,z}\, {\bm\tau}_i\cdot{\bm\tau}_j-
           \tau_{j,z}\, {\bm\tau}_i\cdot{\bm\tau}_k )
           \bfr_{ij}\; f_{ij}(\bfq) \bigg ]
      \ ,  \label{eq:jijklp}
\end{eqnarray}
where the functions $f_{ij}(\bfq)$
are defined in Eq.~(\ref{eq:fijq}).

Finally, note that the present approach can also be used to
derive the currents associated to the the Tucson-Melbourne
(TM)~\cite{Coo79} and Brazil~\cite{Rob} TNI
interaction models, since these can be cast in the form~\cite{Car83} 

\begin{equation}
  V_{j;ki}=F_S(j;ki)\,\{{\bm\tau}_i\cdot{\bm\tau}_j\,,\,
   {\bm\tau}_j\cdot{\bm\tau}_k\} \, + \, 
  F_A(j;ki)\,[{\bm\tau}_i\cdot{\bm\tau}_j\,,\,
  {\bm\tau}_j\cdot{\bm\tau}_k] \ .
  \label{eq:vijkgen}
\end{equation}
For example, the TM model has

\begin{eqnarray}
F_S(j;ki)&=&c_S\,\{X_{ij}\,,\,X_{jk}\} \,+\, B(\bfr_{ij},\bfr_{jk})\,
\{S_{ij}+{\bm\sigma}_i\cdot{\bm\sigma}_j\,,\,
  S_{jk}+{\bm\sigma}_j\cdot{\bm\sigma}_k\,\}
\ , \label{eq:fs}\\
F_A(j;ki)&=&c_A\,[X_{ij}\,,\,X_{jk}] \ , \label{eq:fan}
\end{eqnarray}
where the parameters $c_S$ and $c_A$ have the values $c_S\simeq-0.063$ and
$c_A\simeq-0.018$~\cite{Coo79}, and the function $B(\bfr_{ij},\bfr_{jk})$
depends on a cutoff $\Lambda$, fitted to reproduce the triton binding energy.

\subsection{Summary of three-body current models}
\label{subsec:sumj3}

We summarize in the present subsection the different models for the 
three-body current used in the present study.  

\begin{itemize}

\item Old-TCO model.  The model is that introduced in
Ref.~\cite{Sch92} and subsequently refined in
Ref.~\cite{Mar98}.  As already mentioned, 
it does not satisfy the CCR with the Urbana or
Tucson-Melbourne TNIs.

\item ME-model.  In the case of the Urbana-type TNI,
the three-body current $\bfj^{ME}_{j;ki}(\bfq)$
satisfying the CCR is given by the configuration-space
expression of Eq.~(\ref{eq:jijk}), which can be derived from
Eqs.~(\ref{eq:jpsr}) and~(\ref{eq:jvr}).
For the TM-type TNI, some difficulties arise, since 
the second term of the operator $F_S(j;ki)$, proportional to 
$B(\bfr_{ij},\bfr_{jk})$, 
cannot be simply related to the exchange of a single $\pi$-like
or $\rho$-like meson. 
Therefore, in this case a hybrid approach is used, where 
the current associated with this last term is treated 
within the linear-path MS scheme, while the rest is obtained 
within the ME scheme.

\item Linear path MS ($LP$-MS) model.  
Within the MS scheme, we select again the linear path 
of Eq.~(\ref{eq:slin}) to construct the three-body current, as given
in Eq.~(\ref{eq:jijklp}).

\end{itemize} 

Note that the current corresponding to the
TNI defined in Eq.~(\ref{eq:tnif}) involves a cyclic sum over
$ijk$, i.e.

\begin{equation}
\bfj_{ijk}(\bfq) = \sum_{{\rm cyclic}\, ijk} \bfj_{j;ki}(\bfq) \ .
\label{eq:jijkt}
\end{equation}

Lastly, it is worth remarking here that, at low values 
of the momentum transfer, the contributions associated
with the operators $\bfj^{ME}_{ijk}(\bfq)$ and 
$\bfj^{LP}_{ijk}(\bfq)$ (as well as $\bfj^{MI,{\rm new}}_{ij}(\bfq)$
and $\bfj^{LP}_{ij}(\bfq)$)  are calculated to be
essentially the same for the observables of interest
in the present study.

\section{Wave Functions}
\label{sec:wf}

The trinucleon bound-state and $N-d$ scattering-state wave functions
are obtained  variationally with the 
pair-correlated-hyperspherical-harmonics (PHH) method~\cite{Kie94}. 
Recently, in a series of papers~\cite{Kie01,Kie99,Viv01}, the method 
has been generalized to solve the $N-d$ elastic
scattering problem above the deuteron breakup threshold (DBT), thus
allowing for the study of electromagnetic processes 
at higher energies than previously treated~\cite{Viv96,Viv00}.
For completeness, the method will be reviewed briefly
and a summary of relevant results obtained for 
$N-d$ scattering observables at energies above the DBT 
will be presented.  

The wave function $\Psi_{1+2}^{LSJJ_z}$ for
a $N-d$ elastic scattering state with an incoming relative 
orbital angular momentum $L$, channel spin $S$ ($S=1/2,3/2$) and
total angular momentum $JJ_z$, is written as 

\begin{equation}
   \Psi_{1+2}^{LSJJ_z}  = \Psi_{C}^{JJ_z}
   + \Psi_{A}^{LSJJ_z} \ ,\label{eq:scatte}
\end{equation}
where $\Psi_C^{JJ_z}$ describes the system
in two regions: (i) the ``core'' region where the three particles 
are close to each other and their mutual interactions are large
and (ii) the ``breakup'' region where the three particles
are far from each other.
The other term $\Psi_{A}^{LSJJ_z}$ describes the system
in the $1+2$ ``clusterization'' asymptotic region, where intercluster
nuclear interactions are negligible.
The function $\Psi_{A}^{LSJJ_z}$ (for $p-d$, as an example) 
is given by

\begin{eqnarray}
   \Psi_{A}^{LSJJ_z}   &=& {1\over \sqrt{3}}
    \sum_{{\rm cyclic}\> ijk} \sum_{L^\prime
   S^\prime} \Big[ \lbrack \phi_d({\bfx}_i) \otimes \chi_i
   \rbrack_{S^\prime} \otimes Y_{L^\prime}({\hat {\bfr}}_i) \Big]_{JJ_z}
   \nonumber \\
   &&\times \Bigg\lbrack \delta_{L L^\prime} \delta_{S S^\prime}
    H^{-}_{L^\prime}(\eta,pr_{i})
    - {\cal S}^J_{LS,L^\prime S^\prime}(E)
    H^{+}_{L^\prime}(\eta,pr_{i}) \Bigg\rbrack
   \>\>\>, \label{eq:psia}
\end{eqnarray}
where $\phi_{d}$ is the deuteron wave function, 
$\chi_i$ the spin state of nucleon $i$, 
$\bfx_i$ and $\bfy_i$ the Jacobi vectors
defined, respectively, as $\bfx_i=\bfr_j-\bfr_k$ and
$\bfy_i=( 2\,\bfr_i - \bfr_j-\bfr_k )/\sqrt{3}\equiv \sqrt{4/3}\bfr_i$,
and $p$ is the
magnitude of the relative momentum between
deuteron and proton. The functions $H^\pm$ are defined as

\begin{equation}\label{eq:coul2}
  H_L^\pm(\eta,p r)=
  {(1-{\rm e}^{-\kappa r})^{2L+1}
   G_L(\eta,p r)\pm {\rm i} F_L(\eta,p r) \over p r}\ ,
\end{equation}
where $F_L$ and $G_L$ are the regular and irregular Coulomb functions,
respectively, and $\eta$ is the Sommerfeld parameter.  Note that 
for $n-d$ scattering $\eta=0$, and $F_L(0,x)/x$ and $G_L(0,x)/x$
reduce to the regular and irregular spherical 
Bessel functions. The factor $(1-{\rm e}^{-\kappa r})^{2L+1}$ has 
been introduced to regularize the function $G$ at the origin, 
and  $\kappa$ is taken as a variational parameter.
The complex parameters ${\cal S}^J_{LS,L^\prime
S^\prime}(E)$ are the $S$-matrix elements which determine phase-shifts
and (for coupled channels) mixing angles at the c.m.
energy $E=T_{c.m.}-B_2$ where $B_2=2.225$ MeV is the deuteron binding
energy and

\begin{equation} 
  T_{c.m.}= p^2/(2\mu)  \ ,
  \label{eq:Ecm}
\end{equation}
is the $N-d$ c.m. kinetic energy,
$\mu$ being the $N-d$ reduced mass.  The sum 
over $L^\prime S^\prime$ in Eq.~(\ref{eq:psia}) 
is over all values compatible with a given  $J$ and parity.

The second term $\Psi_C^{JJ_z}$ of the trial wave function must describe
those configurations of the system where the particles are close to
each other. For large interparticle separations and energies below
the DBT, $\Psi_C^{JJ_z}$ goes to zero, whereas for higher
energies it must reproduce an outgoing three particle state. 
In terms of the PHH basis, $\Psi_C^{JJ_z}$ is expanded as~\cite{Kie94},

\begin{equation}\label{eq:chh2}
     \Psi_C^{JJ_z}=  \rho^{-5/2}\sum_{\alpha=1}^{N_c}
      \sum_{K=1}^{N_K(\alpha)}  u_{\a,K}(\rho)  Z_{\alpha K}\ ,
\end{equation}
where $\rho=\sqrt{x_i^2+y_i^2}$ is the hyperradius. The
functions $Z_{\alpha K}$ are antisymmetric under the
exchange of any two pairs of particles 
and account for the angle-spin-isospin
and hyperangle dependence of channel $\alpha,K$.  
The hyperangle is defined as ${\rm cos}\, \phi_i = x_i/\rho$.
The index $\alpha$
denotes collectively the spectator $i$ and pair $jk$ orbital and spin
angular momenta and isospins coupled to produce
a state with total angular momentum and parity
$J^\pi$, 
while the index $K$ specifies the order of the Jacobi
polynomial in the hyperangle.  
The values of $N_c$ and $N_K(\alpha)$ are increased until
the desired degree of convergence in the quantity of interest 
is obtained (see the discussion in Sec.~\ref{subsec:conv}).
In the PHH approach, a correlation factor is 
included in $Z_{\a K}$ in order to better take into account 
those correlations induced by the repulsion of the 
potential at short distances.  In this way, the rate of 
convergence in the $N_K(\alpha)$ expansion is
improved very significantly ($N_K(\alpha)<10$ in all cases). 

The functions
$u_{\a,K}(\rho)$ are the hyperradial functions to be determined by
the variational procedure, once the boundary conditions are specified.
In practice, the functions $u_{\a,K}(\rho)$ are chosen to be regular at the
origin ($u_{\a,K}(0)=0$) and to have the following behavior as
$\rho\rightarrow\infty$

\begin{eqnarray}
  u_{\a,K}(\rho) &\rightarrow & 0\ , \qquad E<0 \ ,\label{eq:asy1} \\
  u_{\a,K}(\rho)&\rightarrow &  \sum_{\alpha'} \sum_{K'}
  \left ( {\rm e}^{-{\rm i} ({n^{-1}\; c}) 
   \ln 2 Q\rho} \right)_{\a K,\a'K'}\;
   {\cal S}^{\;b}_{LS,\a'K'}(E) \; {\rm e}^{{\rm i} Q\rho} \ ,
   \qquad E={Q^2\over m}>0 \ ,
   \label{eq:asy2}
\end{eqnarray}
where ${\cal S}^{\;b}_{LS,\a'K'}$ are the $S$-matrix elements
for the process $1+2\rightarrow 1+1+1$.
The matrices $n$ and $c$ are defined as

\begin{equation}\label{eq:cccc}
    { n}_{\a K,\a'K'}= \lim_{\rho\rightarrow\infty}
    \int d\Omega\;\; Z^\dag_{\alpha K} \, Z_{\alpha' K'} \ , \qquad
  { c}_{\a K,\a'K'}= \lim_{\rho\rightarrow\infty}
    \int d\Omega\;\; Z^\dag_{\alpha K} \,
     \rho V_C \, Z_{\alpha',K'} \ ,
\end{equation}
where $V_C$ is the Coulomb potential energy and 
$d\Omega=(\cos\phi_i)^2(\sin\phi_i)^2d\phi_i\,d\hat\bfx_i\,d\hat\bfy_i$.
Once the above boundary conditions are applied, it has been shown that 
the Kohn variational principle for scattering states is valid also
above the DBT (for more details, see Ref.~\cite{Viv01}). 
This principle can therefore be 
used to compute the matrix elements ${\cal S}^J_{LS,L^\prime S^\prime}(E)$
and ${\cal S}^{\;b}_{LS,\a'K'}(E)$ and the
functions $u_{\alpha K}(\rho)$ occurring in the expansion of $\Psi_C$.
This is achieved in practice by making the functional

\begin{equation}
   [{\cal S}^J_{LS,L^\prime S^\prime}(E)]= 
  {\cal S}^J_{LS,L^\prime S^\prime}(E)
   -\sqrt{3}{\rm i}\;m\; p\; \langle \Psi_{1+2}^{LSJJ_z} | H-E
   | \Psi_{1+2}^{LSJJ_z} \rangle\ , \label{eq:kohn}
\end{equation}
stationary with respect to variations in the
${\cal S}^J_{LS,L^\prime S^\prime}$ and
$u_{\alpha K}$.

Phase-shifts and mixing angles for $n-d$ scattering 
have been obtained from a realistic Hamiltonian model, 
and have been shown to be in excellent agreement with corresponding 
Faddeev results~\cite{Hub95,Kie98}, thus establishing the high accuracy of 
the PHH expansion for this scattering problem.  It is important to 
emphasize that the PHH scheme permits the straightforward
inclusion of Coulomb distortion effects in the $p-d$ channel.
The PHH results for $p-d$ elastic
scattering are as accurate as those for $n-d$ scattering.

\begin{figure}[p]
\includegraphics[height=18cm]{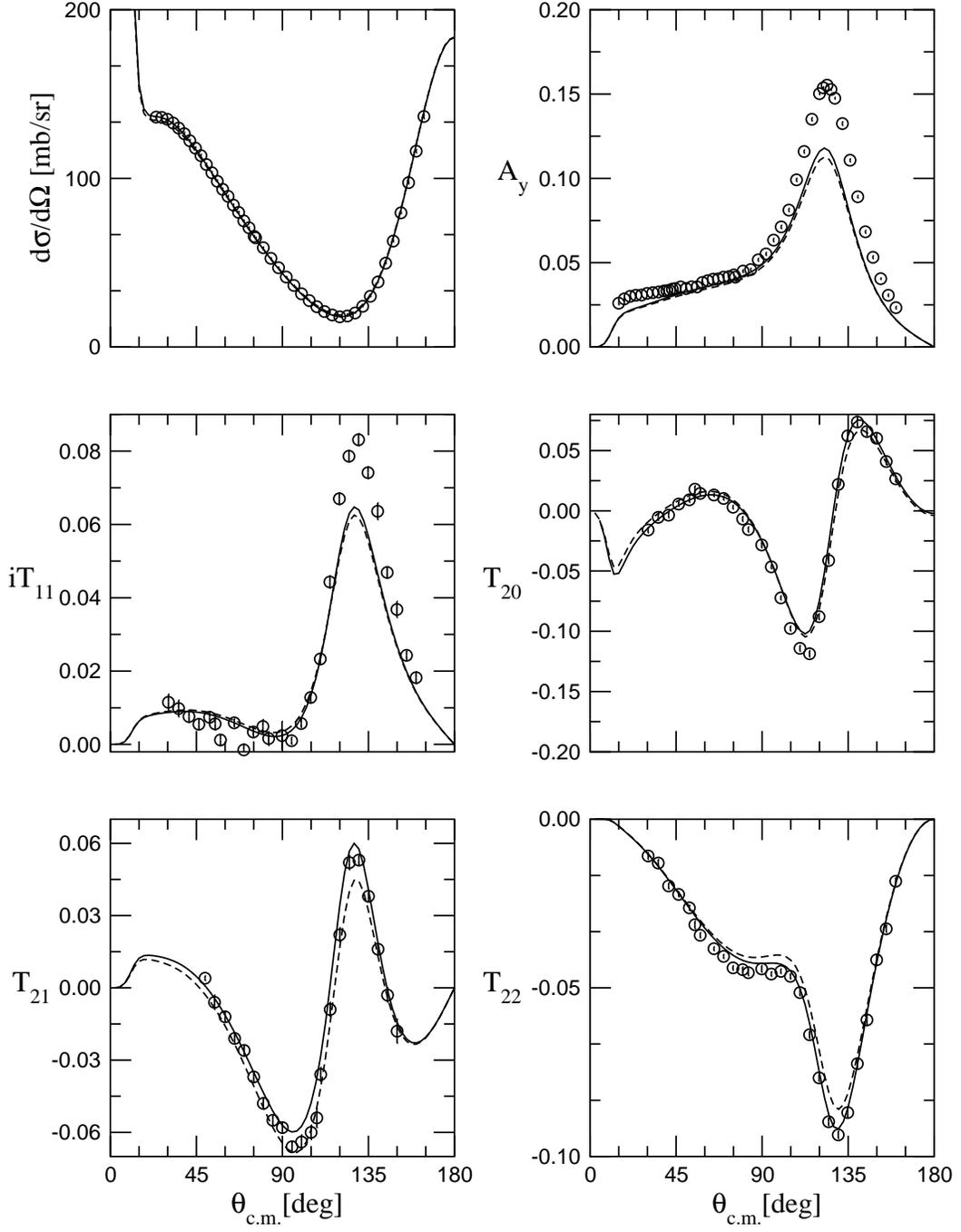}
\caption{Differential cross section, 
proton vector-analyzing power $A_y$, 
and four deuteron tensor polarization observables 
for $pd$ elastic scattering at $T_{c.m.}=$6.66 MeV
as function of the c.m. scattering angle.
The dashed and solid curves are obtained with the AV18
and AV18/UIX interaction models, respectively.
The experimental differential cross section and $A_y$ data
are from Ref.~\protect\cite{sag94}, while the
experimental tensor-analyzing power data are from
Ref.~\protect\cite{gru83}.}
\label{fig:3nobs}
\end{figure}

For example, various $p-d$ observables at $T_{c.m.}=6.66$ MeV
predicted by the AV18/UIX model are shown in Fig.~\ref{fig:3nobs}
and are found to be in good agreement with 
the available experimental data~\cite{sag94,gru83}. The large discrepancy
observed for the $A_y$ and ${\rm i}\, T_{11}$ observables 
(the ``$A_y$-puzzle'') is connected to a not well understood  
deficiency of the nuclear interaction, most likely of present TNI 
models. Resolving this ``$A_y$-puzzle'' is a current and important 
area of research.

The bound-state wave function $\Psi_3^{JJ_z}$ ($J=1/2$)
is just given by the term $\Psi_C^{JJ_z}$,
which is expanded as in Eq.~(\ref{eq:chh2}). In this case,
the functions $u_{\a,K}(\rho)$ are determined by the 
Rayleigh-Ritz variational principle, applying
the boundary conditions $u_{\a,K}(\rho\rightarrow\infty)\rightarrow0$.
In practice, they are expanded in terms of Laguerre 
polynomials multiplied for an exponential 
factor~\cite{Kie93,Kie94}.
The number of channels included in such an expansion will be
denoted by $N_c({\rm b.s.})$ in the following.
The PHH expansion is very accurate
also for bound states, as shown, for example, in Ref.~\cite{nogga}, where 
a very detailed comparison with the results of
the Faddeev calculations of the Bochum 
group has 
been performed.

In the following, it is convenient to use the wave function
$\Psi_{{\bfp},\sigma_2 \sigma}^{(+)}$, where $\sigma_2$ and
$\sigma$ are the spin projections of the $d$ and $N$ 
clusters, and ${\bfp}$ their relative momentum
in the incident channel. It is given by

\begin{equation}
   \Psi_{{\bfp},\sigma_2 \sigma}^{(+)} = 4\pi
   \sum_{SS_z} \langle 1\sigma_2,\frac {1}{2}\sigma |SS_z\rangle
   \sum_{LMJJ_z} {\rm i}^L\> 
    \langle SS_z,LM | JJ_z \rangle Y_{LM}^*({\hat {\bfp}}) 
    {{\rm e}^{ {\rm i} \sigma_{L}} \over 2{\rm i}}
    \Psi_{1+2}^{LSJJ_z}\>\>\>, \label{eq:psiplus}
\end{equation}
where $\sigma_L$ is the Coulomb phase shift.  For a $nd$ state
the factor ${\rm e}^{{\rm i} \sigma_L}$ is omitted.  The
wave function $\Psi_{{\bfp},\sigma_2 \sigma}^{(+)}$ satisfies 
outgoing wave boundary
conditions, and is normalized to unit flux, while the two- and
three-nucleon bound-state wave functions are normalized to one. 

In earlier papers~\cite{Viv96,Viv00}, the sum over $J$
in Eq.~(\ref{eq:psiplus}) was truncated to a given value $J_{\rm max}=
7/2$, since the analysis was limited to study low-energy
radiative capture ($T_{c.m.}\le 2$ MeV). In the present work,
we extend the calculations to higher energy. In this case,
it is necessary to take into account also the contribution 
of higher partial waves. For large values of $J$, and correspondingly
large values of $L$, the centrifugal barrier between the deuteron
and the third nucleon prevents the two clusters to approach each other.
The corresponding $\Psi_{1+2}^{LSJJ_z}$ of either the $nd$ or $pd$
state can therefore be approximated to describe the free or 
Coulomb-distorted motion. For $pd$ as an example,

\begin{equation}
   \Psi_{1+2}^{LSJJ_z} \rightarrow
    \Psi_{1+2,{\rm free}}^{LSJJ_z}= {1\over \sqrt{3}}
    \sum_{{\rm cyclic}\> ijk} 
     \Big[ \lbrack \phi_d({\bfx}_i) \otimes \chi_i
   \rbrack_{S^\prime} \otimes Y_{L^\prime}({\hat {\bfy}}_i) \Big]_{JJ_z}
    { F_{L}(\eta,pr_{i}) \over pr_i}\qquad
     J>J_{\rm max}
    \>\>\>. \label{eq:psib}
\end{equation}
In the calculation of transition matrix elements, we 
found it convenient to divide the sum over $J$ in Eq.~(\ref{eq:psiplus})
as $\sum_J\rightarrow\sum_{J\le J_{\rm max}}+\sum_{J>J_{\rm max}}$.
In the first sum, the wave functions $\Psi_{1+2}^{LSJJ_z}$ are
calculated by taking into account the full PHH expansion
(the effect of increasing the number $N_c({\rm s.s.})$
of scattering state channels is studied for a few selected cases in 
Sec.~\ref{subsec:conv}).
In the second sum, the wave function $\Psi_{1+2}^{LSJJ_z}$
is approximated as in Eq.~(\ref{eq:psib}). In this case,
the sum over $L$ can be evaluated analytically
to  reconstruct the Coulomb distorted ``plane wave'' 
describing $p-d$ motion. In summary,

\begin{equation}
   \Psi_{{\bfp},\sigma_2 \sigma}^{(+)} = 4\pi 
   \sum_{J\le J_{\rm max},J_z} 
   \sum_{SS_z} \langle \frac {1}{2}\sigma,1\sigma_2 |SS_z\rangle
   \sum_{LM} {\rm i}^L\> 
    \langle SS_z,LM | JJ_z \rangle Y_{LM}^*({\hat {\bfp}}) 
    {{\rm e}^{ {\rm i} \sigma_{L}} \over 2{\rm i}}
    \Psi_{1+2}^{LSJJ_z}+ 
    \Delta\Psi_{{\bfp},\sigma_2 \sigma}^{(+)}
    \>\>\>, \label{eq:psiplus2}
\end{equation}
where 

\begin{eqnarray}
      \Delta\Psi_{{\bfp},\sigma_2 \sigma}^{(+)} &=&
      \sum_{{\rm cyclic}\> ijk} 
     \phi_{d}^{1\sigma_2}({\bfx}_i) \chi_{i}^{{1\over 2}\sigma} 
      \; \psi_c^{(+)}({\bfp},{\bfr}_i)
     - 4\pi      \sum_{J\le J_{\rm max},J_z}
      \sum_{SS_z} \langle \frac {1}{2}\sigma,1\sigma_2 |SS_z\rangle\times
      \nonumber \\
     && 
     \times \sum_{LM} {\rm i}^L\> 
      \langle SS_z,LM | JJ_z \rangle Y_{LM}^*({\hat {\bfp}}) 
      {{\rm e}^{ {\rm i} \sigma_{L}} \over 2{\rm i}}
      \Psi_{1+2,{\rm free}}^{LSJJ_z}
     \>\>\>, \label{eq:psiplus3}
\end{eqnarray}
and $\psi^{(+)}_c({\bfp},{\bfr})$ is the solution of the 
three-dimensional Schr\"odinger equation with the pure Coulomb
potential behaving asymptotically as a plane plus a scattered
wave, i.e.

\begin{equation}
   \psi^{(+)}_c({\bfp},{\bfr}) =
     e^{{\rm i} {\bfp}\cdot {\bfr}}
     {\rm e}^{-\pi\eta/2} \Gamma(1+{\rm i}\eta) {}_1F_1(-{\rm i}
     \eta,{\rm i}p r -{\rm i} {\bfp}\cdot {\bfr})
     \>\>\>. \label{eq:psiapw3}
\end{equation}
Here, ${}_1F_1$ is the confluent hypergeometric function. The
function given in Eq.~(\ref{eq:psiapw3}) reduces simply to the plane 
wave for $\eta=0$ ($nd$ case).

\section{Results}
\label{sec:res}

In the present section we report results for the $np$ radiative capture at 
thermal neutron energies and deuteron photodisintegration 
cross section at low energy, the
$nd$ and $pd$ radiative capture reactions at c.m. energy 
$T_{c.m.}=$ 2--20 MeV, and the isoscalar and isovector 
magnetic form factors of $^3$H and $^3$He.  
In the next two subsections, we report the
results for the $pd$ radiative capture at $T_{c.m.}=2.0$
and $3.33$ MeV, for which there are very accurate
cross-section and polarization data~\cite{Goe92,Smi99}.
One reason for doing so is to test the quality of the bound
and scattering wave functions, in particular by studying the
rate of convergence of calculated reduced-matrix elements
(RMEs) with respect to the number of channels included in
the PHH expansions of these wave functions.

The second reason is to make a comparative study of the
different current operator models introduced in the
present work.  Some of these models satisfy the
CCR exactly,  others do so only approximately.
The question is 
how critical is this lack of current conservation and 
how large are the contributions of three-body currents
induced by the trinucleon interaction.

In Ref.~\cite{Viv00} significant deviations were obtained
between the measured and calculated tensor observables
$T_{20}$ and $T_{21}$ in $p$$d$ radiative capture.
That earlier study was carried out with a current
operator including, in addition to one-body, two- and
three-body terms, denoted as old-ME and
old-TCO in the present work.  As shown below,
most of the observed discrepancy between theory and experiment
can be traced back to the fact that the
current of Ref.~\cite{Viv00} was not {\it exactly} conserved.

In the other subsections, the predictions obtained with the
new models of the electromagnetic current will be compared
with data in $A$=2 and 3 nucleon systems.

\subsection{Test of the wave functions}
\label{subsec:conv}

In order to test the PHH wave functions,
we have performed a series of calculations
of the $pd$ capture reaction at $T_{c.m.}=2$ MeV with a Hamiltonian
including the Argonne $v_{18}$ (AV18) two-nucleon~\cite{Wir95}
and Urbana IX (UIX) three-nucleon~\cite{Pud95} interactions
(the AV18/UIX Hamiltonian model).  The model for the electromagnetic
current chosen for this test is the new ``full'' one, i.e.
including the one-body, the new-ME two-body and the ME 
three-body terms. More precisely,

\begin{equation}
   \bfj^{\rm full-new}(\bfq)=\sum_i \bfj_i(\bfq)+\sum_{i<j} [
              \bfj^{MI,{\rm new}}_{ij}(\bfq)+
              \bfj^{MD}_{ij}(\bfq)]+
              \sum_{{\rm cyclic}\ ijk}
              \bfj^{ME}_{j;ki}(\bfq)\ ,            
         \label{eq:current2}
\end{equation}
where $\bfj^{MI,{\rm new}}_{ij}(\bfq)$ and 
$\bfj^{ME}_{j;ki}(\bfq)$ are given in Eqs.~(\ref{eq:jminew})
and~(\ref{eq:jijk}), respectively. The RMEs are computed
from the matrix elements

\begin{equation}
   j_{\sigma_3\lambda J_z}^{LSJ}(p,q)=
  \langle \Psi_3^{{1\over 2}\sigma_3} |
     \hat \epsilon^*_{\lambda}({\bf q}) 
     \cdot {\bf j}^\dagger({\bf q}) | \Psi^{LSJJ_z}_{1+2}
      \rangle \ , \label{eq:j}
\end{equation}
where $\epsilon_{\lambda}({\bf q})$, $\lambda=\pm 1$, are the
spherical components of the photon polarization vector~\cite{Viv96}.

Some of the most relevant RMEs
for $pd$ capture at this energy are the electric dipoles 
$\overline{\cal E}^{LSJ}_1$ induced
by transitions between $p$$d$ states in relative
orbital angular momentum quantum number $L=1,3$
and the $^3$He state. Here $S=1/2,3/2$ are the channel spin
quantum numbers obtained by coupling the spins of the proton and 
deuteron, and $\vec J=\vec L +\vec S$ (the notation and 
definition used for the RMEs are those of Ref.~\cite{Viv00}).
The calculated RMEs are listed in Table~\ref{tab:rme}. In the different
calculations, we  varied the number of channels included
in the bound and scattering PHH wave functions, namely the
value $N_c$ in the first sum of Eq.~(\ref{eq:chh2}).  
The channels are ordered for increasing values of $l_\alpha+L_\alpha$, 
$l_\alpha$ and $L_\alpha$ being the orbital angular momentum of the 
pair and of the third nucleon with respect to the pair, respectively.
In this analysis, the values for $N_K(\alpha)$ were taken large enough
to have full convergence with respect to the order $K$ of Jacobi
polynomials.

\begin{table}
\caption[Table]{RMEs ($\times 10^3$) for $pd$ radiative capture 
at $T_{\rm c.m.}=2$ MeV obtained with the AV18/UIX
Hamiltonian and the new-ME two- and three-body currents. For the
exact definition of the RMEs, see Eq.~(4.26) of 
Ref. \protect\cite{Viv00}. In the table, $N_c({\rm b.s.})$ 
($N_c({\rm s.s.})$) indicates the number of channels included in the
expansion of the bound state (scattering state of given $J$ and
parity $\pi$).}
\begin{tabular}{l@{$\qquad$}l@{$\qquad$}c@{$\qquad$}c@{$\qquad$}
                                        c@{$\qquad$}c@{$\qquad$}}
\hline
\multicolumn{6}{c}{State $J={1\over 2}^-$} \\
\hline
RME & $(N_c({\rm b.s.}),N_c({\rm s.s.}))$ 
  & (12,10) & (18,10) & (18,14) & (18,18)  \\
\hline
$\overline {\cal E}_1^{1{1\over 2}{1\over 2}}$ 
  && 2.699 & 2.701 & 2.693 & 2.689\\
$\overline {\cal E}_1^{1{3\over 2}{1\over 2}}$ 
  &&-0.134 & -0.131 & -0.203 & -0.201 \\
\hline
\multicolumn{6}{c}{State $J={3\over 2}^-$} \\
\hline
RME & $(N_c({\rm b.s.}),N_c({\rm s.s.}))$ 
  & (12,13) & (18,13) & (18,22) & (18,29) \\
\hline
$\overline {\cal E}_1^{1{1\over 2}{3\over 2}}$ 
  && 2.725 & 2.733 & 2.714 & 2.711 \\
$\overline {\cal E}_1^{1{3\over 2}{3\over 2}}$ 
  && 0.103 & 0.103 & 0.089 & 0.089 \\
$\overline {\cal E}_1^{3{3\over 2}{3\over 2}}$ 
  && 0.075 & 0.075 & 0.126 & 0.127 \\
\hline
\end{tabular}
\label{tab:rme}
\end{table}

First, consider the effect of the truncation 
of the PHH expansion in the description of the bound-state.
The calculated binding energy for ${}^3$He with the AV18/UIX 
potential is $7.725$ ($7.741$) MeV after the inclusion
of $N_c=12$ ($18$) channels in the PHH expansion. As can be seen
by inspecting the two columns corresponding to the cases $N_c({\rm b.s.})=
12$ and $18$, the changes in the values of the RMEs is at most 2\%. 
We have checked that the inclusion of additional channels in the 
bound state wave function produces tiny changes in the binding 
energy (less than 10 keV) and negligible changes in all the RMEs. 

Next, consider the convergence with respect to 
$N_c({\rm s.s.})$, the number of channels
in the scattering wave functions.
In general, the dependence of the calculated RMEs on $N_c({\rm s.s.})$
is weak. The only exceptions are the RMEs due to the inhibited $E_1$
transitions proceeding through the spin-channel $S=3/2$ states, which 
require the inclusion of a fairly large number of channels.
In general, the convergence can be checked by looking at
$Nd$ elastic scattering phase-shifts $\delta^{LSJ}$ 
obtained with the given value of $N_c({\rm s.s.})$.
Let us consider, for example,
the transitions to the $J={1\over 2}^-$ $pd$ scattering state.
The elastic phase shift $\delta^{LSJ}$ for the state $L=1$, $S=1/2$ 
and $J=1/2$ were found to have the values
$\delta^{1{1\over2}{1\over2}}=-7.393^\circ$, 
$-7.366^\circ$ and $-7.365^\circ$ 
for  $N_c({\rm s.s.})=10$, $14$ and $18$ channels, respectively.
The corresponding changes in $\overline {\cal E}_1^{1{1\over 2}{1\over 2}}$ 
are given in the first row of Table~\ref{tab:rme} and
are very small. On the other hand, for the $L=1$, $S=3/2$ and $J=1/2$ state,
the elastic phase-shift turns out to be
$\delta^{1{3\over2}{1\over2}}=21.666^\circ$, $22.318^\circ$ and 
$22.319^\circ$ for  $N_c({\rm s.s.})=10$, $14$ and $18$ channels, 
respectively. The elastic phase-shift here has a fairly large change 
passing from $N_c=10$ to $N_c=14$, due to the appearance of important 
channels in the PHH expansion. The corresponding change in 
$\overline {\cal E}_1^{1{3\over 2}{1\over 2}}$ is very significant,
as can be seen by inspecting the second row of Table~\ref{tab:rme}.
However, by adding more channels, the elastic phase-shift and
the corresponding capture RMEs show only tiny changes. A similar check
has been performed for all the other scattering states included 
in the calculation. 

We have verified by direct calculation 
that differences between the observables obtained by using the RMEs 
computed with the two largest values of $N_c$ are completely negligible. 
Note that in the present paper the bound and scattering
wave functions have been obtained on a more extended grid
and with more PHH components than in previous 
publications~\cite{Viv96,Viv00}. However, this better accuracy in the
wave functions has produced negligible changes in the $nd$ and $pd$
capture observables for $T_{c.m.}\le 2$ MeV, which were the focus 
of Refs.~\cite{Viv96,Viv00}.

For the range of energies considered here, the most important 
$Nd$ scattering waves are those with $J^\pi={1\over2}^\pm$, 
${3\over2}^\pm$ and ${5\over2}^\pm$. For these scattering
states a fairly large number of channels has to be included in the
PHH expansion of the ``core'' wave function $\Psi_3^{JJ_z}$. 
Scattering states with higher values of $J^\pi$
give very small contributions. As mentioned before,
we have retained the full PHH expansion in the states up
to $J_{\rm max}=7/2$. For larger values of $J$, the
scattering wave function has been approximated
as in Eq.~(\ref{eq:psib}).

\subsection{Test of the two- and three-body current models}
\label{subsec:comp}

\begin{figure}[p]
\includegraphics[height=18cm]{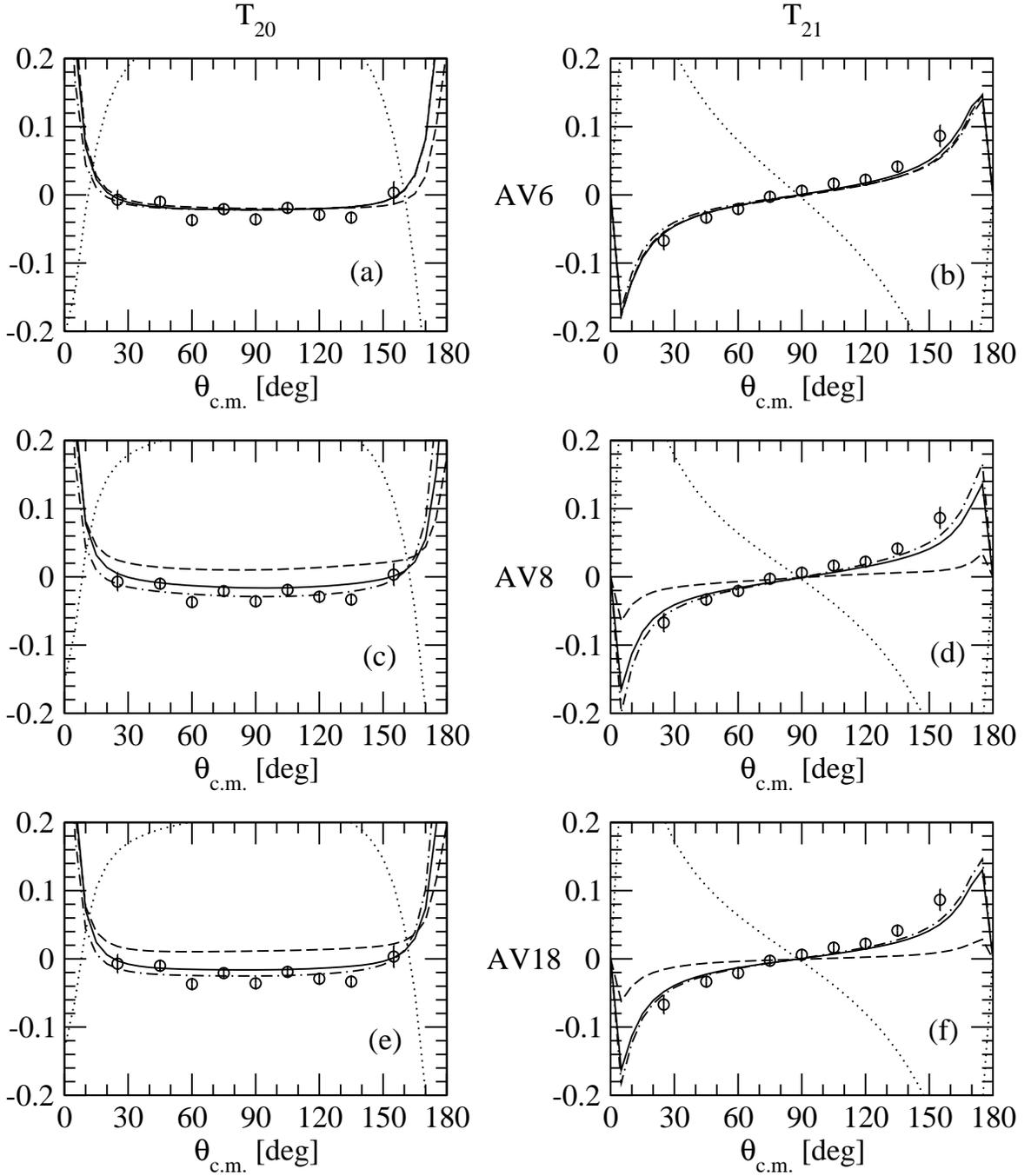}
\caption{Deuteron tensor polarization observables $T_{20}$ and $T_{21}$ 
for $pd$ radiative capture at $T_{c.m.}=$2 MeV 
as function of the c.m. $\gamma$-$p$ scattering angle, 
obtained with the 
AV6, AV8 and AV18 Hamiltonian models. Dotted, dashed and solid lines 
are obtained with only one-body current, with one- and 
old-ME two-body currents and with one- and $LP$-MS 
two-body currents, respectively. 
The results obtained in long-wavelength approximation 
are also shown (dotted-dashed lines). 
The experimental data are from Ref.~\protect\cite{Smi99}.}
\label{fig:t20-t21.1}
\end{figure}

We have calculated the  
$T_{20}$ and $T_{21}$ observables at $T_{c.m.}=$ 2 MeV using the 
Argonne $v_6$ (AV6)~\cite{Wir02}, the Argonne $v_8$ (AV8)~\cite{Pud97}
and AV18 two-nucleon interaction.  The AV6 interaction is 
momentum-independent, while the momentum-dependence of the AV8 
is due only to the spin-orbit operator.
The results are shown in Fig.~\ref{fig:t20-t21.1}. 

First, let us consider the calculations
performed with the AV6 interaction, reported in 
panels (a) and (b). The dotted 
curves are obtained by including only the one-body current 
contributions. The dashed curves are obtained when
the contributions of the old-ME two-body currents of
Refs.~\cite{Viv96,Viv00,Mar98}, see Eq.~(\ref{eq:jmiold}),
are added to the one-body ones. The solid curves are obtained 
including instead the two-body current contributions calculated 
within the MS scheme and using 
the linear path ($LP$-MS, see Sec.~\ref{subsec:j2ms}). 
Finally, the dotted-dashed curves are obtained in the 
long-wavelength-approximation (LWA). 
We observe that in this case
there is no significant difference between the old-ME, $LP$-MS, LWA 
and experimental results.  For this potential $v^p_{ij}=0$, 
and therefore the old-ME and new-ME two-body current models
coincide. We see that the current derived from the 
momentum-independent part of the interaction using
either the ME or $LP$-MS schemes are almost equivalent.
This is not too surprising, given the small
photon energy involved in the process, $q\approx 0.035$ fm$^{-1}$,
see discussion in Sec.~\ref{subsec:j2ms}.

Next, let us consider the calculations
performed with the AV8 and AV18 interactions, reported in 
panels (c)--(f). Now, the solid curves
are obtained using the new-ME scheme, namely
the current $\bfj^{MI,{\rm new}}_{ij}(\bfq)$
given in Eq.~(\ref{eq:jminew}).
Note that the old-ME two-body current model results (dashed lines) 
are in significant disagreement with the LWA ones and the 
experimental data, as can be seen by inspecting panels (c)--(f)
of Fig.~\ref{fig:t20-t21.1}.  This is not the case for the
new-ME results (those obtained with the 
$LP$-MS model have not been reported, since they are practically
coincident with the solid lines).  
This indicates that the current operator $\bfj_{ij}^{ME}(\bfq;v^p)$ used
in Refs.~\cite{Viv96,Viv00,Mar98} and in earlier studies,
which does not satisfy {\it exactly} the CCR with the
momentum-dependent terms of the
two-nucleon interaction, contain spurious contributions. 

This conclusion is supported by another observation.
Using the bound and scattering
wave functions derived from the AV18 interaction, but 
taking into account $\bfj_{ij}^{ME}(\bfq;v^0)$ only, a good 
description of the observables $T_{20}$ and $T_{21}$
is still obtained (see,  for example, Fig.~24 of Ref.~\cite{Golak00}). 
The inclusion of the old-model current
$\bfj_{ij}^{ME}(\bfq;v^p)$ produces
quite large effects on the RMEs and spoils such an
agreement with the data~\cite{Viv00}. On the other hand,
the contribution of the momentum-dependent part of the interaction
is noticeably smaller than that one produced by $v^0_{ij}$,
as can be seen, for example, in studies of the binding energies of the light
nuclei~\cite{Wir91}. Therefore, one can reasonably expect that also
$|\bfj_{ij}(\bfq;v^p)|\ll |\bfj_{ij}(\bfq;v^0)|$.
The current $\bfj_{ij}^{LP}(\bfq;v^p)$, constructed
in order to properly satisfy the CCR with $v^p_{ij}$,
gives correctly a small contribution to the RMEs,
and now the $T_{20}$ and $T_{21}$ at $T_{c.m.}=2$ MeV
are well reproduced. The same happens at higher
energies, as will be shown below.

\begin{figure}[p]
\includegraphics[height=16cm]{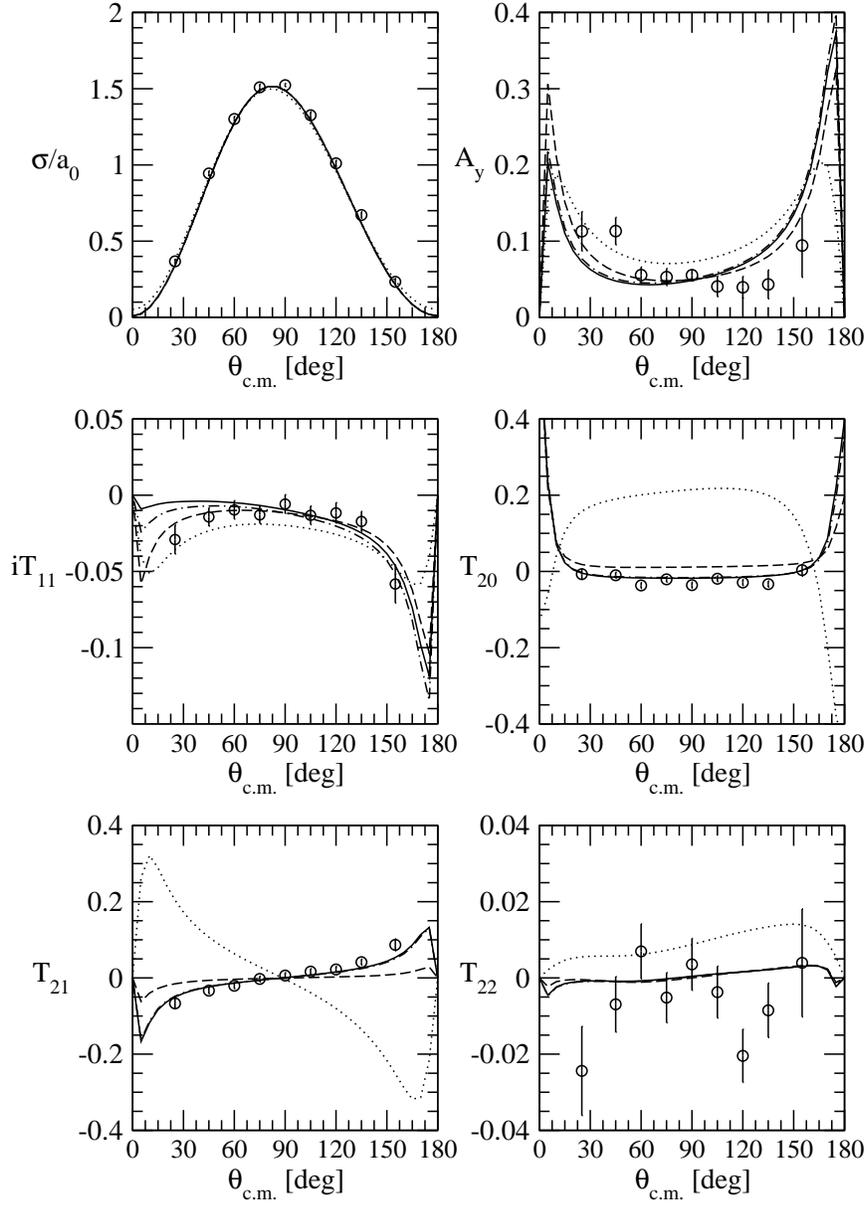}
\caption{Differential cross section, proton vector analyzing power, 
and the four deuteron tensor polarization observables 
for $pd$ radiative capture at $T_{c.m.}=$2 MeV
as function of the c.m. $\gamma$-$p$ scattering angle, 
obtained with the AV18 Hamiltonian model. 
The dotted curves are obtained with only one-body currents, 
the  dashed and dotted-dashed curves retain, in addition,
the contributions from the MI two-body operators 
$\bfj_{ij}^{{\rm MI},{\rm old}}(\bfq)$ 
and $\bfj_{ij}^{{\rm MI},{\rm new}}(\bfq)$
(see Eqs.~(\ref{eq:jmiold}) and~(\ref{eq:jminew})), 
respectively. 
The solid curves are obtained including, in addition to 
$\bfj_{ij}^{{\rm MI},{\rm new}}(\bfq)$, also the 
``model-dependent'' two-body contributions. 
The experimental data are from Ref.~\protect\cite{Smi99}.
In most of the panels,
the dotted-dashed and solid curves are indistinguishable.}
\label{fig:obs.2.00.av18}. 
\end{figure}

\begin{figure}[p]
\includegraphics[height=20cm]{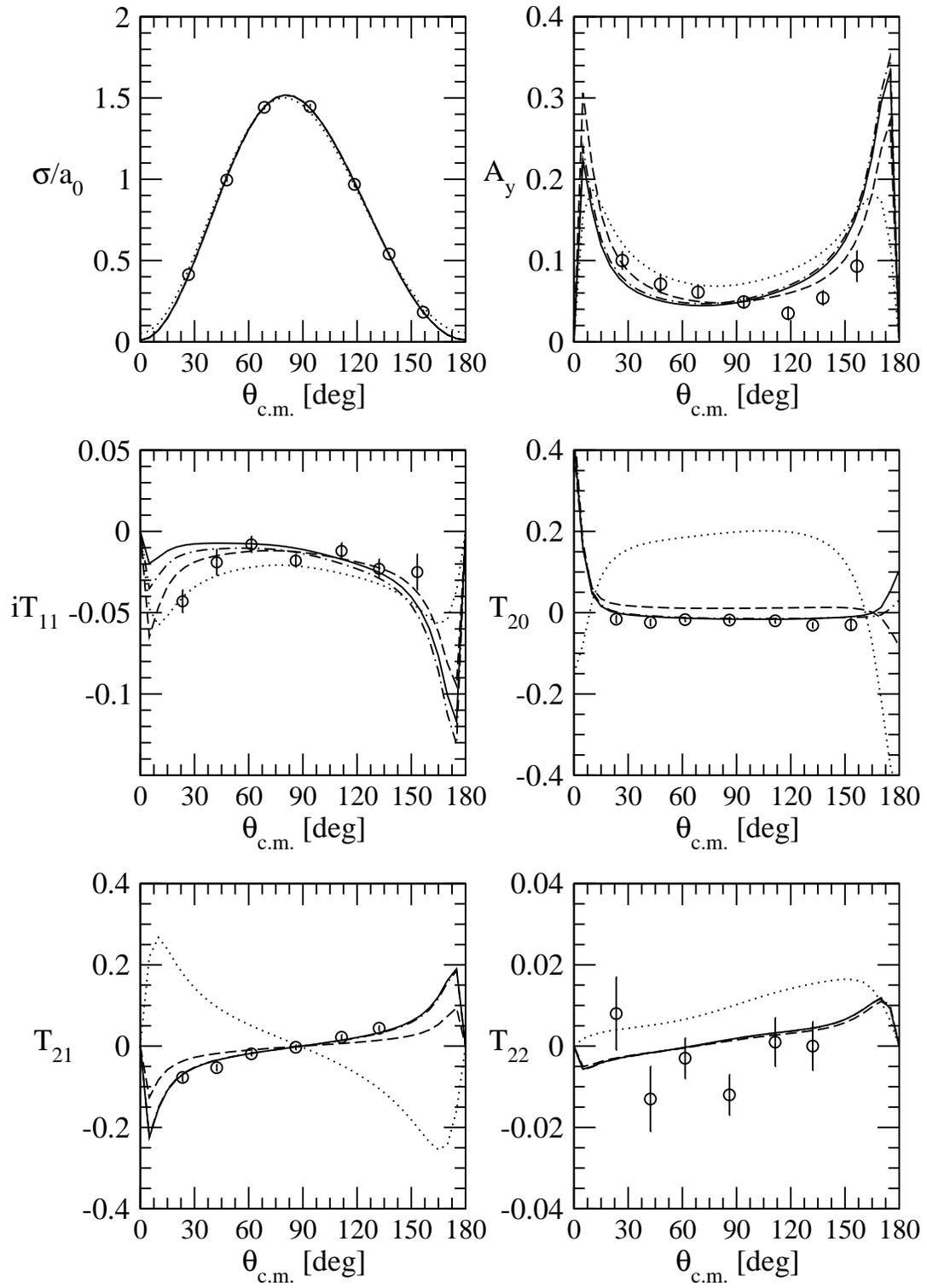}
\caption{Same as Fig.~\protect\ref{fig:obs.2.00.av18} 
at $T_{c.m.}=$3.33 MeV.
The experimental data are from Ref.~\protect\cite{Goe92}.}
\label{fig:obs.3.33.av18}
\end{figure}

In order to verify that the agreement found in Ref.~\cite{Viv00}
for other observables is not spoiled, 
the differential cross section, proton vector analyzing power and 
the four deuteron tensor analyzing powers for $pd$ capture at 
$T_{c.m.}$= 2 and 3.33 MeV, calculated with the 
AV18 two-nucleon interaction,
are compared with the experimental data of Refs.~\cite{Goe92,Smi99} in 
Figs.~\ref{fig:obs.2.00.av18} and~\ref{fig:obs.3.33.av18}.
In the figures, the dotted curves
are obtained with only one-body current contributions, 
the dashed and dotted-dashed curves are obtained using 
the one-body plus ``model-independent'' (MI) contributions 
$\bfj_{ij}^{MI,{\rm old}}(\bfq)$ and 
$\bfj_{ij}^{MI,{\rm new}}(\bfq)$, respectively,
and the solid curve is obtained when, in addition to 
$\bfj_{ij}^{MI,{\rm new}}(\bfq)$, the 
``model-dependent'' (MD) contributions ($\bfj_{ij}^{MD}(\bfq)$), due to the 
$\rho\pi\gamma$ and $\omega\pi\gamma$ transition currents and 
to the current associated with the excitation of one intermediate 
$\Delta$ resonance, are retained (``full'' model). The contributions from 
the isospin-symmetry-breaking operators 
described in Sec.~\ref{subsec:ist} are also included, but have been
found to be completely negligible.

With the new model for the
nuclear current operator, there is an overall good agreement 
between experimental results and theoretical predictions, except 
for the i$T_{11}$ observable at small c.m. angles. 
Furthermore, comparing the solid and dotted-dashed lines, we
conclude that the MD contributions are typically very small, the only
exception being those for i$T_{11}$.  As will be shown below, this 
observable is also influenced by three-body current contributions.
The improved description of the measured $T_{20}$ and $T_{21}$
observables, discussed earlier, is evident.

\begin{figure}[p]
\includegraphics[height=10cm]{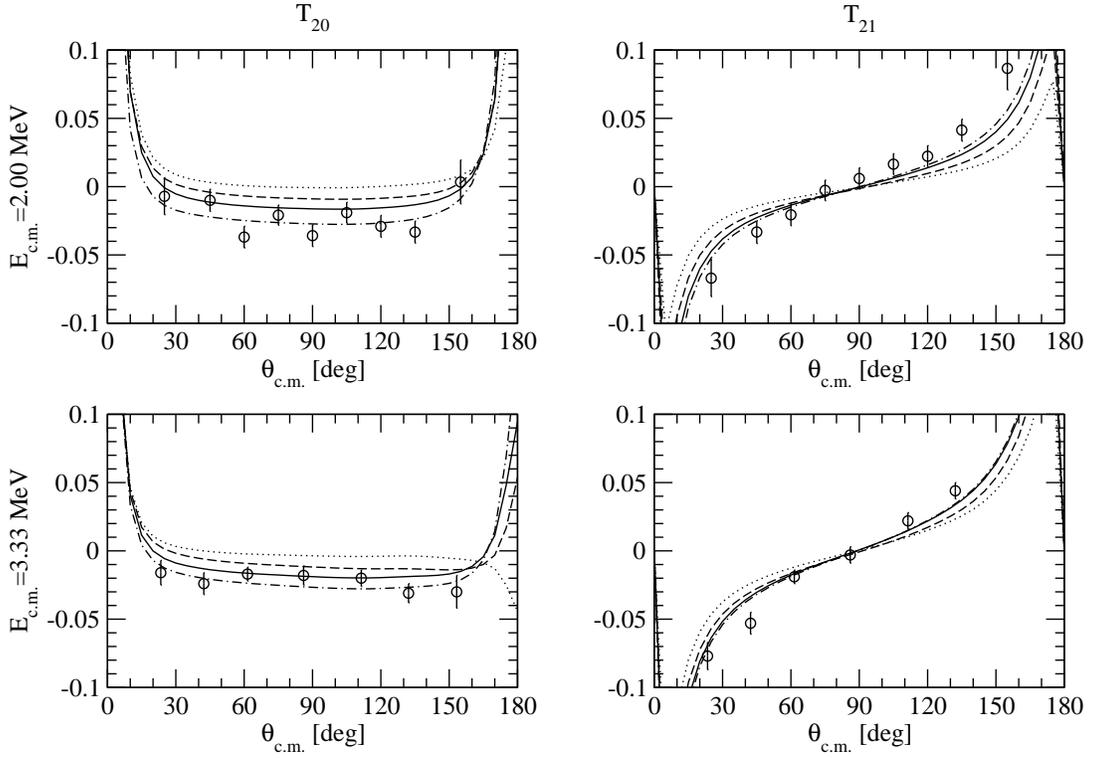}
\caption{Deuteron tensor polarization observables $T_{20}$ and $T_{21}$ 
for $pd$ radiative capture at $T_{c.m.}=$2 and 3.33 MeV
as function of the c.m. $\gamma$-$p$ scattering angle, 
obtained with the 
AV18/UIX Hamiltonian model. The dotted, dashed and solid curves correspond
to the calculation with one- and two-body currents only, with the three-body 
current obtained within the TCO approach, with the three-body current 
$\bfj_{ijk}^{ME}(\bfq)$ obtained within the ME scheme
(see Eqs.~(\protect\ref{eq:jijkt}) and~(\protect\ref{eq:jijk})). 
The results obtained in LWA 
are also shown (dotted-dashed curves).
The experimental data are from Refs.~\protect\cite{Goe92,Smi99}.}
\label{fig:t20-t21.2}
\end{figure}

We now turn our attention to the three-body current. In 
Fig.~\ref{fig:t20-t21.2} the tensor spin observables 
$T_{20}$ and $T_{21}$ for $pd$ radiative capture at $T_{c.m.}=$ 2 and 
3.33 MeV are calculated using the wave functions from 
the AV18/UIX Hamiltonian model.
The dotted curves correspond to calculations
with one- and new-ME two-body currents only. The
dashed curves have been obtained by including, in addition,
the three-body current of Ref.~\cite{Mar98}, obtained within the 
old-TCO approach. Finally, the solid curves correspond
to calculations with one-,  new-ME two-body and
new-ME three-body current 
$\bfj_{ijk}^{ME}(\bfq)$ of Eqs.~(\ref{eq:jijkt}) and~(\ref{eq:jijk}), 
obtained within the ME scheme. The results 
obtained with the three-body current operator calculated within the 
$LP$-MS scheme ($\bfj_{ijk}^{LP}(\bfq)$ of Eq.~(\ref{eq:jijklp}))
are not shown, because they coincide with those obtained with the ME method.
Finally, the dotted-dashed curves
are the LWA results.  Inspection of Fig.~\ref{fig:t20-t21.2}
indicates that, if we use the wave functions
obtained from a Hamiltonian including a TNI {\it but} disregard the 
corresponding three-body current (dotted curves),
there is a significant disagreement with the data. The use of the
old-TCO three-body current improves partially the description of the data,
but only including the new-ME (or, equivalently, the $LP$-MS) 
three-body current
leads to a satisfactory agreement with the experimental data
and LWA results, especially for the $T_{21}$ observable.

\begin{figure}[p]
\includegraphics[height=18cm]{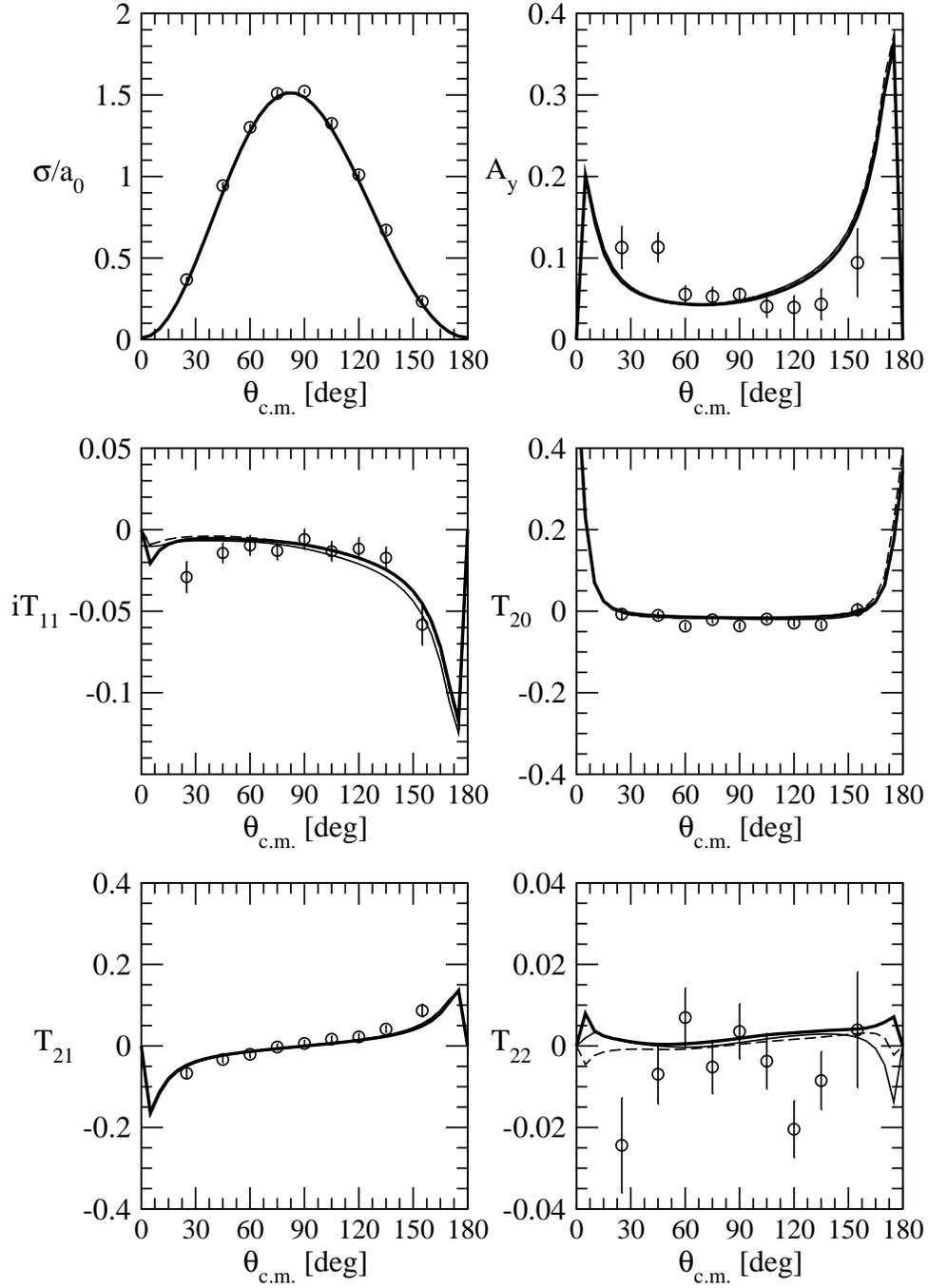}
\caption{Differential cross section, proton vector analyzing power, 
and the four deuteron tensor polarization observables 
for $pd$ radiative capture at $T_{c.m.}=$2 MeV
as function of the c.m. $\gamma$-$p$ scattering angle, 
obtained with the 
AV18 (dashed lines), AV18/TM (thin solid lines) 
and AV18/UIX (thick solid lines) Hamiltonian models. The 
model for the nuclear current operator include one-, two-, 
and three-body contributions and they satisfy the CCR with each given 
Hamiltonian.  
The experimental data are from Ref.~\protect\cite{Smi99}.
In most of the panels,
the three curves are indistinguishable.
}
\label{fig:obs.2.00}
\end{figure}

\begin{figure}[p]
\includegraphics[height=20cm]{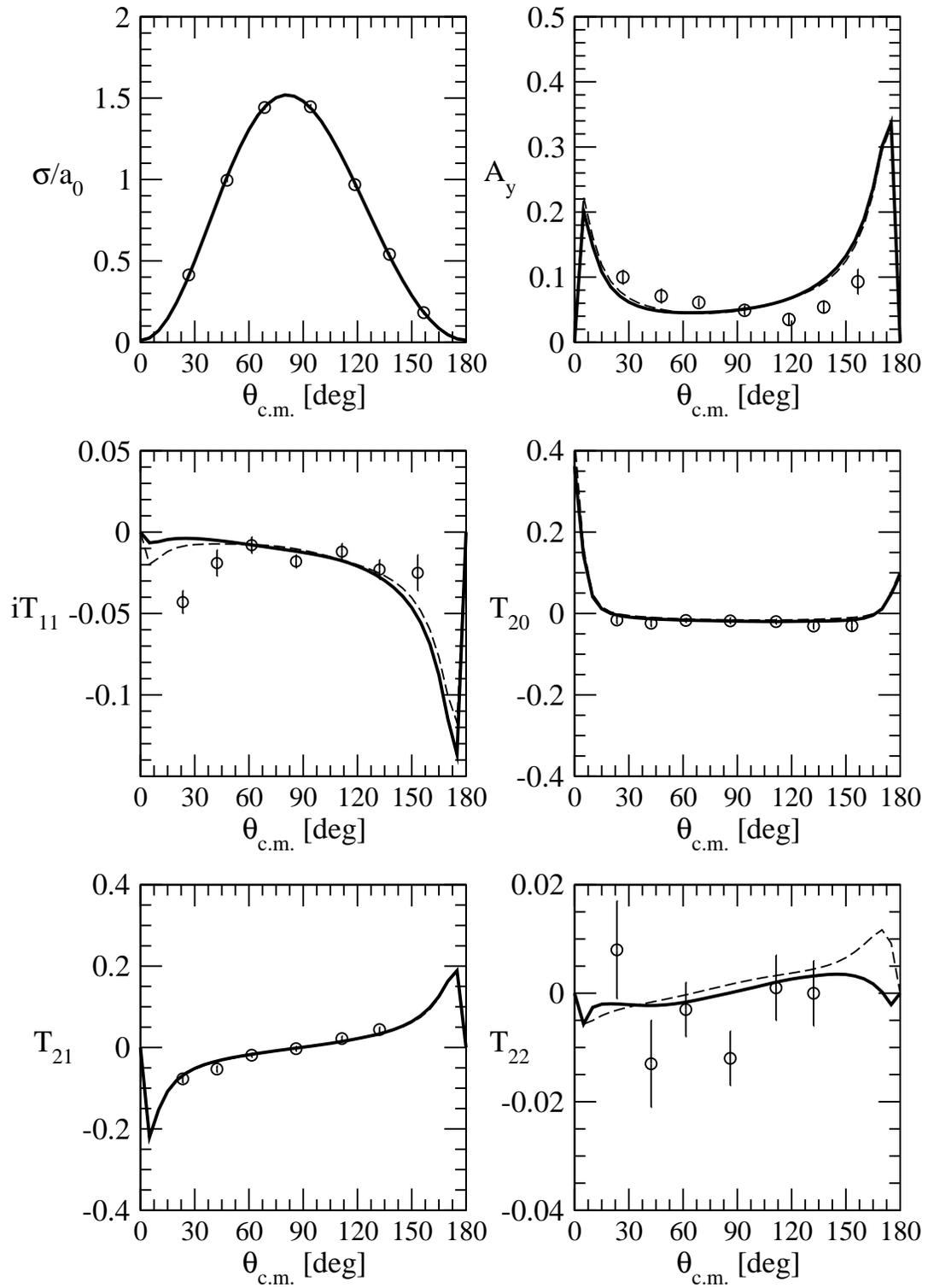}
\caption{Same as Fig.~\protect\ref{fig:obs.2.00} 
at $T_{c.m.}=$3.33 MeV. The dashed and thick solid lines 
are obtained with the AV18 and AV18/UIX Hamiltonian models, respectively. 
The experimental data are from Ref.~\protect\cite{Goe92}.}
\label{fig:obs.3.33}
\end{figure}

Finally, the differential cross section and the spin polarization 
observables for $pd$ capture at 
$T_{c.m.}$= 2 and 3.33 MeV, calculated with the AV18, AV18/UIX and AV18/TM 
Hamiltonian models are compared with the experimental data of 
Refs.~\cite{Goe92,Smi99} in Figs.~\ref{fig:obs.2.00} 
and~\ref{fig:obs.3.33}. The dashed lines are the AV18 results
obtained using the corresponding wave functions and including,
in addition to the one-body current operator, the new-ME current
$\bfj_{ij}^{MI,{\rm new}}(\bfq)$ and the MD current $\bfj_{ij}^{MD}(\bfq)$.
The thin solid curves are obtained with the Hamiltonian
including the AV18 and the Tucson-Melbourne 
(TM)~\cite{Coo79} TNI (AV18/TM model).
In this case, the current includes the one-body, the 
new-ME and MD two-body and the new-ME three-body currents (constructed
to satisfy the CCR with the AV18/TM Hamiltonian).
The thick solid lines are finally obtained with the AV18/UIX
Hamiltonian, and include the corresponding set of one-, two-, and
three-body currents in the new-ME scheme.
As the figures suggest, there are no significant differences 
between the AV18, AV18/TM and AV18/UIX results,
except for some tiny effects in the ${\rm i}T_{11}$ observable.

\subsection{Magnetic structure of $A$=3 nuclei}
\label{subsec:ms3}

\begin{figure}[p]
\includegraphics[height=9cm]{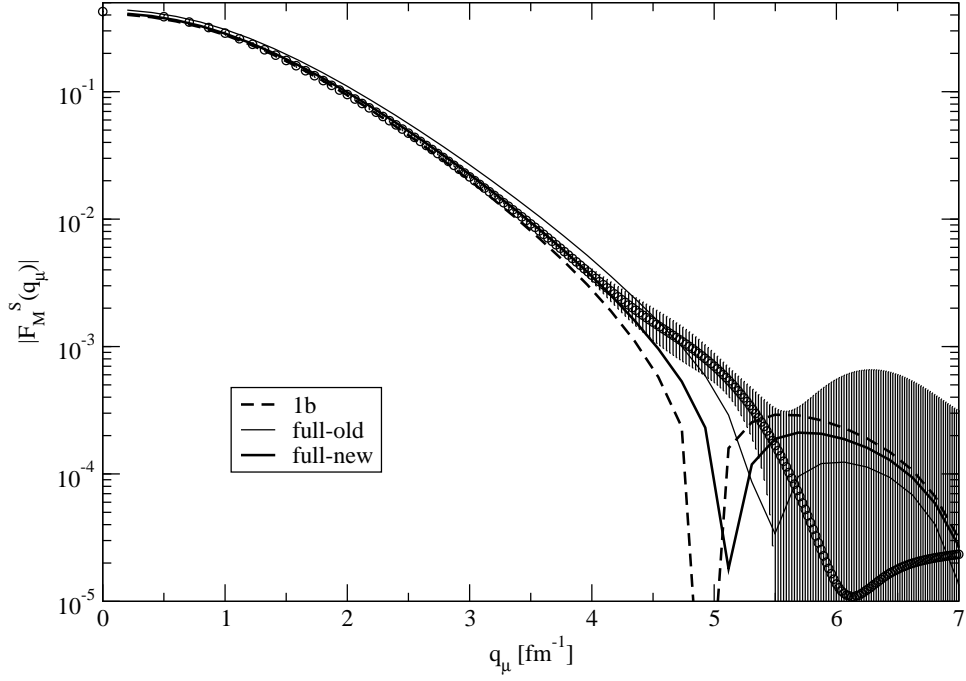}
\caption{The isoscalar combination of the $^3$He and $^3$H 
magnetic form factors, obtained with single nucleon 
currents (1b), and with the inclusion of 
two- and three-body currents in the new model summarized 
in Secs.~\protect\ref{subsec:sumj2} and~\protect\ref{subsec:sumj3} 
(full-new). Also listed are the results obtained with the 
old-ME two-body 
and old-TCO three-body currents of Ref.~\protect\cite{Mar98} (full-old).
The experimental data are from 
Ref.~\protect\cite{Col65,McC77,Sza77,Arn78,Dun83,Ott85,Jus85,Bec87,Amr94}.}
\label{fig:ismff}
\end{figure}

\begin{figure}[p]
\includegraphics[height=9cm]{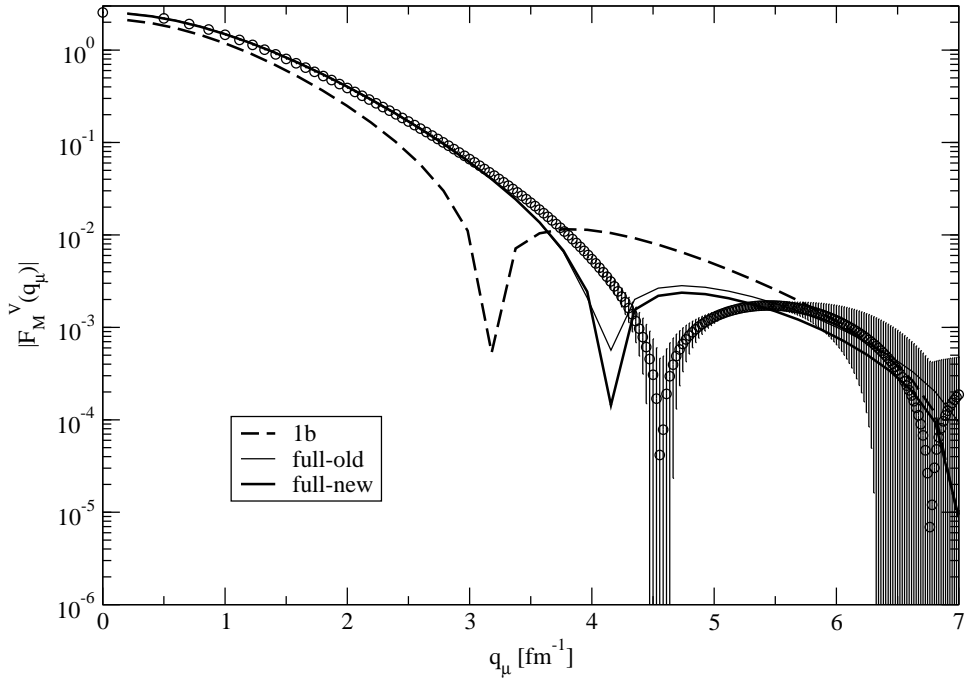}
\caption{Same as Fig.~\protect\ref{fig:ismff}, but 
for the isovector combination of the $^3$He and $^3$H 
magnetic form factors.
The experimental data are from 
Ref.~\protect\cite{Col65,McC77,Sza77,Arn78,Dun83,Ott85,Jus85,Bec87,Amr94}.}
\label{fig:ivmff}
\end{figure}

The isoscalar and isovector combinations of the magnetic moments 
and form factors of 
$^3$He and $^3$H are given in Table~\ref{tab:mm} and 
Figs.~\ref{fig:ismff} and~\ref{fig:ivmff}, respectively. 
The nuclear wave functions have been calculated using the 
the AV18/UIX Hamiltonian model. The results labeled ``1b'' are 
obtained retaining only the one-body current operator,  
those labeled ``full-new'' are obtained including, in addition, 
the new-ME two-body current contributions and the 
three-body current contributions calculated in the ME scheme. 
Also listed, 
are the results obtained with the old-ME two-body and old-TCO 
three-body currents, as in Ref.~\cite{Mar98}, labeled ``full-old''.
These last results are slightly different from those reported in 
Ref.~\cite{Mar98}, due to the present use of more accurate trinucleon 
wave functions. The experimental data are from 
Refs.~\cite{tunl,Col65,McC77,Sza77,Arn78,Dun83,Ott85,Jus85,Bec87,Amr94}.

Note that: (i) the ``full-old'' and ``full-new'' results 
for the isoscalar and isovector magnetic moments 
differ by less than 1 \% and are very close to the experimental data; 
(ii) the experimental results for the isovector magnetic form factor 
are fairly well reproduced for momentum transfer $q_\mu\leq 3.5$ fm$^{-1}$, 
and the ``full-new'' curve is slightly closer to the experimental data 
in the region $q_\mu\geq 4$ fm$^{-1}$ than the ``full-old'' curve; 
(iii) the ``full-new'' curve for the isoscalar magnetic form factor 
is closer to the experimental data than the ``full-old'' curve in the 
region  $q_\mu\leq 4$ fm$^{-1}$; (iv) the ``full-new'' curves 
for the isoscalar and isovector form factors are in disagreement with the 
data for $q_\mu\geq$ 4--4.5 fm$^{-1}$, and the discrepancy 
between theory and experiment in this region remains unresolved. 
However, the experimental data for the isoscalar magnetic form 
factor have large errors at high $q_\mu$ values.

\begin{table}
\caption[Table]{Isoscalar and isovector combinations $\mu_S$ and 
$\mu_V$ of the 
$^3$He and $^3$H magnetic moments, in nuclear magnetons, 
compared with experimental data. The results 
labeled ``1b'' are obtained with single-nucleon currents 
only, those labeled ``full-new'' retain in addition 
two- and three-body currents in the new model summarized 
in Secs.~\protect\ref{subsec:sumj2} and~\protect\ref{subsec:sumj3}. 
Also listed are the results obtained with the old-ME two-body 
and old-TCO three-body currents of Ref.~\protect\cite{Mar98}
(``full-old''). The experimental data are from 
Ref.~\protect\cite{tunl}.}
\begin{tabular}{lll}
\hline
& $\mu_S$ & $\mu_V$\\
\hline
1b &  0.407 & 2.165 \\
Full-new & 0.414 & 2.539 \\
Full-old & 0.442 & 2.557 \\
\hline
Expt. & 0.426 & 2.553 \\
\hline
\end{tabular}
\label{tab:mm}
\end{table}

\subsection{$A$=2 radiative capture reaction and deuteron photodisintegration}
\label{subsec:rcap2}

The calculated values for the $^1$H($n$,$\gamma$)$^2$H cross section 
at thermal neutron energies with the old- and new-ME models 
of the current are listed in Table~\ref{tab:np}. The AV18 two-nucleon 
interaction is used.

\begin{table}
\caption[Table]{Total cross section in mb for the 
$n$$p$ radiative capture, calculated using the AV18 two-nucleon 
interaction.  The results 
labeled ``1b'' are obtained with single-nucleon currents 
only, those labeled ``1b+2b-MI(old-ME)'' and 
``1b+2b-MI(new-ME)'' retain in addition 
model-independent two-body currents in the old-ME and new-ME model 
summarized in Secs.~\protect\ref{subsec:sumj2}.
The results labeled ``full-old'' and ``full-new''
are obtained by adding the contributions of
the model-dependent two-body currents 
to the ``1b+2b-MI(old-ME)'' and ``1b+2b-MI(new-ME)'' results, 
respectively. The experimental value is from
Ref.~\protect\cite{Mug81}.}
\begin{tabular}{ll}
\hline
& $\sigma$(mb)\\
\hline
1b & 304.6 \\
1b+2b-MI(old-ME) & 326.1 \\
1b+2b-MI(new-ME) & 324.7 \\
Full-old & 334.2 \\
Full-new & 332.7 \\
\hline
Expt. & 332.6 $\pm$ 0.7 \\
\hline
\end{tabular}
\label{tab:np}
\end{table}

The small difference between the results obtained with the old- and 
new-ME models is due to differences in the isovector structure of the 
two-body currents from the momentum dependent terms of the AV18. 
The result with the new-ME model happens to be in perfect agreement 
with the experimental value reported in Ref.~\cite{Mug81}.

\begin{figure}[p]
\includegraphics[height=9cm]{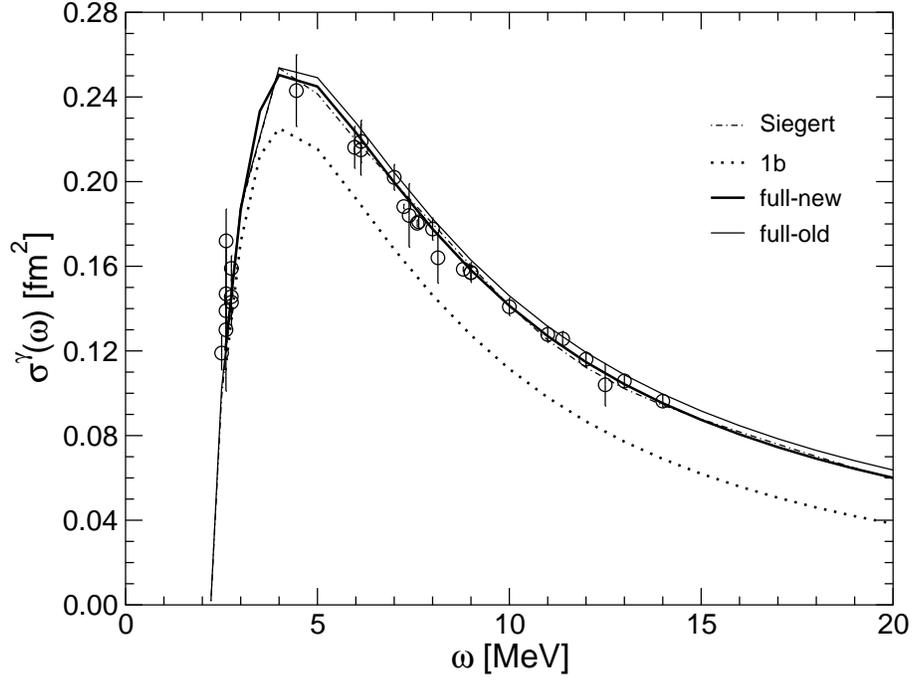}
\caption{The deuteron photodisintegration cross sections 
obtained with one-body current (1b) and with, in addition, 
the old (full-old) or new (full-new) models for two-body 
currents are compared to experimental values. Also shown 
are the results obtained by using the Siegert form for the $E_1$ 
transition.
The experimental data are from 
Refs.~\protect\cite{Bis50,Sne50,Col51,Car51,Bir85,Mor89,DeG92}.}
\label{fig:gd}
\end{figure}

The deuteron photodisintegration cross sections up to 20 MeV 
photon energies obtained in impulse approximation 
and with the full-old and full-new ME current models are shown 
in Fig.~\ref{fig:gd}, along with the experimental 
data~\cite{Bis50,Sne50,Col51,Car51,Bir85,Mor89,DeG92}.
Also shown in Fig.~\ref{fig:gd} are the predictions in which 
the dominant $E_1$ transitions connecting the deuteron and 
$n$$p$ triplet $P$-waves are calculated using the Siegert form 
for the $E_1$ operator, valid in the LWA limit. Again, differences 
between the results obtained with the new- and old-ME current models
is to be attributed mostly to differences in the isovector currents 
originating from the momentum dependence of the AV18. Indeed, 
these terms ensure that the new-ME current is exactly conserved, and 
make the corresponding results essentially identical to the Siegert 
predictions. 

\subsection{$A$=3 radiative capture reactions}
\label{subsec:rcap3}

We report here the results for the radiative capture 
reactions $^2$H($n,\gamma$)$^3$H and $^2$H($p,\gamma$)$^3$He,
obtained with the AV18/UIX Hamiltonian model. 

\subsubsection{The $^2$H($n,\gamma$)$^3$H radiative capture reaction}
\label{subsubsec:nd}

At thermal energies the $nd$ capture reaction proceeds through 
$S$-wave capture predominantly via magnetic dipole transitions
from the initial doublet $J$=1/2 and quartet $J$=3/2 $nd$ scattering
states.  In addition, there is a small contribution due to
an electric quadrupole transition from the initial quartet state.

The results for the thermal energy cross section and photon 
polarization parameter are presented in Table~\ref{tab:ndcapt}, 
along with the experimental data~\cite{JBB82,Kea88}. 
As can be seen by inspection of the table, 
the cross section calculated with single-nucleon 
currents is approximately a factor of 2 smaller
than the measured value. A previous calculation~\cite{Viv96}
gave $\sigma_T({\rm 1b})=0.223$ mb, very close to the
results presented in the first row of Table~\ref{tab:ndcapt}.
Inclusion of the model-independent new-ME two-body
currents leads to a value of $\sigma_T$  10\% smaller
than obtained earlier with the MI old-ME currents
of Ref.~\cite{Viv96}.
By adding the model-dependent two-body current, an estimate
of $\sigma_T=0.523$ mb is obtained. This value is to be
compared with the corresponding result $\sigma_T=0.558$ mb 
obtained in Ref.~\cite{Viv96}. The
use of the present MI two-body current operators 
therefore leads to an estimate closer 
to the experimental datum $\sigma_T=0.508\pm015$ mb~\cite{JBB82}. 
However, the addition of the three-body currents, 
which give a rather sizable contribution
as can be seen from the row labeled ``full-new''
in Table~\ref{tab:ndcapt},
brings the total cross section to $\sigma_T=0.556$ mb.
The 9\% slight overprediction is presumably due to the 
model-dependent currents associated with the $\Delta$
excitations. Fortunately, at $T_{c.m.}>1$ MeV, this MD 
current gives a negligible contribution to the cross section 
and the other polarization observables, as already shown
in Figs.~\ref{fig:obs.2.00.av18} and~\ref{fig:obs.3.33.av18}.

\begin{table}
\caption[Table]{\label{tab:ndcapt}
Cross section (in mb) and photon polarization 
parameter $R_c$ of the reaction $^2$H($n,\gamma$)$^3$H  for the 
AV18/UIX potential model at thermal neutron energy. The results 
labeled ``1b'' are obtained with single-nucleon currents 
only, those labeled ``1b+2b-MI(old-ME)'' and 
``1b+2b-MI(new-ME)'' retain in addition 
MI two-body currents in the old-ME and new-ME 
scheme, respectively.
The results labeled ``$\ldots$+2b-MD'' are obtained by adding 
the model-dependent two-body currents to the single-nucleon
and the MI new-ME two-body currents. Finally, the
results labeled ``full-new'' are obtained by including
to the latter model also the
the contribution of the ME three-body
currents, defined in Sec.~\protect\ref{subsec:sumj3}.
The experimental values for $\sigma_T$ and $R_c$ are from
Refs.~\protect\cite{JBB82,Kea88}, respectively.
}  
\begin{tabular}{l ll }
\tableline
 Current component & $\sigma_T$ & $R_c$  \\
\tableline
1b                & 0.227  & --0.061  \\
1b+2b-MI(old-ME)  & 0.462  & --0.446  \\
1b+2b-MI(new-ME)  & 0.418  & --0.429  \\
$\ldots$+2b-MD    & 0.523  & --0.469   \\ 
Full-new          & 0.556  & --0.476   \\ 
\tableline
Expt.         & $0.508 \pm 0.015$ & --$0.42\pm 0.03 $ \\
\tableline
\end{tabular}
\end{table}

The photon polarization parameter is very sensitive to two-body
currents (for its definition in terms of RMEs, see Ref.~\cite{Viv96}).
For example, for the AV18/UIX Hamiltonian, their inclusion 
produces roughly a six-fold increase, in absolute
value, of the single-nucleon prediction. Also in this case,
we find a 13\% overprediction (in absolute value) of this
parameter. The small reduction of $|R_c|$, found when the
new-ME model for the two-body current is used, is compensated by
the inclusion of the three-body currents.

\begin{figure}[p]
\includegraphics[height=9cm]{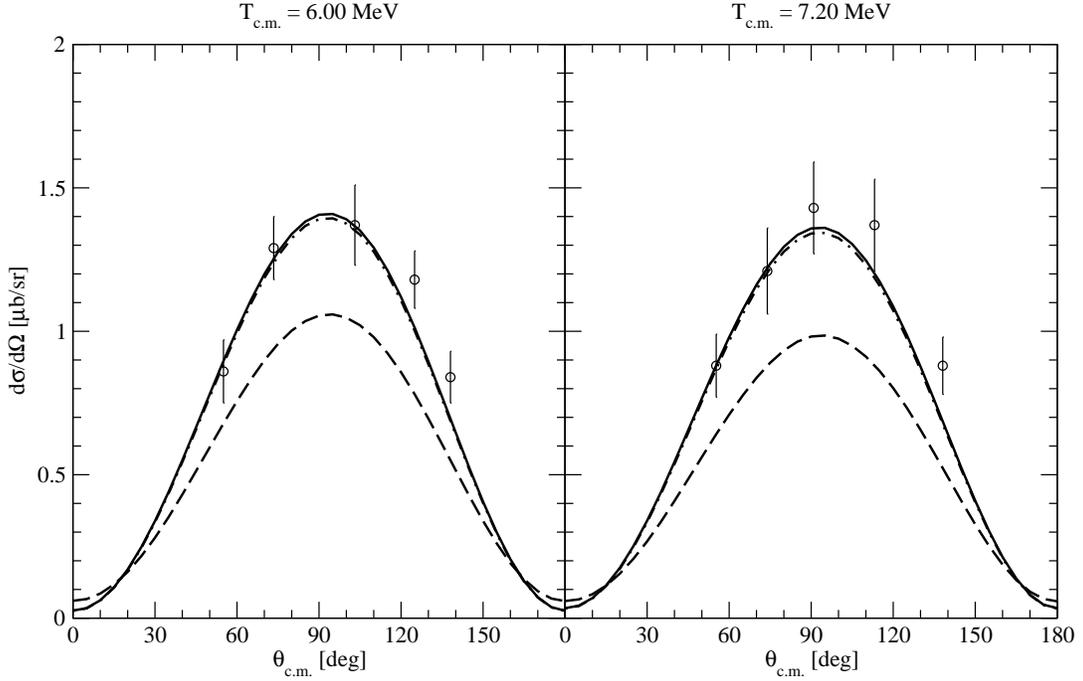}
\caption{Differential cross section in $\mu$b/sr
for $nd$ radiative capture at $T_{c.m.}=$6 MeV
and 7.2 MeV as function of the c.m. $\gamma$-$n$ 
scattering angle, obtained with the 
AV18/UIX Hamiltonian model. The dashes curves have been obtained 
using the single-nucleon current only. The dotted-dashed curves
have been obtained with the inclusion of the new-ME 
two-body currents. Finally, the solid curves
have been obtained by adding the ME three-body current.
The experimental data are from Ref.~\protect\cite{mitev86}.}
\label{fig:ndxs}
\end{figure}

At higher energies, there exist several measurements of the 
unpolarized differential cross section for both the radiative 
capture process $^2$H($n,\gamma$)$^3$H~\cite{mitev86} and 
for the ``time-reversed'' process 
$^3$H($\gamma,n$)$^2$H~\cite{bosch64,faul81,skopik81,koseik66,pfeiffer68}. 
In the c.m. system and
at low energies, the unpolarized cross sections are related 
by the principle of detailed balance

\begin{equation}
   \left({d\sigma_{\rm photo}\over d\Omega}\right)_{c.m.}=
   {3\over 2} \left({p\over q}\right)^2
   \left({d\sigma_{\rm capt}\over d\Omega}\right)_{c.m.}
   \>\>\> ,\label{eq:photo}
\end{equation}
where $q$ and $p$ are the $\gamma$ and the relative $nd$ momenta,
respectively. In Figure~\ref{fig:ndxs}, we compare our
predictions for $(d\sigma_{\rm capt}/d\Omega)_{c.m.}$ 
with the experimental results of Ref.~\cite{mitev86},
which is the only direct measurement of the
$^2$H($n,\gamma$)$^3$H differential cross section.
The dashed curves represent the results obtained with the inclusion 
of the single-nucleon current only, the dotted-dashed curves are obtained 
with one- and new-ME two-body contributions (both MI and MD), and
the solid curves represent the ``full-new'' result, 
obtained including, in addition to the one-body and above-mentioned
two-body currents, also the ME three-body current contributions.
At these energies the process is dominated by
$E_1$ transitions between the $nd$ $P$-wave states
and the ${}^3$H ground state, as can be inferred from the bell-shape
of the curves. The slight distortion of the peak is
due to non-negligible contributions from
$E_2$ RMEs, coming, in particular, from 
the $J=3/2,5/2$ states with $S=3/2$.
Our ``full-new'' calculation reproduces quite well the
experimental data, with some differences  at large
angles. As will be shown below,
this has some consequences for the so-called fore-aft asymmetry,
 discussed in Sec.~\ref{sec:foreaft}.

\begin{figure}[p]
\includegraphics[height=9cm]{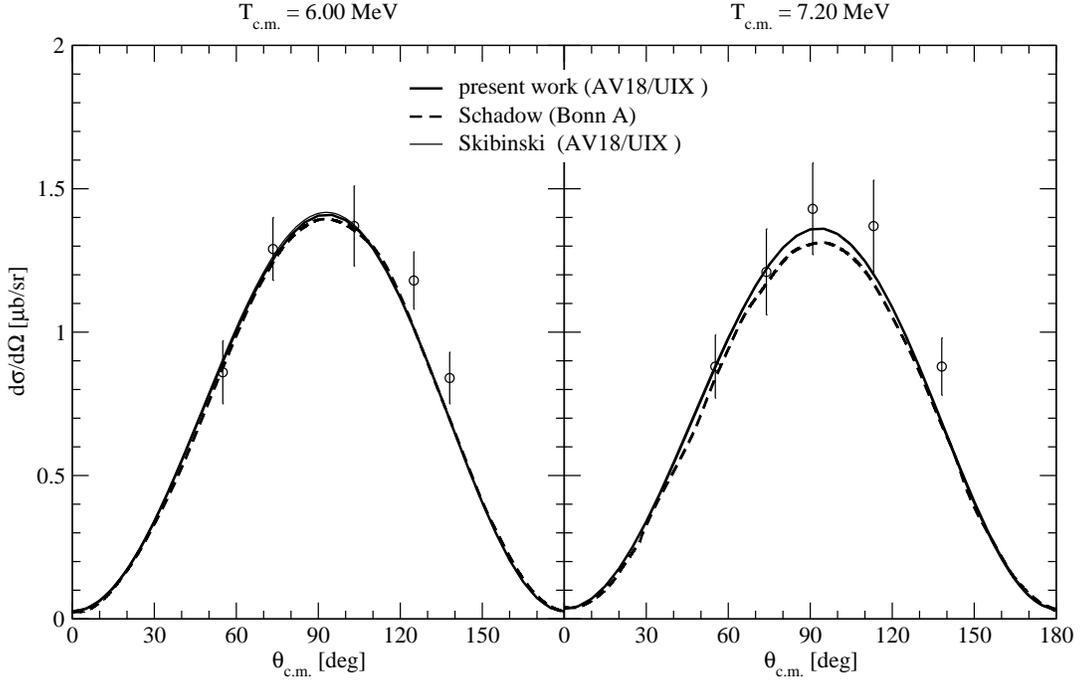}
\caption{Differential cross section in $\mu$b/sr
for $nd$ radiative capture at $T_{c.m.}=$6 MeV
and 7.2 MeV as function of the c.m. $\gamma$-$n$ 
scattering angle, obtained in present work (thick solid curves)
and in Ref.~\protect\cite{Skib03} (thin solid curves)
using the AV18/UIX potential model. The two curves are 
practically indistinguishable. The dashed curves 
are the results of Ref.~\protect\cite{Schad01}
obtained with the Bonn A potential model~\protect\cite{Mac87}.
The experimental data are from Ref.~\protect\cite{mitev86}.}
\label{fig:ndxs_comp}
\end{figure}

In Fig.~\ref{fig:ndxs_comp}, we compare our
``full-new'' results with those obtained 
in Refs.~\cite{Schad01,Skib03}. In Ref.~\cite{Skib03},
the same potential model (AV18/UIX) as in present work
has been used, but the authors adopt a slightly different current model
(they do not consider $\bfj_{ij}^{ME}(\bfq;v^p)$,
$\bfj_{ij}^{MD}(\bfq)$ and the three-body current).
In Ref.~\cite{Schad01}, the exchange currents are taken into account 
using Siegert's theorem, and a different
two-body potential model has been used (Bonn A~\cite{Mac87}),
without any inclusion of three-nucleon forces.
As can be seen by inspecting Fig.~\ref{fig:ndxs_comp},
the three theoretical calculations are practically
the same, with some differences with results
of Ref.~\cite{Schad01} at $T_{c.m.}=7.20$ MeV.
This difference is likely due to the use of a
different potential model, which slightly underestimates
the ${}^3$H binding energy. 

In addition to unpolarized cross sections, there are also
a few analyzing-power angular distribution data~\cite{mitev86},
but they have large error bars, and therefore we have decided 
not to perform a comparison for this observable.

\subsubsection{The $^2$H($p,\gamma$)$^3$He radiative capture reaction}
\label{subsubsec:pd}

\begin{figure}[p]
\includegraphics[height=9cm]{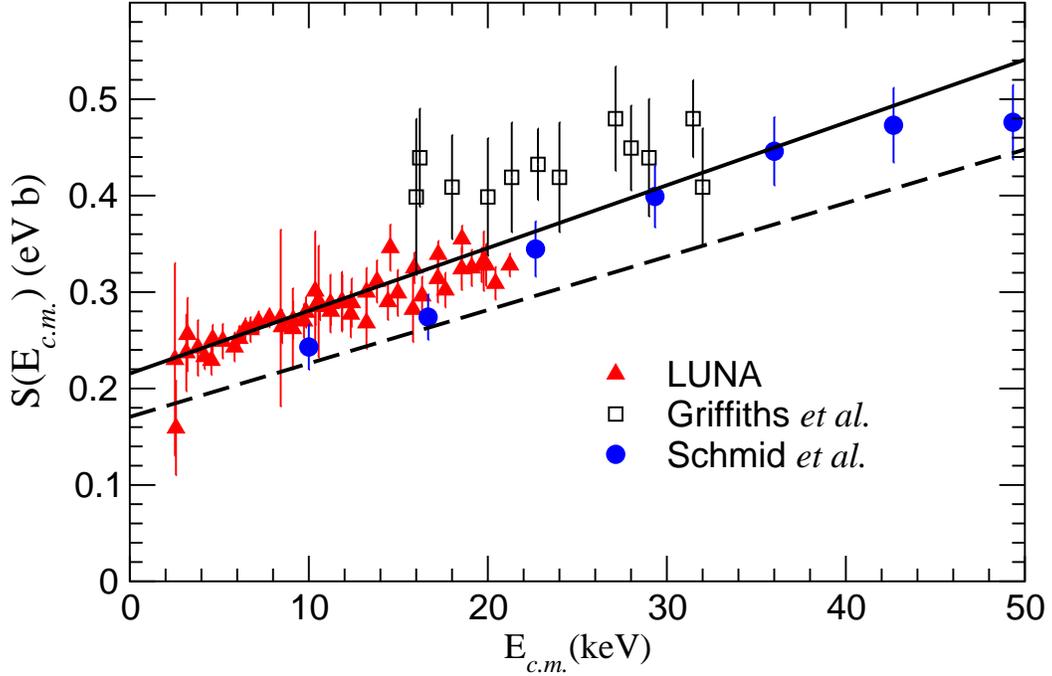}
\caption{The $S$ factor of the $^2$H($p,\gamma$)$^3$He reaction, 
obtained with the AV18/UIX Hamiltonian model and with the 
one-body current only (dashed line), and with the one-, two- and 
three-body current (solid line) is compared with the 
experimental results from Refs.~\protect\cite{Lun02,Gri63,Sch95,Sch96}.}
\label{fig:sfpd}
\end{figure}

In Fig.~\ref{fig:sfpd} we present the results for the 
astrophysical $S$ factor of the $^2$H($p,\gamma$)$^3$He 
radiative capture reaction at thermal energies. 
This quantity is defined as

\begin{equation}
S(T_{c.m.})=T_{c.m.}\,\sigma_T(T_{c.m.})\,{\rm e}^{2\pi\alpha/v_{rel}}\ ,
\label{eq:sf}
\end{equation}
where $T_{c.m.}$ is the $pd$ c.m. kinetic energy, 
$\sigma_T(T_{c.m.})$ is the total cross section, $\alpha$ is 
the fine structure constant, and $v_{rel}$ is the $pd$ relative velocity.
The experimental data are from Refs.~\cite{Lun02,Gri63,Sch95,Sch96}.
The solid curve represents the ``full-new'' result, 
obtained including, in addition to the one-body currents, 
also the new-ME two-body current contributions and the 
three-body current contributions calculated in the ME scheme. 
The dashed curve 
represents the result obtained with the inclusion of the 
single-nucleon current only. Here, no significant difference has been seen 
between the results obtained with the present model for the 
nuclear current operator and the ``old'' one of Refs.~\cite{Viv96,Viv00}. 
The agreement between the theoretical predictions and the 
experimental data, especially the very recent LUNA data~\cite{Lun02}, 
is excellent. In particular, the calculated $S$ factor at 
zero energy is 0.219 eV~b~\cite{Viv00}, in very nice agreement with the 
LUNA result of 0.216$\pm$0.010 eV~b.

\begin{figure}[p]
\includegraphics[height=18cm]{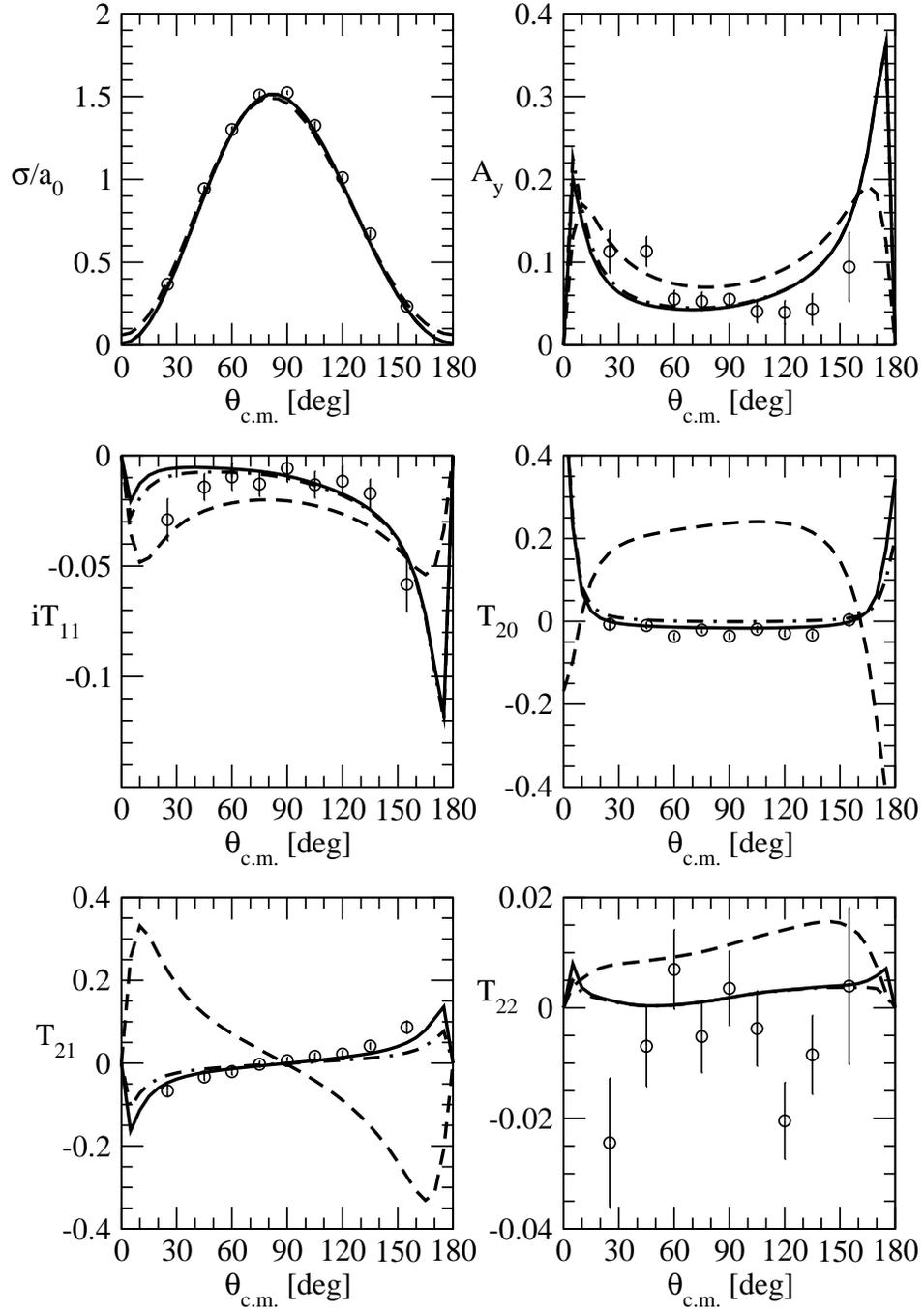}
\caption{Differential cross section, proton vector analyzing power, 
and the four deuteron tensor polarization observables 
for $pd$ radiative capture at $T_{c.m.}=$2 MeV
as function of the c.m. $\gamma$-$p$ scattering angle, 
obtained with the 
AV18/UIX Hamiltonian model. The dashed, dotted-dashed and solid lines 
are obtained with one-body contributions, one- and new-ME two-body 
contributions, and one-, two- and three-body contributions. 
The experimental data are from Ref.~\protect\cite{Smi99}.}
\label{fig:obs.2.00f}
\end{figure}

\begin{figure}[p]
\includegraphics[height=20cm]{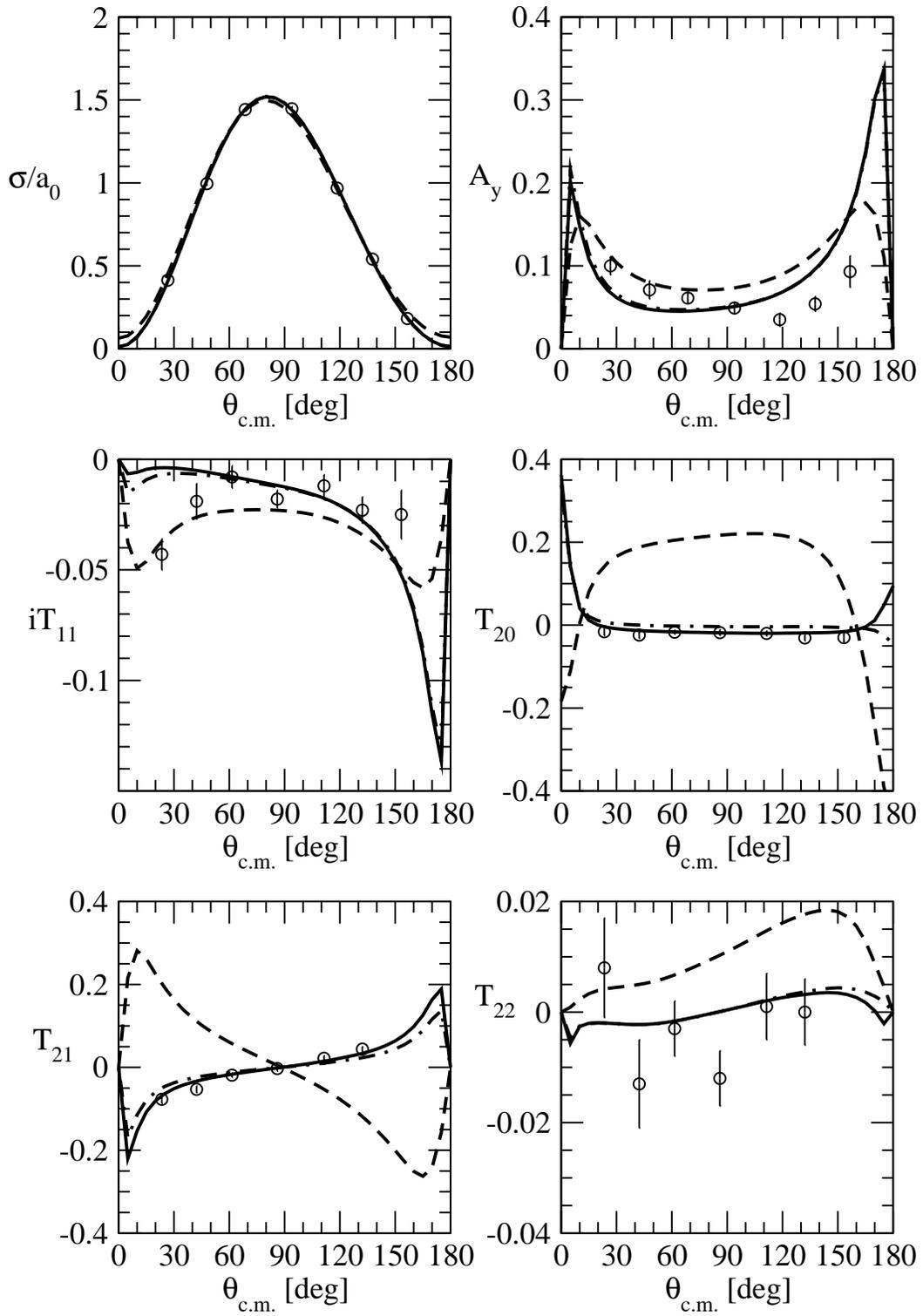}
\caption{Same as Fig.~\protect\ref{fig:obs.2.00f} 
at $T_{c.m.}=$3.33 MeV. 
The experimental data are from Ref.~\protect\cite{Goe92}.}
\label{fig:obs.3.33f}
\end{figure}

\begin{figure}[p]
\includegraphics[height=13cm]{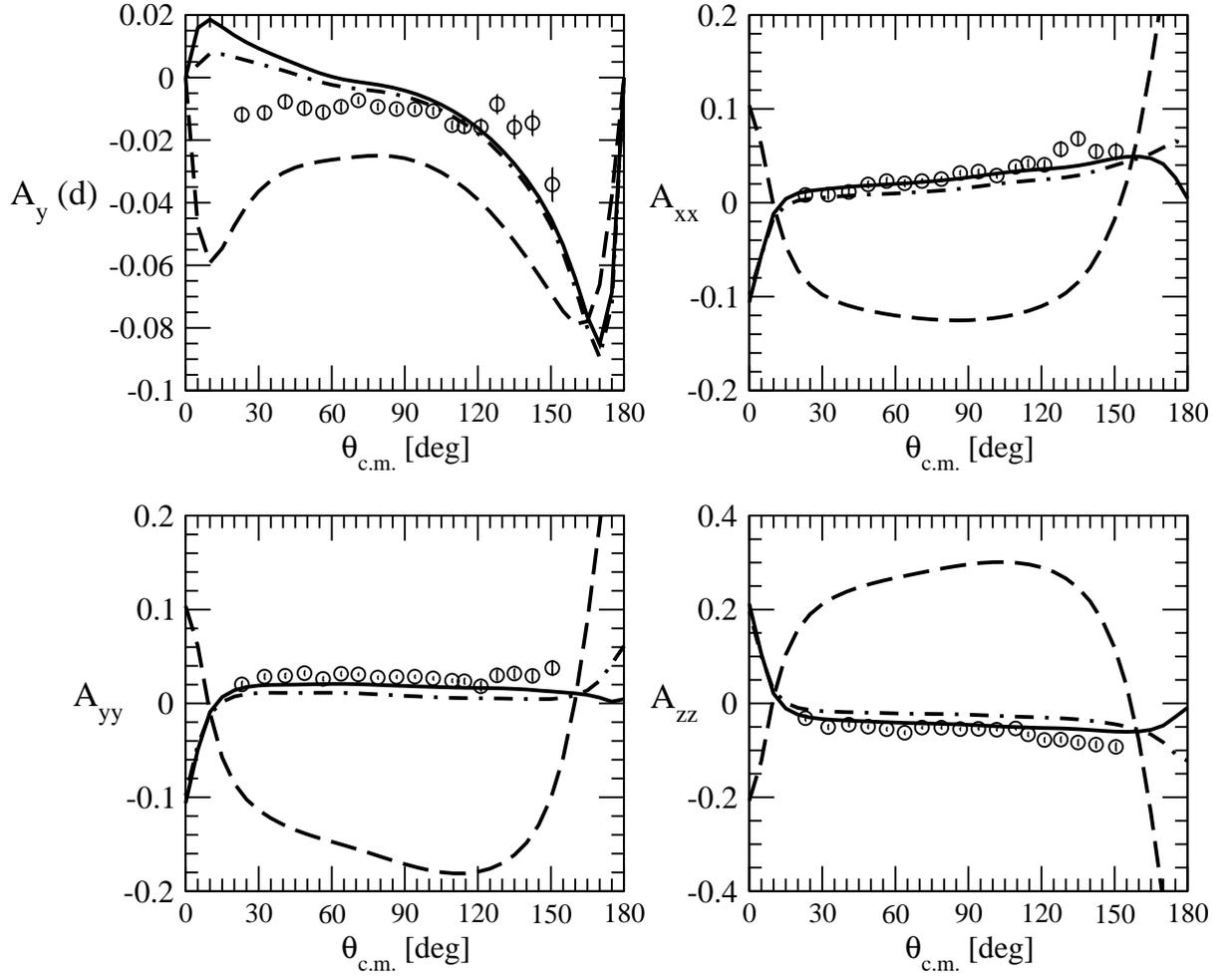}
\caption{Four deuteron tensor polarization observables 
for $pd$ radiative capture at $T_{c.m.}=$5.83 MeV
as function of the c.m. $\gamma$-$p$ scattering angle, 
obtained with the 
AV18/UIX Hamiltonian model. The same notation as in 
Fig.~\protect\ref{fig:obs.2.00f} is used for the different lines. 
The experimental data are from Ref.~\protect\cite{Aki01}.}
\label{fig:obs.5.83f}
\end{figure}

\begin{figure}[p]
\includegraphics[height=15cm]{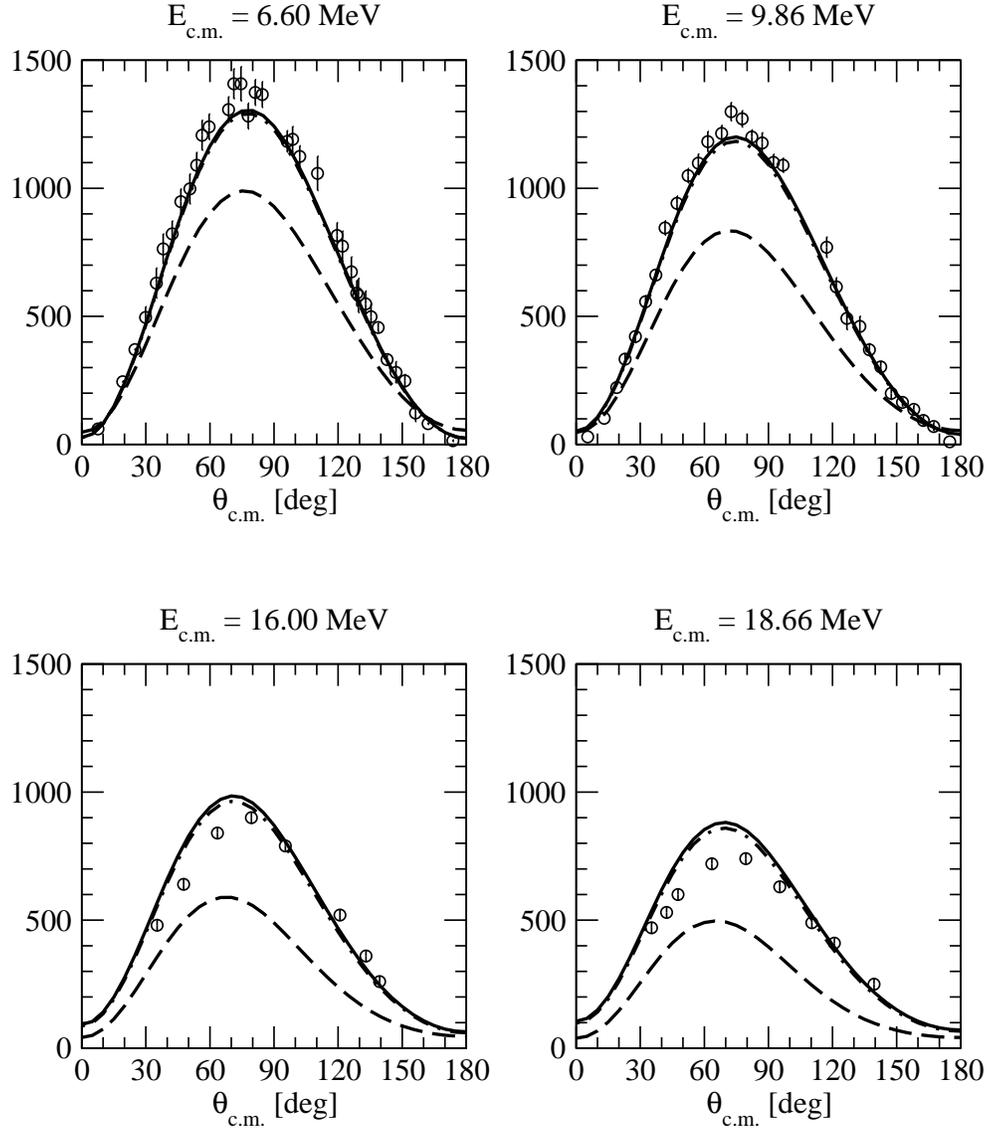}
\caption{Differential cross section in $\mu$b/sr
for $pd$ radiative capture up to $T_{c.m.}=$18.66 MeV
as function of the c.m. $\gamma$-$p$ scattering angle, 
obtained with the 
AV18/UIX Hamiltonian model. The same notation as in 
Fig.~\protect\ref{fig:obs.2.00f} is used for the different lines. 
The experimental data are from Ref.~\protect\cite{Bel70} for 
$T_{c.m.}=$6.60 and 9.86 MeV, and from 
Ref.~\protect\cite{Ang83} for the other cases.}
\label{fig:xs}
\end{figure}

\begin{figure}[p]
\includegraphics[height=15cm]{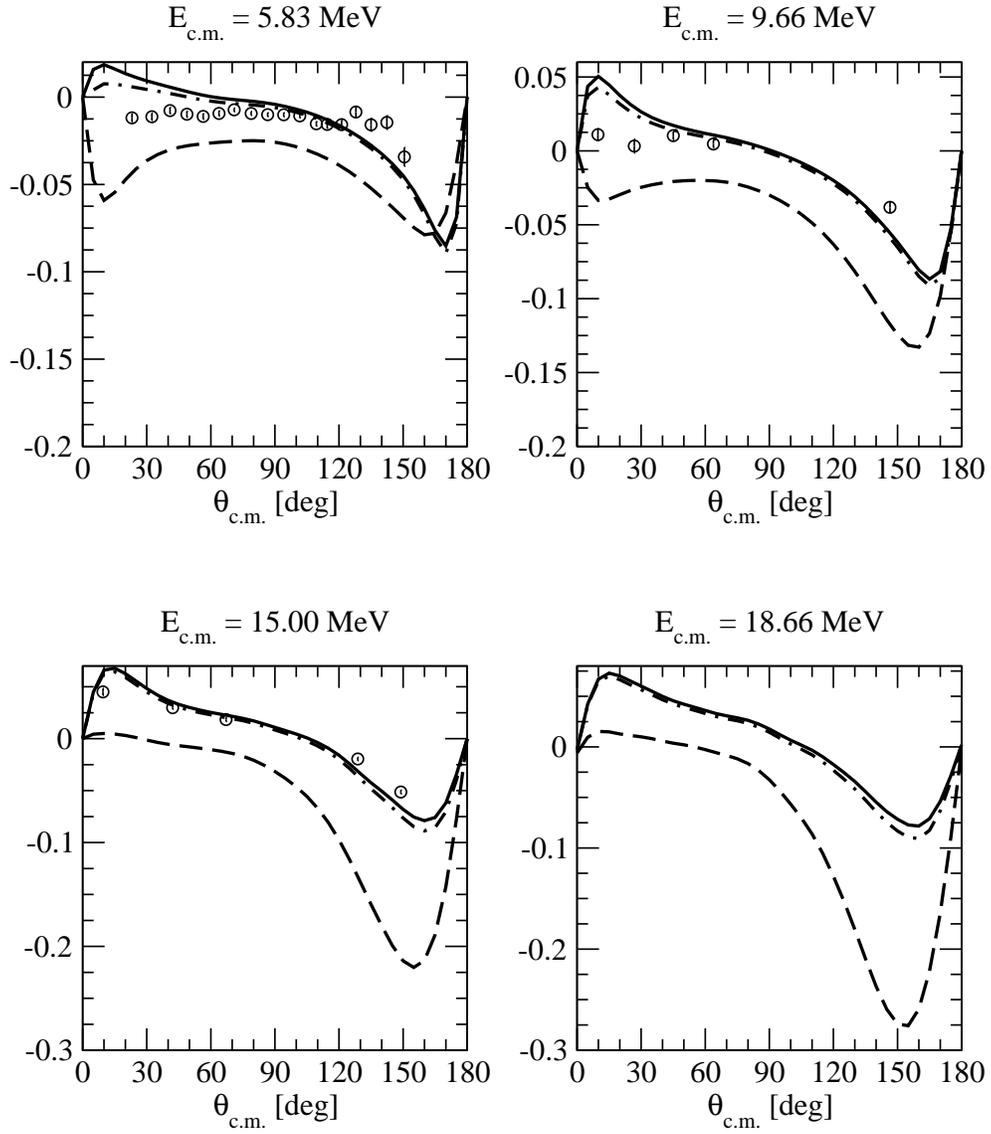}
\caption{Deuteron vector analyzing power $A_y(d)$
for $pd$ radiative capture up to $T_{c.m.}=$18.66 MeV
as function of the c.m. $\gamma$-$p$ scattering angle, 
obtained with the 
AV18/UIX Hamiltonian model. The same notation as in 
Fig.~\protect\ref{fig:obs.2.00f} is used for the different lines. 
The experimental data are from Ref.~\protect\cite{Aki01} for 
$T_{c.m.}=$5.83 MeV and from Refs.~\protect\cite{Jou86,Joupc} for 
$T_{c.m.}=$9.66 and 15.00 MeV. No data at $T_{c.m.}=18.66$ MeV
are available in literature up to now.
}
\label{fig:ayd}
\end{figure}

\begin{figure}[p]
\includegraphics[height=15cm]{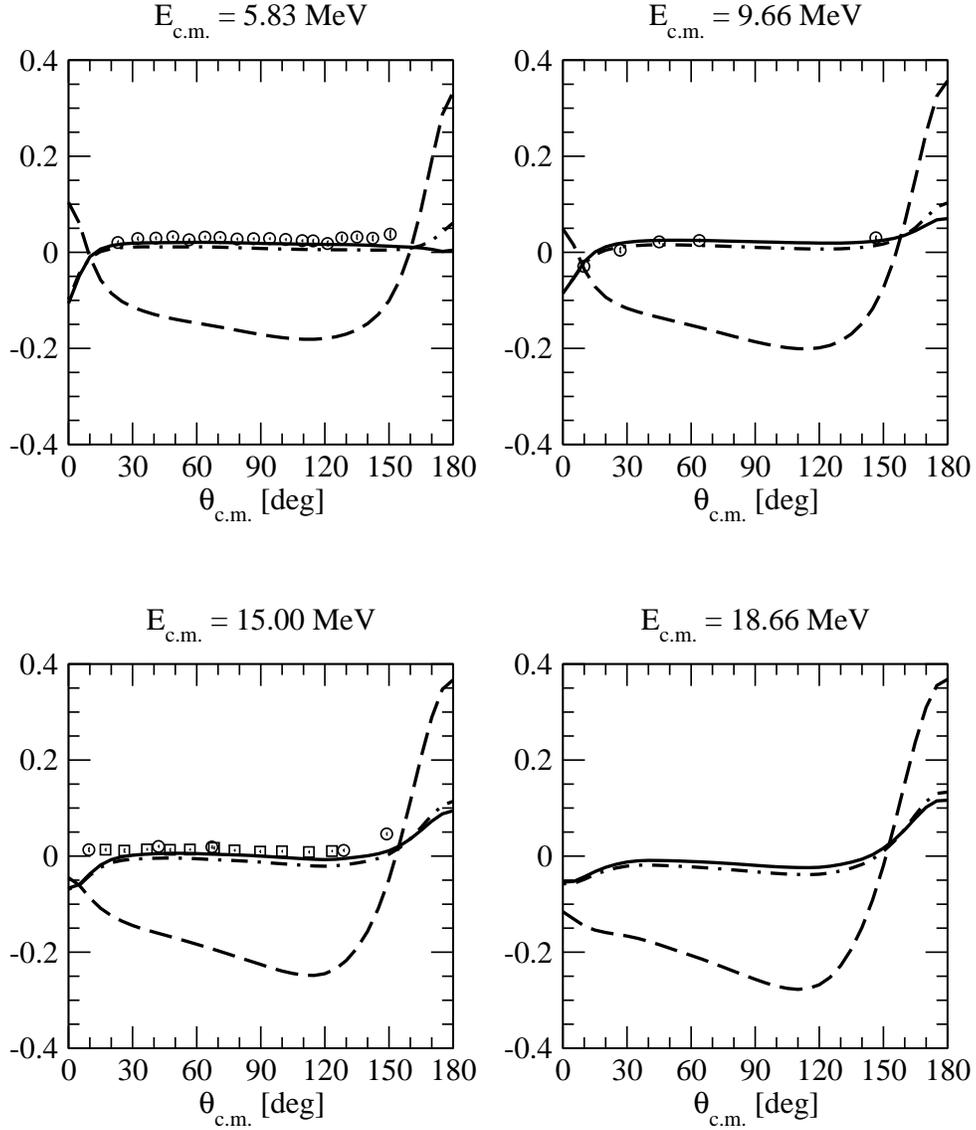}
\caption{Deuteron tensor analyzing power
for $pd$ radiative capture up to $T_{c.m.}=$18.66 MeV
as function of the c.m. $\gamma$-$p$ scattering angle, 
obtained with the 
AV18/UIX Hamiltonian model. The same notation as in 
Fig.~\protect\ref{fig:obs.2.00f} is used for the different lines. 
The experimental data are from Ref.~\protect\cite{Aki01} for 
$T_{c.m.}=$5.83 MeV, from Refs.~\protect\cite{Jou86,Joupc} for 
$T_{c.m.}=$9.66, and from Refs.~\protect\cite{Jou86,Joupc,Ank98} for 
$T_{c.m.}=$15.00 MeV. No data at $T_{c.m.}=18.66$ MeV
are available in literature up to now.
}
\label{fig:ayy}
\end{figure}

There exist several measurements of ${}^2$H($p$,$\gamma$),${}^3$He
observables between $T_{c.m.}=2$ and $20$ MeV. 
In Figs.~\ref{fig:obs.2.00f}--\ref{fig:ayy}, we compare
the predictions obtained with our new model of the
current with a selected set of observables. In all these
figures, the dashed lines are obtained with only one-body contributions, 
the dotted-dashed ones are obtained with one- and new-ME two-body 
contributions (MI+MD), 
the solid curves are the ``full'' results with, in addition, 
also three-body contributions, obtained in the ME scheme.
For completeness, 
the predicted angular distributions of the differential cross section 
$\sigma/a_0$, proton vector analyzing power $A_y$, 
and deuteron vector and tensor analyzing powers i$T_{11}$,  
$T_{20}$, $T_{21}$, $T_{22}$, at $T_{c.m.}=2$ and 
3.33 MeV are again given in Figs.~\ref{fig:obs.2.00f} and~\ref{fig:obs.3.33f}. 
Note that these two 
c.m. energies are just below and above the deuteron breakup 
threshold. We can conclude that: 
(i) an overall nice description for 
all the observables has been obtained, 
with the only exception of the i$T_{11}$ 
deuteron polarization observable at small angles; (ii) some 
small three-body currents effects are noticeable, especially in 
the $T_{20}$ and $T_{21}$ deuteron tensor observables. 

The predicted angular distributions of the 
deuteron vector and tensor analyzing powers $A_{y}$(d),  
$A_{xx}$, $A_{yy}$, $A_{zz}$, at $T_{c.m.}=5.83$ 
are given in Fig.~\ref{fig:obs.5.83f}. The experimental data are 
from Ref.~\cite{Aki01}. Comments similar to those above
can be made in this case too.

The differential cross section $d\sigma/d\Omega$ and the deuteron 
vector- and tensor-analyzing powers $A_{y}$(d) and $A_{yy}$ 
are given 
in Figs.~\ref{fig:xs}, \ref{fig:ayd} and~\ref{fig:ayy} for four 
different c.m. energies. 
The differential cross section is nicely reproduced 
by theory at $T_{c.m.}=$6.60 and 9.86 MeV, while some 
discrepancies are present at $T_{c.m.}=$16.00 and 18.66 MeV. 
However, it should be pointed out that these data sets are 
quite old, and new experimental studies of this 
process in this energy range would be very useful. 
The $A_y$(d) observables are poorly reproduced at
small angles for $T_{c.m.}=$5.83 and 9.66 MeV. However, 
at $T_{c.m.}=$15.00 MeV, the discrepancy between theory and experiment seems 
to disappear. It would be interesting to continue this comparison at 
higher values of $T_{c.m.}$. The $A_{yy}$ observables are 
nicely reproduced in the whole range of $T_{c.m.}$. Some small discrepancies 
are present at small angles for $T_{c.m.}=$15.00 MeV. It is important 
to note, however, that the $A_y$(d) and $A_{yy}$ observables are 
obtained dividing for the differential cross section, which is 
close to zero at small and large values of the c.m. angle. 
In view of this, the agreement between theory and experiment for the 
$A_y$(d) and $A_{yy}$ observables should be considered satisfactory 
in the whole range of c.m. angles.

\begin{figure}[p]
\includegraphics[height=17cm]{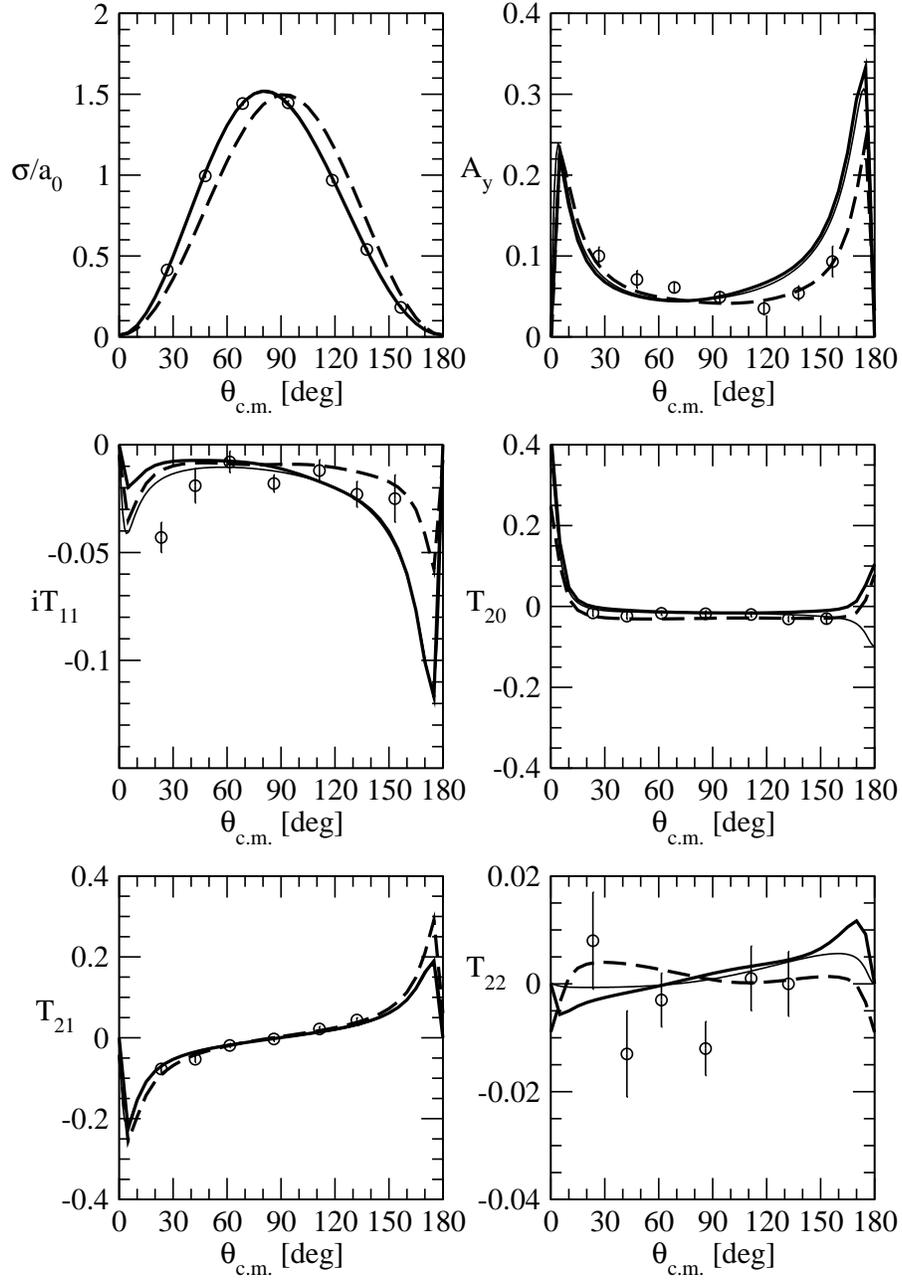}
\caption{Differential cross section, proton vector analyzing power, 
and the four deuteron tensor polarization observables 
for $pd$ radiative capture at $T_{c.m.}=$3.33 MeV
as function of the c.m. $\gamma$-$p$ scattering angle, 
obtained with the 
AV18 Hamiltonian model and the one- plus new-ME two-body 
nuclear current operators. 
The dashed, thin-solid and thick-solid lines 
are obtained with no Coulomb interaction 
in the bound- and scattering-state wave 
functions, with Coulomb interaction only in the bound-state wave 
function, and with inclusion of the Coulomb interaction both in the 
bound- and in the scattering-state wave functions, respectively. 
The experimental data are from Ref.~\protect\cite{Goe92}.}
\label{fig:obs.3.33c}
\end{figure}

Finally, in Fig.~\ref{fig:obs.3.33c} we 
study the importance of including the Coulomb 
interaction in the bound- and scattering-state wave functions. 
In fact, the $T_{c.m.}=$3.33 MeV differential cross section and vector-
and tensor-analyzing powers are calculated using the AV18 
nuclear Hamiltonian and one-body plus new-ME two-body 
currents. The Coulomb interaction is included 
both in the bound- and scattering-state wave functions 
(thick-solid lines), in the bound- but not in the scattering-state 
wave functions (thin-solid lines), and neither in the bound- 
nor in the scattering-state wave functions (dashed lines).
The Coulomb interaction plays a small but significant role, 
particularly in the differential cross section.

\subsubsection{The fore-aft asymmetry}
\label{sec:foreaft}
\begin{figure}[p]
\includegraphics[height=15cm]{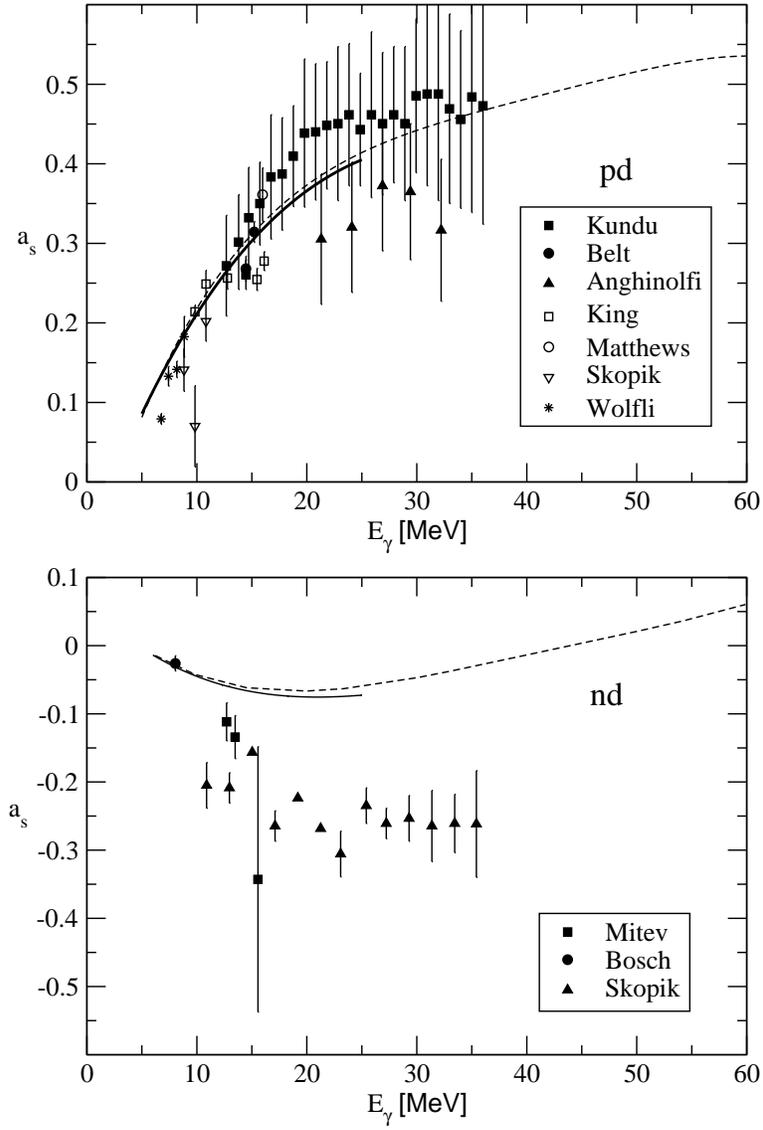}
\caption{Fore-aft asymmetry for $pd$ and
$nd$ radiative capture as function of the
$\gamma$ energy, obtained with the 
AV18/UIX Hamiltonian model and the ``full-new''
current (solid lines). The results of the
calculation of Ref.~\protect\cite{Schad01}
have been also reported (dashed lines).
The $pd$ experimental data are from 
Refs.~\protect\cite{wolfli66,Bel70,kundu71,matt74,skopik79,
Ang83,skopik83,king84},
and the $nd$ experimental data are from Refs.~\protect\cite{bosch64,
skopik81,mitev86}.}
\label{fig:foreaft}
\end{figure}

A quantity of particular interest, both for
historical reasons and for purposes of comparison with data,
is the so-called fore-aft asymmetry in the angular distribution
of the cross section. This quantity is defined according to

\begin{equation}
 a_s={\sigma(54.7^\circ)-\sigma(125.3^\circ) \over
   \sigma(54.7^\circ)+\sigma(125.3^\circ)}
   \>\>\> ,\label{eq:foreaft}   
\end{equation}
where $\sigma(\theta_{c.m.})=(d\sigma_{\rm capt}/d\Omega)_{c.m.}$ and
$\theta_{c.m.}$ is the $\gamma$-$N$ scattering angle. 
The $^2$H($p,\gamma$)$^3$He and $^2$H($n,\gamma$)$^3$H
asymmetries resulting from our calculation and from that of 
Ref.~\cite{Schad01} are compared with existing data in 
Fig.~\ref{fig:foreaft}.

The $pd$ asymmetry
is well reproduced by the calculation (it is a consequence of the
good agreement between the theoretical and
experimental cross section angular distributions,
shown in Fig.~\ref{fig:xs}). On the other hand, there are
large differences between the theoretical and
experimental $nd$ asymmetries. These differences could be
due to problems in the analysis of the data to extract the
experimental values of $a_s$. For example,
the experimental asymmetry comes out mainly from the
measurement taken at the largest angle (see Fig.~\ref{fig:ndxs}).
Note that for other values of $\theta_{c.m.}$, good agreement 
is obtained between theory and experiment. There are also inconsistencies 
between the asymmetries given in Refs.~\cite{skopik81,mitev86},
therefore it is likely that these discrepancies are due to 
experimental problems. However, if experimentally confirmed,
this problem could be of relevance, since the present calculations
seem unable to predict $|a_s(nd)|>0.1$.

\section{Summary and Conclusions} 
\label{sec:con}

We have investigated two different approaches for constructing
conserved two- and three-body electromagnetic currents: one is based on 
meson-exchange mechanisms, while the other uses minimal 
substitution in the explicit and implicit--via
the isospin exchange operator--momentum dependence of the 
two- and three-nucleon interactions.
In the meson exchange model developed by 
Riska and collaborators~\cite{Ris85a,Sch89,Ris85b,Ris85c} 
and used in earlier studies~\cite{Viv96,Viv00,Car90,Sch89,Sch91}, 
some of the terms associated with the isospin-
and momentum-dependent
components of the two-nucleon interaction were ignored, 
since, due to their short range character, they were expected 
to give negligible contributions. The resulting currents, however, 
were not strictly conserved. This limitation is removed in 
the present work. We have also shown how the two-body currents 
obtained in the meson-exchange framework (more
precisely, their longitudinal part) can be derived in 
the minimal-substitution scheme, originally developed 
by Sachs~\cite{Sac48}, by selecting a specific path in the 
space-exchange operator.  Lastly, we have constructed a 
realistic model for the three-body electromagnetic current 
satisfying the current conservation relation with the 
Urbana or Tucson-Melbourne three-nucleon interactions.

A variety of observables have been calculated to 
test the present model of nuclear current operator. 
In particular, for the $A$=3 nuclear systems, 
cross sections as well as polarization observables 
have been calculated and compared with the corresponding 
experimental results in the energy range 0--20 MeV. 

In general, the contributions due to the new two- and three-body 
currents---from the momentum dependence of the two-nucleon 
interaction and from the three-nucleon interaction, 
respectively---are found to be numerically small.  However, they 
resolve the discrepancies between theory and experiment obtained 
in earlier studies~\cite{Viv00} for some of the polarization 
parameters measured in $p$$d$ radiative capture, specifically 
the tensor polarizations $T_{20}$ and $T_{21}$. These contributions 
also reduce the overprediction of the $n$$d$ radiative
capture cross section 
at thermal neutron energy from the 15 \% obtained in Ref.~\cite{Viv96}
to the current 9 \%. 

In conclusion, (i) the predictions for the $n$$p$ radiative 
capture and low-energy deuteron photodisintegration~\cite{Sch04}, 
and for the magnetic form factors of $^3$He and $^3$H, have remained 
essentially unchanged from those reported in previous 
studies~\cite{Mar98}; (ii) a satisfactory agreement 
between theory and experiment for $p$$d$ radiative capture observables 
above deuteron breakup threshold up to 20 MeV has been found, 
in particular for the tensor observables.  Some discrepancies,
however, persist in the vector polarization observables 
at forward angles. 
\section*{Acknowledgments}
The authors thank J.\ Jourdan for useful discussions and 
for letting them use his data prior to publication.
R. S. thanks the INFN, Pisa branch, for financial support
during his visit. He also gratefully acknowledges the
support of the U.S. Department of Energy under contract 
number DE-AC05-84ER40150.

\appendix
\section{}
\label{app:path}

In this appendix we show how the meson-exchange (ME) 
two-body currents (rather, their longitudinal components) 
from the static part of the two-nucleon interaction 
can be derived in the minimal substitution (MS) 
scheme by selecting a specific path in the space-exchange 
operator. To this end, we first write 

\begin{eqnarray}
v_{2,ij}\, \bm\tau_i\cdot\bm\tau_j
&=&\sum_a(v^a_{PS,ij}+v^a_{V,ij}+v^a_{VS,ij})\, 
\bm\tau_i\cdot\bm\tau_j \ ,
\label{eq:v2ijs}\\
v^a_{PS,ij}&=&-f^2_{PS,a}
({\bm\sigma}_i\cdot\bna_i) ({\bm\sigma}_j\cdot\bna_j)\,Y_a(r)\ ,
\label{eq:vpsa} \\
v^a_{V,ij}&=&-f^2_{V,a}
({\bm\sigma}_i\times\bna_i)\cdot({\bm\sigma}_j\times\bna_j)\,Y_a(r)\ ,
\label{eq:vva} \\
v^a_{VS,ij}&=& f^2_{VS,a}\,Y_a(r) \ ,
\label{eq:vvsa} \\
Y_a(r)&=&\frac{{\rm e}^{-m_ar}}{4\pi\,r} \ , 
\label{eq:ya}
\end{eqnarray}
where $f_{PS,a}$, $f_{V,a}$, $f_{VS,a}$ 
and $m_a$ are appropriate parameters, see 
Eqs.~(\ref{eq:vpsdef})--(\ref{eq:vvsdef}). 
The two-body current which satisfies the 
current conservation relation with $v_{2,ij}$ is 
then written in MS scheme as 

\begin{equation}
\bfj_{ij}(\bfq)=\sum_a\bigg[\bfj^a_{ij}(\bfq;PS)+\bfj^a_{ij}(\bfq;V)
  +\bfj^a_{ij}(\bfq;VS)\bigg] \ ,
\label{eq:jijq}
\end{equation}
where each term of the sum is given in Eq.~(\ref{eq:jq}), 
with $v_{2,ij}$ replaced by the corresponding  
$PS$, $V$ or $VS$ potential.  
First, we consider the simple spin-independent current $\bfj^a_{ij}(\bfq;VS)$: 
by choosing $\gamma_{ij}=-\gamma^\prime_{ji}$ and using Eq.~(\ref{eq:tauz}), 
$\bfj^a_{ij}(\bfq;VS)$ can be written as

\begin{eqnarray}
\bfj^a_{ij}(\bfq;VS)&=&f^2_{VS,a}\,Y_a(r)\, \bfj^a_{ij}(\bfq) \ ,
\nonumber\\
\bfj^a_{ij}(\bfq)&=&G_E^V(q_\mu^2)\,({\bm \tau}_i\times{\bm \tau}_j)_z\,
\int_{\gamma_{ij}}
d\bfs\; {\rm e}^{{\rm i}\bfq\cdot\bfs} \ ,
\label{eq:jija}
\end{eqnarray}
where the path $\gamma_{ij}$ is selected to be 

\begin{eqnarray}
  \bfs&=&
  \bfR_{ij}-x\bfr_{ij}+{\rm i}\,r_{ij}\frac{{\hat \bfq}}{q}
  \bigg[L_a(x)-m_a\bigg]\ , 
  \hspace{1cm} -\frac{1}{2}\leq x\leq \frac{1}{2} \ ,
  \nonumber \\
  \bfR_{ij}&=&\frac{1}{2}(\bfr_i+\bfr_j) \ , 
  \qquad \bfs\bigg(x=-{1\over 2}\bigg)=\bfr_i\ ,
  \qquad \bfs\bigg(x=+{1\over 2}\bigg)=\bfr_j\ ,
  \nonumber \\
  L_a(x)&=&\sqrt{m_a^2+q^2(\frac{1}{4}-x^2)} \ .
  \label{eq:sriska}
\end{eqnarray}
It is straightforward to verify that 
$\sum_a\bfj^a_{ij}(\bfq;VS)$ reduces to  $\bfj_{ij}(\bfq;VS)$ of 
Eq.~(\ref{eq:jvs}), obtained in the ME scheme. In this case, both the 
longitudinal and transverse components of $\bfj_{ij}(\bfq;VS)$  
are exactly reproduced.

Next, using Eq.~(\ref{eq:jq}), the $PS$ current reads 

\begin{equation}
\bfj_{ij}^a (\bfq;PS)= 
-f_{PS,a}^2\,\bfj_{ij}^a(\bfq) ({\bm\sigma}_i\cdot{\bm\nabla}_i)\,
({\bm\sigma}_j\cdot{\bm\nabla}_j)Y_a(r) \ ,
\label{eq:jpsa}
\end{equation}
which can be rewritten as 

\begin{eqnarray}
\bfj_{ij}^a(\bfq) ({\bm\sigma}_i\cdot\bna_i)\,
({\bm\sigma}_j\cdot\bna_j)Y_a(r)&=&
({\bm\sigma}_i\cdot\bna_i)\,({\bm\sigma}_j\cdot\bna_j)\,
\bfj_{ij}^a(\bfq)\,Y_a(r) \nonumber \\
&-& 
[({\bm\sigma}_i\cdot\bna_i)\bfj_{ij}^a(\bfq)]\,
[({\bm\sigma}_j\cdot\bna_j)Y_a(r)] \nonumber \\
&-&
[({\bm\sigma}_j\cdot\bna_j)\bfj_{ij}^a(\bfq)]\,
[({\bm\sigma}_i\cdot\bna_i)Y_a(r)] \nonumber \\
&-&Y_a(r)
({\bm\sigma}_i\cdot\bna_i)\,({\bm\sigma}_j\cdot\bna_j)\,
\bfj_{ij}^a(\bfq) \ .
\label{eq:jss}
\end{eqnarray}
The $a$-meson in-flight current is again exactly reproduced 
(first line of Eq.~(\ref{eq:jss})). However, only the 
longitudinal components of the contact terms are 
reproduced by Eq.~(\ref{eq:jss}) (second and third lines), 
since, via Eq.~(\ref{eq:intq}), 

\begin{eqnarray}
({\bm\sigma}_i\cdot\bna_i)\,\bfq\cdot\bfj_{ij}^a(\bfq)&=&
-{\bm\sigma}_i\cdot\bfq\,{\rm e}^{{\rm i}\bfq\cdot\bfr_i} \ , 
\label{eq:jsi} \\
({\bm\sigma}_j\cdot\bna_j)\,\bfq\cdot\bfj_{ij}^a(\bfq)&=&
{\bm\sigma}_j\cdot\bfq\,{\rm e}^{{\rm i}\bfq\cdot\bfr_j} \ , 
\label{eq:jsj}
\end{eqnarray}
and the last term of Eq.~(\ref{eq:jss}) vanishes when dotted with $\bfq$.
Therefore, using Eqs.~(\ref{eq:jss})--(\ref{eq:jsj}), 
Eq.~(\ref{eq:jpsa}) can be rewritten as 

\begin{eqnarray}
\bfj^a_{ij}(\bfq;PS)&=&-f^2_{PS,a}
\bigg[
({\bm\sigma}_i\cdot\bna_i) ({\bm\sigma}_j\cdot\bna_j) \bfj^a_{ij}(\bfq)
\,Y_a(r)
\nonumber \\
&+&G_E^V(q_\mu^2)\,({\bm \tau}_i\times{\bm \tau}_j)_z\,
[{\rm e}^{{\rm i}\bfq\cdot\bfr_i} {\bm \sigma}_i\,({\bm\sigma}_j\cdot\bna_j)
-{\rm e}^{{\rm i}\bfq\cdot\bfr_j} {\bm \sigma}_j\,({\bm\sigma}_i\cdot\bna_i)]
\,Y_a(r) \bigg] \nonumber \\
&+&{\rm additional}\,\,{\rm transverse}\,\,{\rm terms} \ .
\label{eq:jpsaf}
\end{eqnarray}
It would be interesting to evaluate the contributions of these 
additional transverse terms. 

Similar considerations are valid for $\bfj_{ij}^a(\bfq;V)$, 
which can be written as 

\begin{eqnarray}
\bfj^a_{ij}(\bfq;V)&=&-f^2_{V,a}
\bigg(
[({\bm\sigma}_i\times\bna_i)\cdot({\bm\sigma}_j\times\bna_j)]\,\, 
\bfj^a_{ij}(\bfq)\,Y_a(r) \nonumber \\
&&\phantom{f^2_{V,a}}
-({\bm\sigma}_i\times\bna_i)\,\,
[({\bm\sigma}_j\times\bna_j)\cdot\bfj^a_{ij}(\bfq)]\,Y_a(r)
\nonumber \\
&&\phantom{f^2_{V,a}}
-({\bm\sigma}_j\times\bna_j)\,\,
[({\bm\sigma}_i\times\bna_i)\cdot\bfj^a_{ij}(\bfq)]\,Y_a(r)
\nonumber \\
&-&G_E^V(q_\mu^2)\,({\bm \tau}_i\times{\bm \tau}_j)_z\,
[{\rm e}^{{\rm i}\bfq\cdot\bfr_i} 
{\bm \sigma}_i\times({\bm\sigma}_j\times\bna_j)
\nonumber \\
&&\phantom{G_E^V(q_\mu^2)\,({\bm \tau}_i\times{\bm \tau}_j)_z}
-{\rm e}^{{\rm i}\bfq\cdot\bfr_j} 
{\bm \sigma}_j\times({\bm\sigma}_i\times\bna_i)]
\,Y_a(r) \bigg) \nonumber\\
&+&{\rm additional}\,\,{\rm transverse}\,\,{\rm terms} \ .
\label{eq:jvaf}
\end{eqnarray}

\section{}
\label{app:l2so2}
We list here the two-body currents obtained with the MS method
from the quadratic-momentum-dependent terms of the interaction. 
To this end, the $(\bfL\cdot \bfS)^2$ term is written as 

\begin{equation}
(\bfL\cdot \bfS)^2=\frac{1}{2} \bfL^2
-\frac{1}{2}\,\bfL\cdot\bfS
+\frac{1}{4}\,\biggl[({\bm \sigma}_i\cdot\bfL)\,
({\bm \sigma}_j\cdot\bfL)
+({\bm \sigma}_j\cdot\bfL)\,
({\bm \sigma}_i\cdot\bfL)\biggr] \ ,
\label{eq:ls2}
\end{equation}
and the potential functions associated with the 
$\bfL\cdot \bfS$ and $\bfL^2$ operators are 
re-defined accordingly, for example

\begin{eqnarray}
{\hat v_{b}}(r)&=&v_{b}(r)-\frac{1}{2}\,v_{bb}(r)\ , \\
{\hat v_{q}}(r)&=&v_{q}(r)+\frac{1}{2}\,v_{bb}(r)\ ,
\end{eqnarray}
and similarly for $v_{b\tau}(r)$ and $v_{q\tau}(r)$.
The spin-orbit currents are then those given in Eqs.~(\ref{eq:jsoq})
and~(\ref{eq:jsotl}) with $v_{b}(r) , v_{b\tau}(r)$ replaced
by ${\hat v_{b}}(r) , {\hat v_{b\tau}}(r)$.

For the $\bfL^2$ terms, given by

\begin{equation}
v_q=\bfL^2 \, [{\hat v_q}(r)
+{\hat v_{q\sigma}}(r)\, {\bm\sigma}_i\cdot{\bm\sigma}_j
+{\hat v_{q\tau}}(r)\, {\bm\tau}_i\cdot{\bm\tau}_j 
+{\hat v_{q\sigma\tau}}(r)\, {\bm\sigma}_i\cdot{\bm\sigma}_j
{\bm\tau}_i\cdot{\bm\tau}_j ] \ ,
\end{equation}
using the linear path of Eq.~(\ref{eq:slin}), we find

\begin{eqnarray}
\bfj_{ij}(\bfq;LL)&=& 
\biggl[ {\hat v_{q}}(r) + {\hat v_{q\sigma}}(r)\,
{\bm\sigma}_i\cdot{\bm\sigma}_j
\biggr]
\biggl[ \bigg({\rm i}\bfr - \bfr\times\bfL \bigg) P_-
+\frac{1}{4} (\bfr\times\bfq)\times\bfr\, P_+ \biggr] \ , 
\label{eq:jllq} \\
\bfj_{ij}^{LP}(\bfq;LL\tau)&=& \frac{1}{2} 
\biggl[ {\hat v_{q\tau}}(r) + {\hat v_{q\sigma\tau}}(r)\,
{\bm\sigma}_i\cdot{\bm\sigma}_j \biggr]
\biggl[ \bigg({\rm i}\bfr - \bfr\times\bfL\bigg) R_-
+\frac{1}{4} (\bfr\times\bfq)\times\bfr R_+  \nonumber \\
&+&{\rm i}\, G_E^V(q_\mu^2) \,({\bm\tau}_i\times{\bm \tau}_j)_z
\, \{ \bfL^2\, , \,\bfr f_{ij}(\bfq) \} \biggr] \ ,
\label{eq:jllqt} 
\end{eqnarray}
where $\bfr\equiv\bfr_{ij}$, $\{\cdots\}$ denotes the anticommutator, 
and $f_{ij}(\bfq)$ is defined in Eq.~(\ref{eq:fijq}).  We have also defined

\begin{eqnarray}
P_\pm&=&\epsilon_i{\rm e}^{ {\rm i}\bfq\cdot\bfr_i} 
    \pm \epsilon_j{\rm e}^{ {\rm i}\bfq\cdot\bfr_j} \ ,
\label{eq:ppm}\\
R_\pm&=&\eta_i{\rm e}^{ {\rm i}\bfq\cdot\bfr_i} 
    \pm \eta_j{\rm e}^{ {\rm i}\bfq\cdot\bfr_j} \ ,
\label{eq:rpm}
\end{eqnarray}
with $\epsilon_i$, $\epsilon_j$, $\eta_{i}$ and $\eta_{j}$ listed
in Sec.~\ref{sec:j2b}. 

For the quadratic spin-orbit terms, given by 

\begin{eqnarray}
{\hat v_{bb}}&=&\frac{1}{4} v_{bb}(r)\{ {\bm \sigma}_i\cdot\bfL
\, , \, {\bm \sigma}_j\cdot\bfL \} \nonumber \\
&+&\frac{1}{8} v_{bb\tau}(r)\{ \{ {\bm \sigma}_i\cdot\bfL\, , \, 
{\bm \sigma}_j\cdot\bfL \}\, , 
{\bm\tau}_i\cdot{\bm\tau}_j \} \ ,
\label{eq:vls2}
\end{eqnarray}
again using the linear path, we find 

\begin{eqnarray}
\bfj_{ij}(\bfq;bb)&=& \frac{1}{8} v_{bb}(r) 
\biggl \{ P_-\, , \, 
[({\bm\sigma}_i\times\bfr)\,({\bm\sigma}_j\cdot\bfL)\,
 +\,({\bm\sigma}_j\times\bfr)\, ({\bm\sigma}_i\cdot\bfL)\, 
] \biggr \}
\label{eq:jbbq} \\
\bfj_{ij}^{LP}(\bfq;bb\tau)&=& \frac{1}{8} v_{bb\tau}(r)
\biggl[ \frac{1}{4} R_+ 
\biggl[({\bm\sigma}_j\times\bfr)\,\,
[\bfq\cdot({\bm\sigma}_i\times\bfr)]
+({\bm\sigma}_i\times\bfr)\,\,
[\bfq\cdot({\bm\sigma}_j\times\bfr)]\biggr]
\nonumber \\
&+&R_-\biggl[
({\bm\sigma}_j\times\bfr)\,\,({\bm\sigma}_i\cdot\bfL)+
({\bm\sigma}_i\times\bfr)\,\,({\bm\sigma}_j\cdot\bfL)+
{\rm i}({\bm\sigma}_i\cdot{\bm\sigma}_j)\bfr-
\frac{\rm i}{2}[{\bm\sigma}_i\,({\bm\sigma}_j\cdot\bfr)
+{\bm\sigma}_j\, ({\bm\sigma}_i\cdot\bfr) ]\biggr]
\nonumber \\
&+&{\rm i}\, G_E^V(q_\mu^2)
({\bm\tau}_i\times{\bm \tau}_j)_z
\biggl\{ 
\{ {\bm\sigma}_i\cdot\bfL\, ,\,  {\bm\sigma}_j\cdot\bfL
\}\, , \, \bfr f_{ij}(\bfq) \biggr\} \biggl] \ .
\label{eq:jbbqt} 
\end{eqnarray}

\section{}
\label{app:j3br}
Using Eq.~(\ref{eq:jrs}), the configuration-space expressions 
for the exchange currents $\bfj_{ij}^{II}(\bfk_i,\bfk_j;B)$
of Sec.~\ref{subsec:j3me}, $B=PS$ or $V$, are given by:

\begin{eqnarray}
\bfj_{ij}^{II}(\bfq; PS)&=&G_E^V(q_\mu^2)
({\bm\tau}_i\times{\bf T}_j)_z\nonumber \\
&\Biggl[ &
{\rm e}^{{\rm i}\bfq\cdot\bfr_i}\,g_{PS}(r){\bm\sigma}_i\,
(\bfss_j\cdot{\hat\bfr})\,+\,
{\rm e}^{{\rm i}\bfq\cdot\bfr_j}\,g_{PS}(r)\bfss_j\,
({\bm\sigma}_i\cdot{\hat\bfr})\nonumber \\
&+&{\rm e}^{{\rm i}\bfq\cdot\bfR}\,
\bigg[\frac{G_{PS,1}(\bfr)}{r^2}\,
\bigg({\bm\sigma}_i\,(\bfss_j\cdot{\hat\bfr})
+\bfss_j\,({\bm\sigma}_i\cdot{\hat\bfr})
+{\hat\bfr}\,({\bm\sigma}_i\cdot\bfss_j)\bigg)\nonumber \\
&&\>\>\>\>\>\>\>\>\>\>\>\>
+{\rm i}\frac{G_{PS,2}(\bfr)}{r}\,{\bm\sigma}_i\,(\bfss_j\cdot\bfq)
-{\rm i}\frac{G_{PS,3}(\bfr)}{r}\,\bfss_j\,({\bm\sigma}_i\cdot\bfq)
\nonumber \\
&&\>\>\>\>\>\>\>\>\>\>\>\>
-{\rm i}\frac{G_{PS,4}(\bfr)}{r}\,{\hat\bfr}\,
({\bm\sigma}_i\cdot{\hat\bfr})(\bfss_j\cdot\bfq)
+{\rm i}\frac{G_{PS,5}(\bfr)}{r}\,{\hat\bfr}\,
({\bm\sigma}_i\cdot\bfq)(\bfss_j\cdot{\hat\bfr})
\nonumber \\
&&\>\>\>\>\>\>\>\>\>\>\>\>
-G_{PS,6}(\bfr)\,{\hat\bfr}\,
({\bm\sigma}_i\cdot\bfq)(\bfss_j\cdot\bfq)
-\frac{G_{PS,7}(\bfr)}{r^2}\,{\hat\bfr}\,
({\bm\sigma}_i\cdot{\hat\bfr})(\bfss_j\cdot{\hat\bfr})
\bigg]\,\Biggr ] \ ,
\label{eq:jpsr}
\end{eqnarray}

\begin{eqnarray}
\bfj_{ij}^{II}(\bfq; V)&=&G_E^V(q_\mu^2)
({\bm\tau}_i\times{\bf T}_j)_z\nonumber \\
&\Biggl[&{\rm e}^{{\rm i}\bfq\cdot\bfr_i}\,g_{V}(r)
{\bm\sigma}_i\times(\bfss_j\times{\hat\bfr})\,+\,
{\rm e}^{{\rm i}\bfq\cdot\bfr_j}\,g_{V}(r)
\bfss_j\times({\bm\sigma}_i\times{\hat\bfr})\nonumber \\
&-&{\rm e}^{{\rm i}\bfq\cdot\bfR}\,
\bigg[\frac{G_{V,1}(\bfr)}{r^2}\,
\bigg((\bfss_j\times{\hat\bfr})\times{\bm\sigma}_i
+({\bm\sigma}_i\times{\hat\bfr})\times\bfss_j
+2{\hat\bfr}\,({\bm\sigma}_i\cdot\bfss_j)\bigg)\nonumber \\
&&\>\>\>\>\>\>\>\>\>\>\>\>
+{\rm i}\frac{G_{V,2}(\bfr)}{r}\,
(\bfss_j\times\bfq)\times{\bm\sigma}_i
  -{\rm i}\frac{G_{V,3}(\bfr)}{r}\,
({\bm\sigma}_i\times\bfq)\times\bfss_j
\nonumber \\
&&\>\>\>\>\>\>\>\>\>\>\>\>
-{\rm i}\frac{G_{V,4}(\bfr)}{r}\,{\hat\bfr}\,
({\bm\sigma}_i\times{\hat\bfr})\cdot(\bfss_j\times\bfq)
+{\rm i}\frac{G_{V,5}(\bfr)}{r}\,{\hat\bfr}\,
({\bm\sigma}_i\times\bfq)\cdot(\bfss_j\times{\hat\bfr})
\nonumber \\
&&\>\>\>\>\>\>\>\>\>\>\>\>
-G_{V,6}(\bfr)\,{\hat\bfr}\,
({\bm\sigma}_i\times\bfq)\cdot(\bfss_j\times\bfq)
-\frac{G_{V,7}(\bfr)}{r^2}\,{\hat\bfr}\,
({\bm\sigma}_i\times{\hat\bfr})\cdot(\bfss_j\times{\hat\bfr})
\bigg]
\nonumber \\
&-&\frac{1}{2}{\rm e}^{{\rm i}\bfq\cdot\bfR}\,
\bigg[G_{V,2}(\bfr)\,
(\bfss_j\times\bfq)\,\,{\bm\sigma}_i\cdot(\bfq\times{\hat\bfr})
+G_{V,3}(\bfr)\,
({\bm\sigma}_i\times\bfq)\,\,\bfss_j\cdot(\bfq\times{\hat\bfr})
\nonumber \\
&&\>\>\>\>\>\>\>\>\>\>\>\>
+{\rm i}\frac{G_{V,4}(\bfr)+G_{V,5}(\bfr)}{r}\,
\bigg((\bfss_j\times{\hat\bfr})\,\,
{\bm\sigma}_i\cdot({\hat\bfr}\times\bfq)
-({\bm\sigma}_i\times{\hat\bfr})\,\,
\bfss_j\cdot({\hat\bfr}\times\bfq)\bigg)\nonumber \\
&&\>\>\>\>\>\>\>\>\>\>\>\>
+{\rm i}\frac{G_{V,2}(\bfr)+G_{V,3}(\bfr)}{r}\,
\bigg(\bfss_j\times({\bm\sigma}_i\times\bfq)
-{\bm\sigma}_i\times(\bfss_j\times\bfq)\bigg)
\bigg]
\Biggr]\ ,
\label{eq:jvr}
\end{eqnarray}
where $\bfr=\bfr_i-\bfr_j$, ${\hat\bfr}=\bfr/r$, 
$\bfR=\frac{1}{2}(\bfr_i+\bfr_j)$, and

\begin{eqnarray}
g_{PS}(r)&=&-\frac{1}{3\,r^2}\bigg[\int_r^\infty\,dr'\,r'^2\,
v_{\sigma\tau}^{II}(r')\,+\,2\,r^3\int_r^\infty\,dr'\,
\frac{v_{t\tau}^{II}(r')}{r'}\bigg] \ , \label{eq:gps}\\
g_{V}(r)&=&\frac{1}{3\,r^2}\bigg[\int_r^\infty\,dr'\,r'^2\,
v_{\sigma\tau}^{II}(r')\,-\,r^3\int_r^\infty\,dr'\,
\frac{v_{t\tau}^{II}(r')}{r'}\bigg] \ , \label{eq:gv}\\
G_{B,1}(\bfr)&=&\int_{-1/2}^{1/2}dx\,{\rm e}^{-{\rm i}x\bfq\cdot\bfr}
\bigg[E_B(x;r)\,-\,r\,\frac{d}{dr}\,E_B(x;r)\bigg] \ , \label{eq:ga1}\\
G_{B,2}(\bfr)&=&\int_{-1/2}^{1/2}dx\,{\rm e}^{-{\rm i}x\bfq\cdot\bfr}
(\frac{1}{2}+x)\,E_B(x;r) \ , \label{eq:ga2}\\
G_{B,3}(\bfr)&=&\int_{-1/2}^{1/2}dx\,{\rm e}^{-{\rm i}x\bfq\cdot\bfr}
(\frac{1}{2}-x)\,E_B(x;r) \ , \label{eq:ga3}\\
G_{B,4}(\bfr)&=&\int_{-1/2}^{1/2}dx\,{\rm e}^{-{\rm i}x\bfq\cdot\bfr}
(\frac{1}{2}+x)\,
\bigg[E_B(x;r)\,-\,r\,\frac{d}{dr}\,E_B(x;r)\bigg] \ , \label{eq:ga4}\\
G_{B,5}(\bfr)&=&\int_{-1/2}^{1/2}dx\,{\rm e}^{-{\rm i}x\bfq\cdot\bfr}
(\frac{1}{2}-x)\,
\bigg[E_B(x;r)\,-\,r\,\frac{d}{dr}\,E_B(x;r)\bigg]\ , \label{eq:ga5}\\
G_{B,6}(\bfr)&=&\int_{-1/2}^{1/2}dx\,{\rm e}^{-{\rm i}x\bfq\cdot\bfr}
(\frac{1}{4}-x^2)\,E_B(x;r) \ , \label{eq:ga6}\\
G_{B,7}(\bfr)&=&\int_{-1/2}^{1/2}dx\,{\rm e}^{-{\rm i}x\bfq\cdot\bfr}
\bigg[3\,E_B(x;r)\,-\,3r\,\frac{d}{dr}\,E_B(x;r)
\,+\,r^2\,\frac{d^2}{dr^2}\,E_B(x;r)\bigg] \ . \label{eq:ga7}
\end{eqnarray}
The functions $E_B(x;r)$ are defined as

\begin{eqnarray}
E_B(x;r)&=&\sum_{a=1}^{N}\,\frac{g_{B,a}}{4\pi}\,{\rm e}^{-r\,L_a(x)} \ ,
\label{eq:ea} \\
L_a(x)&=&\sqrt{m_a^2\,+\,\frac{q^2}{4}\,\left(1-4x^2\right)} \ .
\label{eq:li}
\end{eqnarray}
The coefficients $g_{B,a}$ are obtained by fitting the functions 
$v_B^{II}(k)$ of Sec.~\ref{subsec:j3me}, $B=PS,V$,
with $\sum_{a=1}^N\,g_{B,a}/(k^2+m_a^2)$~\cite{Sch89}.


%
%
%
%
%
\newpage


\begin{thebibliography}{100}
\bibitem{Car98} J.\ Carlson and R.\ Schiavilla,
                Rev.\ Mod.\ Phys.\ {\bf 70}, 743 (1998).
%
\bibitem{Viv96} M.\ Viviani, R.\ Schiavilla, and A.\ Kievsky,
                Phys.\ Rev.\ C {\bf 54}, 534 (1996).
%
\bibitem{Viv00} M.\ Viviani, A.\ Kievsky, L.E.\ Marcucci, S.\ Rosati, 
		and R.\ Schiavilla,
                Phys.\ Rev.\ C {\bf 61}, 064001 (2000).
%
\bibitem{Mar98} L.E.\ Marcucci, D.O.\ Riska, and R.\ Schiavilla,
                Phys.\ Rev.\ C {\bf 58}, 3069 (1998).
%
\bibitem{Kie93} A.\ Kievsky, M.\ Viviani, and S.\ Rosati,
                Nucl.\ Phys.\ {\bf A551}, 241 (1993).
%
\bibitem{Kie94} A.\ Kievsky, M.\ Viviani, and S.\ Rosati,
                Nucl.\ Phys.\ {\bf A577}, 511 (1994).
%
%
\bibitem{Kie01} A.\ Kievsky, M.\ Viviani, and S.\ Rosati,
                Phys.\ Rev.\ C {\bf 64}, 024002 (2001).
%
\bibitem{Wir95} R.B.\ Wiringa, V.G.J.\ Stoks, and R.\ Schiavilla,
                Phys.\ Rev.\ C {\bf 51}, 38 (1995).
%
\bibitem{Pud95} B.S.\ Pudliner, V.R.\ Pandharipande, J.\ Carlson, 
		and R.B.\ Wiringa,
                Phys.\ Rev.\ Lett.\ {\bf 74}, 4396 (1995).
%
\bibitem{Ris85a} D.O.\ Riska,
                Phys.\ Scr.\ {\bf 31}, 107 (1985).
%
\bibitem{Car90} J.\ Carlson, D.O.\ Riska, R.\ Schiavilla, and R.B.\ Wiringa,
                Phys.\ Rev.\ C {\bf 42}, 830 (1990).
%
\bibitem{Sch89} R.\ Schiavilla, V.R.\ Pandharipande, and D.O.\ Riska,
                Phys.\ Rev.\ C {\bf 40}, 2294 (1989).
%
\bibitem{Sch91} R.\ Schiavilla and D.O.\ Riska,
                Phys.\ Rev.\ C {\bf 43}, 437 (1991).
%
\bibitem{Sch92} R.\ Schiavilla, R.B.\ Wiringa, V.R.\ Pandharipande, 
		and J.\ Carlson,
                Phys.\ Rev.\ C {\bf 45}, 2628 (1992).
%
\bibitem{Mar04} L.E.\ Marcucci, K.M.\ Nollett, R.\ Schiavilla, 
                and R.B.\ Wiringa, 
		{\tt ArXiv:nucl-th/0402078}.
%
\bibitem{Lun02} The LUNA Collaboration,
                Nucl.\ Phys.\ {\bf A706}, 203 (2002).
%
\bibitem{Viv03} M.\ Viviani, L.E.\ Marcucci, A.\ Kievsky, R.\ Schiavilla, 
                and S.\ Rosati, 
		Eur.\ Phys.\ J.\ A {\bf 17}, 483 (2003).
%
\bibitem{Mar03a} L.E.\ Marcucci, M.\ Viviani, A.\ Kievsky, S.\ Rosati, 
                and R.\ Schiavilla, 
		Few-Body Syst. Suppl. {\bf 14}, 319 (2003).
%
\bibitem{Mar03b} L.E.\ Marcucci, M.\ Viviani, A.\ Kievsky, S.\ Rosati, 
                and R.\ Schiavilla, 
		Few-Body Syst. Suppl. {\bf 15}, 87 (2003).
%
\bibitem{Goe92} F.\ Goeckner, W.K.\ Pitts, and L.D.\ Knutson, 
                Phys.\ Rev.\ C {\bf 45}, R2536 (1992).
%
\bibitem{Buc85} A.\ Buchmann, W.\ Leidemann, and H.\ Arenh\"ovel, 
                 Nucl.\ Phys.\ A {\bf 443}, 726 (1985).
%
\bibitem{Aren00} H. Arenh\"ovel, F. Ritz, and T. Wilbois,
                Phys. Rev. C {\bf 61}, 034002 (2000).
%
\bibitem{Aren04} H. Arenh\"ovel, A. Fix, and M. Schwamb,
                 Phys. Rev. Lett. {\bf 93}, 202301 (2004).
%
\bibitem{Golak00} J. Golak {\it et al.},
                  Phys. Rev. C {\bf 62}, 054005 (2000).
%
\bibitem{Skib03} R. Skibinski {\it et al.},
                 Phys. Rev. C {\bf 67}, 054001 (2003);
                 {\it ibid.}, 054002 (2003).
%
\bibitem{Schad01} W. Schadow, O. Nohadani, and W. Sandhas,
                  Phys. Rev. C {\bf 63}, 044006 (2001).
%
\bibitem{Sieg37} A. J. F. Siegert, Phys. Rev. {\bf 52}, 787 (1937).
%
\bibitem{Delt04} A. Deltuva, L. P. Yuan, J. Adam, Jr., 
                  A. C. Fonseca, and P. U. Sauer,
                  Phys. Rev. C {\bf 69}, 034004 (2004).
%
\bibitem{BS04} H. Sadeghi and S. Bayegan, {\tt ArXiv:nucl-th/0411114}.
%
\bibitem{ELOT00} V. D. Efros, W. Leidemann, G. Orlandini, 
                 and E. L. Tomusiak,
                 Phys.\ Lett.\ {\bf B484}, 223 (2000);
                 Phys.\ Rev.\  {\bf C69}, 044001 (2004).
%
\bibitem{Ris89} D.O.\ Riska,
                Phys.\ Rep.\ {\bf 181}, 207 (1989).
%
\bibitem{Ris85b} D.O.\ Riska, Phys.\ Scr.\ {\bf 31}, 471 (1985).
%
\bibitem{Ris85c} D.O.\ Riska and M.\ Poppius, 
                 Phys.\ Scr.\ {\bf 32}, 581 (1985).
%
\bibitem{Sac48} R.G.\ Sachs, Phys.\ Rev.\ {\bf 74}, 433 (1948).
%
\bibitem{Nym67} E.M.\ Nyman, Nucl.\ Phys.\ {\bf B1}, 535 (1967).
%
\bibitem{Wir84} R.B.\ Wiringa, R.A.\ Smith, and T.L.\ Ainsworth,
                Phys.\ Rev.\ C {\bf 29}, 1207 (1984).
%
\bibitem{Mac01} R.\ Machleidt,
                Phys.\ Rev.\ C {\bf 63}, 024001 (2001).
%
\bibitem{Sch04} R.\ Schiavilla, J.\ Carlson, and M.\ Paris,
                Phys.\ Rev.\ C {\bf 70}, 044007 (2004).
%
\bibitem{Coo79} S.A.\ Coon, {\it et al.},
                Nucl.\ Phys.\ {\bf A317}, 242 (1979).
%
\bibitem{Rob} M.R.\ Robilotta\ and\ H.T.\ Coelho, 
              Nucl.\ Phys. {\bf A460}, 645 (1986).
%
\bibitem{Car83} J.\ Carlson, V.R. Pandharipande, and R.B.\ Wiringa,
                Nucl.\ Phys.\ {\bf A401}, 59 (1983).
%
\bibitem{Kie99}  A. Kievsky, S. Rosati, and M. Viviani,
               Phys. Rev. Lett. {\bf 82}, 3759 (1999)
%
\bibitem{Viv01} M. Viviani, A. Kievsky, and  S. Rosati,
                Few--Body Systems {\bf 30}, 39 (2001)
%
\bibitem{Hub95} D. H\" uber {\it et al.}, Few-Body Systems,
                {\bf 19}, 175 (1995).
%
\bibitem{Kie98} A. Kievsky {\it et al.},
                Phys.\ Rev.\ C {\bf 58}, 3085 (1998).
%
\bibitem{sag94} K. Sagara {\it et al.}, 
                Phys.\ Rev.\ C {\bf 50}, 576 (1994);
                K. Sagara (private communication).
%
\bibitem{gru83} W. Gr\"uebler {\it et al.},
                Nucl.\ Phys.\ {\bf A398}, 445 (1983);
                F. Sperisen {\it et al.},
                {\it ibid.}\ {\bf A422}, 81 (1984).
%
\bibitem{nogga} A. Nogga {\it et al.}, 
                Phys.\ Rev.\ C {\bf 67}, 034004 (2003).
%
%
\bibitem{Smi99} M.K.\ Smith and L.D.\ Knutson, 
                Phys.\ Rev.\ Lett.\ {\bf 82}, 4591 (1999).
%
\bibitem{Wir02} R.B.\ Wiringa and S.C.\ Pieper,
                Phys.\ Rev.\ Lett.\  {\bf 89}, 182501 (2002). 
%
\bibitem{Pud97} B.S.\ Pudliner, V.R.\ Pandharipande, J.\ Carlson, 
		S.C.\ Pieper, and R.B.\ Wiringa,
                Phys.\ Rev.\ C {\bf 56}, 1720 (1997).
%
\bibitem{Wir91} R.B.\ Wiringa,
                Phys.\ Rev.\ C {\bf 43}, 1585 (1991).
%
\bibitem{tunl}  D.R.\ Tilley, H.R.\ Weller, and H.H.\ Hasan,
                Nucl. Phys. {\bf A474}, 1 (1987).
%
\bibitem{Col65} H.\ Collard {\it et al.}, 
                Phys.\ Rev.\ {\bf 138}, B57 (1965).
%
\bibitem{McC77} J.S.\ McCarthy, I.\ Sick, and R.\ Whitney, 
                Phys.\ Rev.\ C {\bf 15}, 1396 (1977).
%
\bibitem{Sza77} Z.M.\ Szalata {\it et al.},
                Phys.\ Rev.\ C {\bf 15}, 1200 (1977).
%
\bibitem{Arn78} R.G.\ Arnold {\it et al.},
                Phys.\ Rev.\ Lett.\ {\bf 40}, 1429 (1978).
%
\bibitem{Dun83} P.C.\ Dunn {\it et al.}, 
                Phys.\ Rev.\ C {\bf 27}, 71 (1983).
%
\bibitem{Ott85} C.R.\ Ottermann {\it et al.}, 
                Nucl.\ Phys.\ {\bf A435}, 688 (1985).
%
\bibitem{Jus85} F.P.\ Juster {\it et al.},
                Phys.\ Rev.\ lett.\ {\bf 55}, 2261 (1985).
%
\bibitem{Bec87} D.H.\ Beck {\it et al.},
                Phys.\ Rev.\ Lett. {\bf 59}, 1537 (1987).
%
\bibitem{Amr94} A.\ Amroun {\it et al.}, 
                Nucl.\ Phys.\ {\bf A579}, 596 (1994).
%
\bibitem{Mug81} S.F.\ Mughabghab, M.\ Divadeenam, and N.E.\ Holden, 
                {\it Neutron Cross Sections from Neutron Resonance Parameters 
                and Thermal Cross Sections} (Academic Press, London, 1981), 
                http://isotopes.lbl.gov/ngdata/sig.htm.
%
\bibitem{Bis50} G.R.\ Bishop {\it et al.}, 
                Phys.\ Rev.\ {\bf 80}, 211 (1950).
%
\bibitem{Sne50} A.H.\ Snell, E.C.\ Barker, and R.L.\ Sternberg,
                Phys.\ Rev.\ {\bf 80}, 637 (1950).
%
\bibitem{Col51} S.A.\ Colgate, Phys.\ Rev.\ {\bf 83}, 1262 (1951).
%
\bibitem{Car51} J.H.\ Carver {\it et al.}, 
                Nature (london) {\bf 167}, 154 (1951).
%
\bibitem{Bir85} Y.\ Birenbaum, S.\ Kahane, and R.\ Moreh, 
                Phys.\ Rev.\ C {\bf 32}, 1825 (1985).
%
\bibitem{Mor89} R.\ Moreh, T.J.\ Kennett, and W.V.\ Prestwich,
                Phys.\ Rev.\ C {\bf 39}, 1247 (1989).
%
\bibitem{DeG92} A.\ De Graeve {\it et al.},
                Phys.\ Rev.\ C {\bf 45}, 860 (1992).
%
\bibitem{JBB82} E. T. Jurney, P. J. Bendt, and J. C. Browne,
                Phys.\ Rev.\ C {\bf 25}, 2810 (1982).
%
\bibitem{Kea88} M. W. Konijnenberg {\it et al.},
                Phys.\ Lett.\ B {\bf 205}, 215 (1988).
%
\bibitem{mitev86} G. Mitev, P. Colby, N. R. Roberson, H. R. Weller, and
                D. R. Tilley,
                Phys.\ Rev.\ C {\bf 34}, 389 (1986);
                Helv.\ Phys.\ Acta\ {\bf 48}, 753 (1965).
%
\bibitem{wolfli66} W. W\"olfli, R. B. B\"osch, J. Lang, and R. M\"uller,
        Phys.\ Lett.\ {\bf 22}, 75 (1966);
        W. W\"olfli, R. B. B\"osch, J. Lang, R. M\"uller, and
         P. Marmier,
        Helv.\ Phys.\ Acta\ {\bf 40}, 946 (1967).
%
\bibitem{Bel70} B.D.\ Belt, C.R.\ Bingham, 
                M.L.\ Halbert, and A.\ van der Woude,
		Phys.\ Rev.\ Lett.\ {\bf 24}, 1120 (1970).
%
\bibitem{kundu71} S. K. Kundu, Y. M. Shin, and G. D. Wait,
                  Nucl.\ Phys.\ {\bf A171}, 384 (1971).
%
\bibitem{matt74} J. L. Matthews, T. Kruse, M. E. Williams, R. O.
                 Owens, and W. Savin,
                 Nucl.\ Phys.\ {\bf A223}, 221 (1974).
%
\bibitem{skopik79} D. M. Skopik, H. R. Weller, N. R. Roberson, and
                 S. A. Wender,
                 Phys.\ Rev.\ C {\bf 19}, 601 (1979).
%
\bibitem{Ang83} M.\ Anghinolfi, P.\ Corvisiero, M.\ Guarnone, 
                G.\ Ricco, and A.\ Zucchiati,
                Nucl.\ Phys.\ {\bf A410}, 173 (1983). 
%
\bibitem{skopik83} D. M. Skopik {\it et al.},
                 Phys.\ Rev.\ C {\bf 28}, 52 (1983).
%
\bibitem{king84} S. King, N. R. Roberson, H. R. Weller, and
                 D. R. Tilley,
                 Phys.\ Rev.\ C {\bf 30}, 21 (1984).
%
\bibitem{bosch64} R. B\"osch {\it et al.},
                Phys.\ Lett.\ {\bf 8}, 120 (1964).
%
\bibitem{faul81} D. Faul {\it et al.},
                 Phys.\ Rev.\ C {\bf 24}, 849 (1981).
%
\bibitem{skopik81} D. M. Skopik {\it et al.},
                 Phys.\ Rev.\ C {\bf 24}, 1791 (1981).
%
\bibitem{koseik66} R. Koseik {\it et al.},
                Phys.\ Lett.\ {\bf 21}, 199 (1966).
%
\bibitem{pfeiffer68} R. Pfeiffer,
                Z.\ Phys.\ {\bf 208}, 129 (1968).
%
\bibitem{Mac87} R. Machleidt, K. Holinde, and Ch. Elster,
                Phys.\ Rep.\ {\bf 149}, 1 (1987).
%
\bibitem{Gri63} G.M.\ Griffiths, M.\ Lal, and C.D.\ Scarfe, 
                Can.\ J.\ Phys.\ {\bf 41}, 724 (1963).
%
\bibitem{Sch95} G.J.\ Schmid {\it et al.}, 
                Phys.\ Rev.\ C\ {\bf 52}, R1732 (1995).
%
\bibitem{Sch96} G.J.\ Schmid {\it et al.}, 
                Phys.\ Rev.\ Lett.\ {\bf 76}, 3088 (1996).
%
\bibitem{Aki01} H.\ Akiyoshi {\it et al.}, 
                Phys.\ Rev.\ C {\bf 64}, 034001 (2001).
%
\bibitem{Jou86} J.\ Jourdan {\it et al.},
                Nucl.\ Phys.\ {\bf A453}, 220 (1986).
%
\bibitem{Joupc} J.\ Jourdan, private communications.
%
\bibitem{Ank98} H.\ Anklin {\it et al.},
                Nucl.\ Phys.\ {\bf A636}, 189 (1998).
%
%
\end{thebibliography}
\end{document}